\documentclass[11pt]{article}
\usepackage{xr}
\usepackage{tikz}
\usetikzlibrary{positioning}
  \usepackage{pgf}
  \usepackage{colortbl}
\usepackage{amsmath,amssymb}
\usepackage{float}
\usepackage{bbm}
\usepackage{xcolor}
\usepackage{enumerate}
\usepackage[authoryear]{natbib}
\usetikzlibrary{arrows,positioning,automata,calc,fit,shapes.geometric,arrows.meta}
\usepackage{geometry}   

\usepackage{rotating,multirow,makecell,array}
\usepackage{caption}
\usepackage{subcaption}
 \usepackage{bbm}

\usepackage{subfiles}
\usepackage{scalerel,stackengine}
\usepackage{lipsum}
\stackMath
\newcommand\reallywidehat[1]{%
\savestack{\tmpbox}{\stretchto{%
  \scaleto{%
    \scalerel*[\widthof{\ensuremath{#1}}]{\kern-.6pt\bigwedge\kern-.6pt}%
    {\rule[-\textheight/2]{1ex}{\textheight}}
  }{\textheight}%
}{0.5ex}}%
\stackon[1pt]{#1}{\tmpbox}%
}
\usepackage[english]{babel}
\usepackage{amsthm}
\usepackage{bbm}
\usepackage{blindtext}
\usepackage{dsfont}
\usepackage{algorithm}
\usepackage{algpseudocode}
\usepackage{amsmath,amssymb}
\usepackage{multirow,makecell}

\usepackage[utf8x]{inputenc}

\usepackage{tabstackengine}

\makeatletter
\def\BState{\State\hskip-\ALG@thistlm}
\makeatother
\usepackage{amssymb}
\usepackage[inline]{enumitem}
\usepackage{url} 

\newcommand{\norma}[1]{\left\lVert#1\right\rVert}

\usepackage[pdftex,colorlinks]{hyperref}

  \hypersetup{
colorlinks =true,
citecolor    = blue,
citebordercolor = violet,
filebordercolor=red,
linkbordercolor=blue
}
\usepackage{authblk}
\usetikzlibrary{shapes.geometric}

\usetikzlibrary{shapes,snakes}
\makeatletter
\newcommand*{\addFileDependency}[1]{
  \typeout{(#1)}
  \@addtofilelist{#1}
  \IfFileExists{#1}{}{\typeout{No file #1.}}
}
\makeatother
 
\def\spacingset#1{\renewcommand{\baselinestretch}%
{#1}\small\normalsize} \spacingset{1}
\newcommand{\eqn}{\begin{eqnarray}}
\newcommand{\ee}{\end{eqnarray}}
\newcommand{\eqnn}{\begin{eqnarray*}}
\newcommand{\een}{\end{eqnarray*}}
\newcommand{\ea}{\end{align}}
\newcommand{\be}{\begin{eqnarray}}

\newcommand{\ba}{\begin{align}}
\newcommand{\expit}{\text{expit}}

\def\bSig\mathbf{\Sigma}

\DeclareMathOperator{\logit}{logit}

\title{Doubly Robust Estimation of the Hazard Difference for Competing Risks Data}{}

\author
{
Denise Rava, drava@ucsd.edu \\
Department of Mathematics, University of California, San Diego
\and
Ronghui Xu, rxu@health.ucsd.edu \\
Department of Mathematics and Herbert Wertheim School of Public Health\\
 University of California, San Diego
}

\date{}

\begin{document}

\maketitle

\begin{abstract}
We consider the conditional treatment effect for competing risks data in observational studies. While it is described as a constant difference between the hazard functions given the covariates, we do not assume specific functional forms  for the covariates. We derive the efficient score for the treatment effect using modern semiparametric theory, as well as two doubly robust scores with respect to 1) the assumed propensity score for treatment and the censoring model, and 2) the outcome models for the competing risks. An important asymptotic result regarding  the estimators is rate double robustness, in addition to the classical model double robustness. 
Rate double robustness enables the use of machine learning and nonparametric methods in order to estimate the nuisance parameters, while preserving the root-$n$ asymptotic normality of the estimators for inferential purposes. We study the performance of the estimators using simulation. The estimators are applied to the data from a cohort of Japanese men in Hawaii followed since 1960s in order to study the effect of mid-life drinking behavior on late life cognitive outcomes. 

\end{abstract}

%



%
\newpage
\section{Introduction}
\label{s:intro}

Our work was motivated by data from the linked epidemiological projects Honolulu Hearth Program (HHP) and Honolulu-Asia Aging Study (HAAS). HHP was established in 1965 as an epidemiological study of rates and risk factors for heart disease and stroke in men of Japanese ancestry living in Oahu and born between 1900 and 1919. 
HAAS was established in 1991 
 as a continuation of the HHP with a shift of focus on brain aging, Alzheimer's disease, vascular dementia, other causes of cognitive and motor impairment, stroke, and the common chronic conditions of late life. 
During  the HAAS period  neuropsychological assessments were performed every 2-3 years until 2012. 
In particular, we are interested in the 
 effect of mid-life exposures captured during HHP on late life cognitive outcomes collected in HAAS. 
 For this cohort of participants then, death presents a competing risk for the cognitive outcomes. 
 This is commonly referred as truncation by death when the cognitive outcomes are analyzed as longitudinal repeated measures \citep{ding2011identifiability,tchetgen2014identification,estes2016time,yang2016using}, and competing risk when the cognitive outcomes are time to events, which is the case here.

In observational studies such as the above, it is necessary to control for confounding in order to study the causal effects of exposures such as alcohol. One way to control for confounding is through including the covariates in a regression model. This leads to the so-called  conditional treatment effect. 
For the analysis of time-to-event data, 
 the additive hazards model \citep{Aalen,aalen1989linear} 
  has recently been considered in this context \citep{Dukes:19, hou2019estimating}.
 For a binary treatment, this conditional treatment effect is the hazard difference given the covariates under the additive hazards model. 
Note that misspecification of the functional form of the covariates in the hazard regression model can lead to  bias in the estimation of the treatment effect of interest.

To alleviate the reliance on the correct specification of the covariate forms which are  `nuisance' themselves, flexible modeling  such as nonparametric approaches might be considered. However, they are often inefficient and lead to slower rates of convergence of the estimated treatment effect; this is  the `curse of dimensionality' problem discussed in \cite{robins:ritov:coda}. 
 Alternatively, 
 there has been a growing literature on doubly robust estimators that protect against misspecification of the `nuisance' parts of the model \citep{RobinsRotnitzky95,robins1995analysis,scharfstein1999adjusting, robins2000marginal,robins2000marginalsnm,robins2000profile,van2003unified,RobinsRotnitzky01,bang2005doubly,tsiatis2007semiparametric,tchetgen2010doubly}. 
In the survival context \cite{zhang2012contrasting}, \cite{bai2017optimal} and \cite{sjolander2017doubly} derived doubly robust estimators for the treatment effect defined as, or equivalent to, a contrast between expectations of functions of the potential failure times, 
 i.e. the failure time that would be observed if the same subject were treated or untreated, respectively, regardless of the actual treatment received. 
 \cite{yang2020semiparametric} developed  a doubly robust estimator for the structural accelerated failure time models.
\cite{petersen2014targeted} and \cite{zheng2016doubly} derived targeted maximum likelihood estimators that are doubly robust 
after discretizing  time and recasting the failure event as a binary outcome. 



In the absence of competing risks, doubly robust estimators for the hazard difference have been proposed by \cite{Dukes:19} and \cite{hou2019estimating}.
\cite{Dukes:19} considered low dimensional setting, i.e.~all the nuisance parameters are estimated parametrically or semiparametrically at root-$n$ rate, but the treatment might be continuous.  \cite{hou2019estimating} considered high dimensional setting, using regularization methods with LASSO as a specific case for their theoretical as well as empirical investigation.

In the following we first derive the semiparametrically efficient score for the cause-specific hazard difference under competing risks. We then propose two doubly robust estimators 
with respect to two sets of models. The first set contains the treatment assignment model, also called the propensity score model,  and the model for the censoring distribution.  The second set contains the cause-specific hazard models for the competing risks. 
The proposed estimators are both model  doubly robust and rate doubly robust. 
Model doubly robust refers to the property that the estimators of treatment effects are consistent and asymptotically normal, as long as any 
 one of the two sets of the models are correctly specified, and that the correctly specified models are estimated at root-$n$ rate. This is doubly robust in the classical sense.
 
 Rate doubly robust refers to the property that the estimators of treatment effects are consistent and asymptotically normal, when both sets of the models estimate the truth but at possibly slower than root-$n$ rate, as long as their product rate is faster than root-$n$. Rate doubly robust property enables the use of modern machine learning or other nonparametric methods, which substantially broadens the range of estimators to be used for the  nuisance parameters. With these methods model specification might become much less an issue than previously, so that one might be less concerned about having at least one of the two sets of models correctly specified. 
We note that  rate double robustness  was  not considered in  \cite{Dukes:19}.
 Meanwhile, although rate double robustness was established in \cite{hou2019estimating} for the LASSO estimators, our results in this paper can be much more broadly applied to potentially many machine learning and other nonparametric methods.
In the process we also weaken the censoring assumption as required in \cite{Dukes:19} and \cite{hou2019estimating}.

The rest of the paper is organized as follows: after formally defining the parameter of interest, in Section \ref{eff} we derive the semiparametrically efficient and Section \ref{drs} the two doubly robust scores. In Section \ref{inference} we describe the implementation of the doubly robust estimators, and
  derive the asymptotic distribution of the estimated treatment effects.
   We study the finite sample performance of the proposed estimators through extensive simulations in Section \ref{simulation}, and  apply them to the HHP-HAAS data in order to estimate the effect of alcohol exposure on  cognitive impairment in Section \ref{application}. We conclude with discussion in the last section.



 \newtheorem{theorem}{Theorem}
 \newtheorem{lemma}{Lemma}
  \newtheorem{assumption}{Assumption}

\subsection{Model and Notation}
\label{s:model}

Denote $T$ 
time to failure, and  $\epsilon=1,\ldots,J$  the type of failure. Let
 $C$ be the censoring random variable,   $X=\min(T,C)$ be the observed (and possibly censored) failure time, and $\delta=\mathbbm{1}\{T \leq C\}$  the event indicator. 
 Let $A= 0$, 1 be a binary treatment, and $Z$ be a vector of baseline covariates. 
 We assume $\tau<\infty$ to be an upper limit of follow-up time.
 
 A commonly used approach for competing risks data is to model the cause-specific hazard function for each type of failure \citep{holt1978competing, 
 kalbfleisch2011statistical}. 
The cause-specific hazard functions are the quantities `just identified' by such data, in the sense that any other quantity that can be identified from the competing risks data, can be expressed as a function of the cause-specific hazard \citep{kalbfleisch2011statistical}.
 We assume that the conditional cause-specific hazard function, $h_j(t | A, Z)=\lim_{\Delta t \rightarrow 0+ }\frac{1}{\Delta t}P(t\leq T < t+\Delta t, \epsilon=j | T\geq t, A, Z)$,
  for $j=1,\ldots,J$, 
 satisfies:  
\be\label{model}
h_j(t | A, Z)= \beta_j A + \lambda_{j}(t , Z), 
\ee
where $\lambda_{j}(t , Z)$, representing the effect of the covariates on the hazard, is left unspecified.
This 
is a key difference from the more traditional cause-specific additive hazards model that assumes linear effects of  both $A$ and $Z$; see for example, \cite{shen1999confidence}. 
From model \eqref{model} then, 
$
\beta_j= h_j(t | A=1, Z)- h_j(t | A=0, Z)
$
is the difference between the conditional cause-specific hazard functions of the two treatment groups.

In the following we assume that $C \perp T  | (A,Z)$, where `$\perp$'  indicates statistical independence. This is a standard assumption  in the analysis of time-to-event data, and it relaxes the stricter $C \perp (A, T)  | Z$ assumption imposed by both \cite{hou2019estimating} and \cite{Dukes:19}.
We will also use the counting process and the at-risk process notation: $N_{j}(t)=\mathbbm{1}\left\{X\leq t, \; \delta=1,\;\epsilon=j\right\}$ and $Y(t)=\mathbbm{1}\left\{X\geq t\right\}$.
Under model \eqref{model}, $M_{j}(t)=N_{j}(t)-  \int_0^t Y(u) h_j(u | A, Z)du$ is a 
local square-integrable martingale with respect to the filtration \\$\mathcal{F}_t=\sigma\left\{N_{j}(s), Y(s+), A, Z : j=1,\ldots,J, \; 
0<s<t\right\}$.
In addition, the predictable covariation process $< M_l, M_j>(t) =0$ for $l \neq j$ because
with absolutely continuous distributions of the event times, 
the probability that the competing events happen at the same time equals zero.

\section{Semiparametrically efficient score for $\beta$}\label{eff}


The derivation below follows the modern semiparametric theory as described in \cite{tsiatis2007semiparametric}. We provide a sketch for the readers here and leave all the details  
to the Supplementary Materials.

Under the semiparametric  model \eqref{model}, 
the parameter of interest is $\beta=[\beta_1,\ldots,\beta_J]^\top$ and the nuisance parameter is $\eta=[\lambda_1(t, z), \ldots, \lambda_J(t, z), \lambda_{c}(t | a,z), P(a|z), f(z)]^\top$, where $\lambda_c(t | a,z)$ is the conditional hazard function for $C$ given $A$ and $Z$, $P(a| z)$ is the conditional probability of $A$ given $Z$, and $f(z)$ is the density or probability function of the covariates $Z$.
The likelihood for a single copy of the data takes  the following form:
\eqnn
L&=&\prod_{j=1}^J\left\{\beta_j A + \lambda_j(X,Z)\right\}^{\mathbbm{1}\{\delta=1, \epsilon=j\}}\exp\left\{-\beta_j At-\Lambda_j(X,Z)\right\} \\
&& \times \left\{\lambda_c(X | A,Z)\right\}^{1-\delta}\exp\left\{-\Lambda_c(X | A,Z)\right\}
P(A|Z) f(Z),
\een
where $\Lambda_j(t,z)=\int_0^t\lambda_j(u,z)du$ for $j=1,\ldots,J$ and $\Lambda_c(t| a,z)=\int_0^t\lambda_c(u| a,z)du$.
From the likelihood, one can derive the score for the parameter of interest.
In the Supplementary Materials we show that under model \eqref{model}, 
\be\label{eq:Sbeta}
S_\beta = \frac{\partial \log L }{ \partial\beta} =\left\{ \int_0^\tau \frac{A \cdot {dM_j(t)} }{ h_j(t | A,Z)} \right\}_{j=1}^J.
\ee
In addition, if $\eta$ were finite dimensional, the score for the nuisance parameter would be $S_\eta=\partial \log L /\partial \eta$. The nuisance tangent space, denoted by $\mathbf\Lambda$,  is the space spanned by the nuisance score. When $\eta$ has infinite dimension, as in our case, the notion of nuisance tangent space can be extended through the definition of parametric submodels. We leave the technicality of this definition to Chapter 4 of \cite{tsiatis2007semiparametric}. 

An estimator $\hat \beta$ is asymptotically linear if there exists a function of the data $\varphi$, such that $\sqrt{n}(\hat \beta-\beta_0)= \sum_{i=1}^n\varphi_i / \sqrt{n} +o_p(1)$. The function $\varphi$, named influence function, has mean zero and finite variance, thus guarantees the asymptotic normality of the estimator $\hat \beta$. Such estimators are therefore desirable and they are uniquely defined by their influence functions. 
Every influence function belongs to the orthogonal complement of the nuisance tangent space \citep[Theorem 4.2]{tsiatis2007semiparametric}. 
This space, denoted by $\mathbf\Lambda^\perp$, is therefore the starting point to define semiparametric estimators for $\beta$ that are consistent and asymptotically normal.

The space $\mathbf\Lambda^\perp$ is also important because it allows one to find orthogonal scores.
 A score $\psi(\beta, \eta)$ is orthogonal if 
\be\label{eq:ortho}
\left. \frac{\partial}{\partial r}{E}\left\{\psi(\beta_0 ; \eta_0 + r(\eta-\eta_0))\right\} \right|_{r=0}=0,
\ee
where we use the subscript `0' to indicate the true value of the parameters.
Orthogonal scores are invariant to small perturbations of the nuisance parameter around the truth and so the estimation of the nuisance parameter may not greatly affect the estimation of the treatment effect \citep{bickel1993efficient,newey1990semiparametric,newey1994asymptotic}. 
It is shown in the Supplementary Materials (Lemma \ref{lemmaorth}) that an estimating function belongs to $\mathbf\Lambda^\perp$ if and only if it is an orthogonal score.

The following lemma gives the form of the orthogonal complement of the nuisance tangent space under model \eqref{model}.
\begin{lemma}\label{ortlemma}
Under model \eqref{model}, the orthogonal complement of the nuisance tangent space takes the following form:
\be\label{ortcm}
\mathbf\Lambda^\perp&=&\left\{\sum_{j=1}^J\int_0^\tau \left[g_j(t,A,Z)-\frac{{E}\left\{g_j(t,A,Z)h_j^{-1}(t | A,Z)S_{c}(t | A,Z) e^{-\sum_{l=1}^J\beta_{l}At} | Z\right\}}{{E}\left\{h_j^{-1}(t | A,Z)S_{c}(t | A,Z) e^{-\sum_{l=1}^J\beta_{l}At} | Z\right\}}\right] \right.\nonumber\\
&&\left. \times  \frac{dM_j(t)}{h_j(t | A,Z)}: \;\;\:\mbox{for\;all }\;g_j(t,A,Z), \  j=1, ..., J
\right\}.
\ee
\end{lemma}
The proofs of all results can be found in the Supplementary Materials.

Among all the semiparametric asymptotically linear estimators of $\beta$,
the efficient score is $S_\beta- \Pi\{S_\beta | \mathbf\Lambda\} = \Pi\{S_\beta | \mathbf\Lambda^\perp\}$ \citep{tsiatis2007semiparametric}, where $\Pi\{S_\beta | \mathbf\Lambda \}$ and  $\Pi\{S_\beta | \mathbf\Lambda^\perp\}$ are the projections of  $S_\beta$ onto $\mathbf\Lambda$ and $\mathbf\Lambda^\perp$, respectively.
 From \eqref{eq:Sbeta} it can be seen that this would be the element of $\mathbf\Lambda^\perp$ that corresponds to $g_j(t,A,Z)=Ae_j$, 
where $e_j$ is a vector with 1 at the $j^{th}$ position and 0 elsewhere, $j=1, ..., J$. 
Therefore we have the following result.
\begin{theorem}\label{efflemma}
Under model \eqref{model} the efficient score has the following form:
\be\label{efsc}
S_{eff} (\beta) &=&\left\{\int_0^\tau \left[A -\frac{{E}\left\{A
h_j^{-1}(t | A,Z)S_{c}(t | A,Z) e^{-\sum_{l=1}^J\beta_{l}At} | Z\right\}}
{{E}\left\{ h_j^{-1}(t | A, Z)S_{c}(t | A,Z) e^{-\sum_{l=1}^J\beta_{l}At} | Z\right\}}\right]\frac{dM_j(t)}
{h_j(t | A,Z)}\right\}_{j=1}^J.
\ee
\end{theorem}
\underline{\it Remark 1}: If we make the stronger assumption of $C \perp (A,T) | Z$ as in \cite{Dukes:19} and \cite{hou2019estimating}, $S_{c}(t | A,Z)=S_{c}(t | Z)$, and so the efficient score simplifies to:
\eqnn
S_{eff} (\beta) &=&\left\{\int_0^\tau \left[A -\frac{E\left\{Ah_j^{-1}(t | A,Z)e^{-\sum_{l=1}^J\beta_{l}At} | Z\right\}}{E\left\{h_j^{-1}(t | A, Z)e^{-\sum_{l=1}^J\beta_{l}At} | Z\right\}}\right]\frac{dM_j(t)}{h_j(t | A,Z)}\right\}_{j=1}^J.
\een
In this case, $S_c$ is therefore no longer needed for the estimation of $\beta$. 
If further $J=1$, the efficient score in \cite{Dukes:19} is recovered. 

\noindent\underline{\it Remark 2}: Traditionally, estimation of parameters from competing risks data allows estimating the parameters for one type of failure at a time.
 This is the case  for the widely used cause-specific Cox model and the traditional additive cause-specific hazards model \cite[]{shen1999confidence}. 
 However, for both the efficient score and the doubly robust scores below, 
the components of $\beta$ are estimated jointly from a multi-dimensional score.

The above score is locally efficient in the sense that its asymptotic variance attains the semiparametric efficency bound when $\lambda_j(t,z)$, $S_c(t | a, z)$ and $P(a | z)$ are known or correctly estimated  \citep[Theorem 4.1]{tsiatis2007semiparametric}.
Unfortunately, since $h_j(t | A, Z)$ in \eqref{efsc} is unknown and estimators for it are not readily available, the efficient score may not be directly used in practice. 
We will however exploit both \eqref{ortcm} and \eqref{efsc} to derive two doubly robust scores for the estimation of $\beta$.

\section{Doubly robust scores}\label{drs}

\subsection*{Doubly robust Score 1}

Denote $\pi(Z)=P(A=1 | Z)$. 
Inspired by \cite{hou2019estimating}, we choose in \eqref{ortcm}
$g_j(t,A,Z)= e^{\sum_{l=1}^J\beta_{l}At} S^{-1}_{c}(t | A,Z) \left\{A-\pi(Z)\right\} h_j(t | A,Z)$,
 and obtain
\be\label{MarquisScore}
{S}_1(\beta ; S_c, \pi, \Lambda)=\left\{\int_0^\tau e^{\sum_{l=1}^J \beta_l A t}S^{-1}_c(t | A,Z)\left\{A -\pi( Z)\right\}dM_j(t; \beta_j, \Lambda_j) \right\}_{j=1}^J, 
\ee
where $ \Lambda = (\Lambda_1, ..., \Lambda_J)^\top$, and  $dM_j(t ; \beta_j, \Lambda_j)= dN_j(t)-Y(t)\beta_j A dt -Y(t) d\Lambda_j(t,Z)$.
We note that $M_j(t ; \beta_0,\Lambda_0) $  is a martingale under model \eqref{model}, where again the subscript `0'  indicates the true value.
The main difference between \eqref{MarquisScore} when $J=1$ and the score from \cite{hou2019estimating} is 
the incorporation of the censoring distribution $S_c$, so that we do not need the stronger assumption $C \perp (T,A) | Z$.
We note also that \cite{hou2019estimating} 
directly constructed their score  as a member of  $\mathbf \Lambda^\perp$ using definition \eqref{eq:ortho}.

\subsection*{Doubly robust Score 2}

The second approach removes the unknown hazard weights  from the efficient score \eqref{efsc}, 
as done in 
\cite{ly} for the additive hazards regression model, and we have
\be\label{score11}
{S}_2(\beta ; S_c, \pi, \Lambda)=\left\{\int_0^\tau \left\{A -{\cal E}(t; \beta, S_c,\pi)\right\}dM_j(t ; \beta_j, \Lambda_j)\right\}_{j=1}^J,
\ee
where:
\eqnn
{\cal E}(t; \beta, S_c, \pi)&=&\frac{{E}\left[A e^{-\sum_{j=1}^J\beta_jAt}S_c(t | A,Z) | Z \right]}{{E}\left[e^{-\sum_{j=1}^J\beta_jAt}S_c(t | A,Z ) | Z\right]}\\&=&\frac{e^{-\sum_{j=1}^J\beta_jt}S_c(t | A=1,Z)\pi(Z)}{e^{-\sum_{j=1}^J\beta_jt}S_c(t | A=1,Z)\pi(Z )+S_c(t | A=0,Z)\left\{1-\pi(Z)\right\}}.
\een
We note that Score 2 in \eqref{score11} is completely new to our best knowledge, 
even in the absence of competing risks.

Since both scores \eqref{MarquisScore} and \eqref{score11} belong to $\mathbf\Lambda^\perp$, they are orthogonal scores. 
In addition,  they are doubly robust with respect to the estimation of both $S_c$ and $\pi$,  and that of $\Lambda$, as stated in the theorem below. 
\begin{theorem}\label{dr2}
$
{E}\left\{{S}_{1}(\beta_0; S_c, \pi, \Lambda)\right\}= 
{E}\left\{{S}_{2}(\beta_0 ; S_c, \pi, \Lambda)\right\}=0$
if either 
$S_c = S_{c0} $ and $\pi = \pi_0 $, or $\Lambda = \Lambda_{0} $, 
where  subscript `0'  indicates the true quantities.
\end{theorem}


Finally, if we are willing to make the stronger assumption $C \perp (T,A) | Z$, the above two scores simplify to:
\be\label{score1nc}
\tilde{S}_1(\beta ;  \pi, \Lambda)&=&\left\{\int_0^\tau e^{\sum_{j=1}^J \beta_j A t}\left\{A -\pi( Z)\right\}dM_j(t; \beta, \Lambda) \right\}_{j=1}^J,
\\
\label{score2nc}
\tilde{S}_2(\beta ;  \pi, \Lambda)&=&\left\{\int_0^\tau \left\{A -\frac{e^{-\sum_{j=1}^J\beta_jt}\pi(Z)}{e^{-\sum_{j=1}^J\beta_jt}\pi(Z )+\left\{1-\pi(Z)\right\}}\right\}dM_j(t ; \beta, \Lambda)\right\}_{j=1}^J,
\ee
where $S_c$ is no longer involved. We will consider the implementation of these two simplified scores in the simulation below as well. 

\section{Estimation and inference}\label{inference}

Given a random sample of size $n$ we write
\be\label{score2}
{S}_{1,n}(\beta ; S_c, \pi, \Lambda):=\frac{1}{n}\sum_{i=1}^n{S}_{1i}(\beta;  S_c, \pi, \Lambda),
\ee
and
\be\label{score1}
{S}_{2,n}(\beta ; S_c, \pi, \Lambda):=\frac{1}{n}\sum_{i=1}^n{S}_{2i}(\beta ; S_c, \pi, \Lambda).
\ee
Both  \eqref{score2} and \eqref{score1} depend on the quantities $S_c, \pi$ and $\Lambda$ that need to be estimated. 

For estimation of the propensity score $\pi(\cdot)$ and the censoring model $S_c(\cdot| \cdot,\cdot)$,  we leave it to the users to choose any working model as long as some mild  assumptions, given later, are satisfied.
From here on we use $\hat S_c, \hat \pi$ and $\hat \Lambda$ to denote  estimators of the nuisance parameters $S_c, \pi$ and $\Lambda$; 
note that the estimator  for $\Lambda$ may also depend on $\beta$ as described below.

For the estimation of $\Lambda$  we consider here the usual linear working models: 
$ \Lambda_{j}(t,Z; G_j, \gamma_j)=G_j(t)+\gamma_j^\top Zt$, $j=1, ..., J$.
The parameters $\gamma= (\gamma_1,\ldots,\gamma_J)^\top$ and $G= (G_1,\ldots,G_J)^\top$ can be estimated using the approach of 
 \cite{shen1999confidence}, which is equivalent to applying the estimating equations of \cite{ly} separately to each failure type:
\be\label{tradgamma}
\hat \gamma_j=\left[\sum_{i=1}^n \int_0^\tau Y_i(t)\{Z_i-\bar Z(t)\}^{\otimes 2}dt\right]^{-1}\left[\sum_{i=1}^n \int_0^\tau Y_i(t)\{Z_i-\bar Z(t)\}dN_{ji}(t)\right],
\ee
where $\bar{Z}(t)=\left\{\sum_{i=1}^n Y_i(t)\right\}^{-1}\sum_{i=1}^n Y_i(t)Z_i$, $Z^{\otimes 2}=Z Z^\top$, and 
\be\label{estbas1}
\hat G_j(t; \beta_j, \gamma_j)=\int_0^t\frac{\sum_{i=1}^n \left\{dN_{ji}(u)-Y_i(u)\beta_jA_idu-Y_i(u)\gamma_j^\top Z_idu\right\}}{\sum_{i=1}^nY_i(u)}.
\ee

For the estimation of $G_j$, following \cite{hou2019estimating} we consider also the weighted  Breslow estimator:
\be\label{estbas2}
\tilde G_j(t; \beta_j, \gamma_j, S_c,\pi)=\int_0^t\frac{\sum_{i=1}^n w_i(S_c,\pi)\left\{dN_{ji}(u)-Y_i(u)\beta_jA_idu-Y_i(u)\gamma_j^\top Z_idu\right\}}{\sum_{i=1}^n w_i(S_c,\pi)Y_i(u)},
\ee
where
$
w_i(S_c,\pi)=A_i\left\{1-\pi(Z_i)\right\} S_c^{-1}(u | A_i,Z_i).
$
The advantage of using \eqref{estbas2} is that it leads to 
the  closed-form solution to $S_{1,n}$ for $j=1, ..., J$: 
\be\label{score1s}
\hat\beta^{(1)}_j = - 
\left\{ \sum_{i=1}^n \int_0^\tau \hat S_c^{-1}(t | A_i,Z_i)(1-A_i)\hat \pi(Z_i)Y_i(t)dt \right\}^{-1} 
\times \sum_{i=1}^n\int_0^\tau \hat S_c^{-1}(t | A_i,Z_i) \nonumber\\
\cdot(1-A_i)\hat \pi(Z_i) 
\cdot\left(dN_{ji}(t)-Y_i(t)\left[\hat \gamma_j^\top \left\{Z_i-\bar{Z}(t; \hat S_c, \hat \pi)\right\}dt+d\bar{N}_j(t; \hat S_c, \hat \pi)\right]\right),
\ee
where
\eqnn
\bar{Z}(t; \hat S_c, \hat \pi)=\frac{\sum_{i=1}^nY_i(t)Z_iw_i(\hat S_c, \hat \pi)}{\sum_{i=1}^nY_i(t)w_i(\hat S_c, \hat \pi)},\quad
d\bar{N}_j(t; \hat S_c, \hat \pi)=\frac{\sum_{i=1}^ndN_{ji}(t)w_i(\hat S_c, \hat \pi)}{\sum_{i=1}^nY_i(t)w_i(\hat S_c, \hat \pi)}.
\een
We show below that the asymptotic distribution of the solution to $S_{1,n}$ 
does not depend on the specific estimator of $\Lambda$, as long as  the nuisance parameters are consistently estimated at a certain rate. 
Therefore, the choice of \eqref{estbas2} is purely due to its numerical advantage.

For $S_{2,n}$, 
using \eqref{estbas1} 
and after some  algebra, we have:
\be\label{score2s}
{S}_{2,n}(\beta; \hat S_c, \hat \pi,\hat  \Lambda)&=&\left\{\frac{1}{n}\sum_{i=1}^n\int_0^\tau \left\{{A}_i-{\cal E}_{i}(t; \beta, \hat S_c, \hat \pi)-\bar{A}(t)+\bar{\cal E}(t) \right\}\right.\\&&\left. \cdot \left\{dN_{ji}(t)-Y_i(t)\left(\beta_j A_i+\hat \gamma_j^\top Z_i\right)dt\right\}\right\}_{j=1}^J,\nonumber
\ee
where
\eqnn
\bar{A}(t)=\frac{\sum_{i=1}^nY_i(t)A_i}{\sum_{i=1}^nY_i(t)},\quad
\bar{\cal E}(t)=\frac{\sum_{i=1}^nY_i(t){\cal E}_{i}(t; \beta, \hat S_c, \hat \pi)}{\sum_{i=1}^nY_i(t)}.
\een
Once the estimators $\hat \pi, \hat S_c$ and $\hat \Lambda$ are available, we define 
$\hat \beta^{(2)}$ to be the root of ${S}_{2,n}(\beta ; \hat S_c, \hat\pi, \hat\Lambda)$.
 
We  study the asymptotic properties of these estimators below. For ease of notation we  assume $J=2$;  extension to  $J>2$ should be straightforward.
We need the following main assumption concerning the convergence of the nuisance parameter estimators. 
\begin{assumption}\label{assumption} 
There exist $S^{*}_c(\cdot| \cdot,\cdot), \pi^{*}(\cdot),\Lambda^{*}(\cdot,\cdot)$ such that: 
\eqnn
\sup_{t \in [0,\tau], z\in \mathcal{Z}, a=0,1}\left|\hat S_c(t | a,z)- S^{*}_c(t | a,z)\right|&=&O_p(a_n),
\\
\sup_{z\in \mathcal{Z}}\left|\hat \pi(z)- \pi^{*}(z)\right|&=&O_p(b_n),
\\
\sup_{t \in [0,\tau], z\in \mathcal{Z}}\left|\hat \Lambda_j(t,z ; \beta_{j0})- \Lambda^{*}_j(t,z)\right|&=&O_p(c_n),
\een
for some $a_n=o(1)$, $b_n=o(1)$, $c_n=o(1)$ and for $j=1,2$, where $ \mathcal{Z}$ is the sample space of $Z$.
\end{assumption}

\subsection{Asymptotic properties using Score 1}


Under Assumption \ref{assumption} and additional General Assumptions 
in the Supplementary Materials,  
under case a), b) or c) listed below, 
$\hat \beta^{(1)}-\beta_{0}=o_p(1)$ and
\eqnn \sqrt{n}\left(\hat \beta^{(1)} - \beta_{0}\right)\overset{D}{\rightarrow}\mathcal{N}(0,\Sigma), 
\een
where:

(a)  
$\Sigma  = \Sigma^{(a)}$ given in the Supplementary Materials,  as long as 
$S^{*}_c =S_{c0} $ and $\pi^{*} =\pi_0 $, 
 $a_n=b_n=n^{-1/2}$,
and Assumptions A\ref{convn}-A\ref{ifc} in the Supplementary Materials hold;

(b) $\Sigma  = \Sigma^{(b)}$ given in the Supplementary Materials,  as long as 
$\Lambda^{*} =\Lambda_0 $, 
 $c_n=n^{-1/2}$,
and Assumptions B\ref{if1nn}-B\ref{ifl} in the Supplementary Materials hold; 

(c) $\Sigma  = \Sigma^{(c)}$ given in the Supplementary Materials,  as long as 
$S^{*}_c =S_{c0} $, $\pi^{*} =\pi_0 $ and 
$\Lambda^{*} =\Lambda_0 $, $a_nc_n=o(n^{-1/2})$ and $b_nc_n=o(n^{-1/2})$, and Assumptions C\ref{ass2n}, C\ref{ass5n} in the Supplementary Materials hold.
In this case a consistent estimator of $\Sigma^{(c)}$ is also given in the Supplementary Materials. 

In the above (a) and (b) are known as {\it model double robustness}, and (c) is known as {\it rate double robustness} \citep{smucler2019unifying,hou2019estimating}.

For application of the above asymptotic results the user may choose from a variety of estimators for the nuisance parameters. 
In simulations below we show  that the variance estimator 
derived in case (c) above 
  is somehow robust to model misspecification. Alternatively we may use nonparametric bootstrap,
the validity of which is guaranteed by the fact that the estimated treatment effect is asymptotically linear.
In the Supplementary Materials 
we also show that the explicit form of the asymptotic variance can be derived for specific working models in the cases (a) and (b). In particular, we illustrate with 
proportional hazards modeling for $S_c$ and logistic regression for $\pi$ in case (a), and 
additive hazards modeling for $\Lambda_j$ in case (b). 
However, 
due to their complex forms and also because in practice one does not know which model is correct,
 we do not derive an estimator for the asymptotic variance. 


\subsection{Asymptotic properties using Score 2}

We obtain similar asymptotic results using Score 2.
Under Assumption \ref{assumption} and additional General Assumptions 
in the Supplementary Materials,  
under case a), b) or c) listed below, 
$\hat \beta^{(2)}-\beta_{0}=o_p(1)$.
In addition, 
\eqnn \sqrt{n}\left(\hat \beta^{(2)} - \beta_{0}\right)\overset{D}{\rightarrow}\mathcal{N}(0,\Gamma), 
\een
where:

(a)  
$\Gamma  = \Gamma^{(a)}$ given in the Supplementary Materials,  as long as 
$S^{*}_c =S_{c0} $, $\pi^{*} =\pi_0 $, 
$a_n=b_n=n^{-1/2}$,
and Assumption A'\ref{ifD} in the Supplementary Materials hold
(model double robustness 1).

(b) $\Gamma  = \Gamma^{(b)}$ given in the Supplementary Materials,  as long as 
 $\Lambda^{*} =\Lambda_0 $, 
and Assumption B'\ref{ifE} in the Supplementary Materials hold
 (model double robustness 2).

(c) $\Gamma  = \Gamma^{(c)}$ given in the Supplementary Materials,  as long as 
$S^{*}_c =S_{c0} $, $\pi^{*} =\pi_0 $ and $\Lambda^{*} =\Lambda_0 $, $a_n c_n=o(n^{-1/2})$ and $b_nc_n=o(n^{-1/2})$, and Assumptions C'\ref{ass2nd}, \ref{ass5nd}  in the Supplementary Materials hold (rate double robustness).
In this case a consistent estimator of $\Gamma^{(c)}$ is also given in the Supplementary Materials.







\section{Simulation experiments}\label{simulation}

In this section we investigate the performance of the proposed estimators on a series of simulated data sets.
For each scenario, we simulate 500 data sets of 1000 observations.
True $\beta_1= \beta_2 =0.1$. 
 The percentage of treated subjects is $40\%-50\%$ and the percentage of censored subjects is $10\%-30\%$.
For both estimators $\hat \beta^{(1)}$ and $\hat \beta^{(2)}$, model-based standard errors 
are used to construct $95\%$ confidence intervals. 
As illustration, in one of the scenarios 
we also report the nonparametric  bootstrap  standard error based on the first 100 simulations due to the intensive computation demand, 
where we draw 100 resamples with replacement from $(X_i, \epsilon_i, \delta_i,A_i,Z_i)$, $i=1,\ldots,n$. 


We consider separately independent and dependent censoring. 
For independent censoring, 
estimation of the censoring distribution is not required and we  use
 the simplified scores \eqref{score1nc} and \eqref{score2nc}.

\subsection{Independent censoring}

Here the censoring variable $C$ is simulated independently of $T,A,Z$. 
We consider four different simulation scenarios described in Table \ref{scenario1}. 
For estimation of the propensity score, the  working models $\mathcal{A}_{log}$, $\mathcal{A}_{log}^{*}$ and  $\mathcal{A}_{tw}$ are given in the footnote of the table, which are, respectively, logistic regression, logistic regression with interaction, and 
 the R package `twang'  implementing  
gradient boosted models  for estimation of the propensity score  \citep{cefalu2021package}.
For the competing risks, we 
fit as working models $\mathcal{B}$ the  semiparametric additive hazards model.

Both $\hat \beta^{(1)}$ and $\hat \beta^{(2)}$ are consistent and asymptotically normal as long as one of the 
 $\mathcal{A}$ or $\mathcal{B}$ working models is correct in low dimensions, i.e.~when one of the working models is parametric or semiparametric and correctly specified. 
 For comparison we also report the estimate of $\beta$ under `Regression' from fitting $\mathcal{B}$, which is valid when $\mathcal{B}$ is correctly specified. 


The results of the simulations are reported in Table \ref{res1}. 
It can be seen that when model $\mathcal{B}$ is misspecified as in Scenarios 3 and 4, the direct regression estimator of $\beta$ is severely biased with poor coverage of the confidence intervals (CI).
The estimators from both Scores 1 and 2 have little bias  in Scenarios 1, 2 and 3, 
with good coverage of CI's  using model-based standard errors (SE), even when one of the models is wrong.
In Scenario 4 where the competing risks are generated by the Cox-Aalen model, the model-based SE's underestimate the SD's, and the CI's undercover. 
Bootstrap increases the SE and hence the coverage of the CI's. 
In addition  in Scenario 4 when `twang' is used, there is no guarantee according to 
 our theory 
and the estimation bias is much more substantial, although this seems less an issue in Scenario 3 when `twang' is used. 

\begin{table} [h] 
\caption{Data-generating mechanisms of Scenarios 1-4; $Z=[Z_1,Z_2]^\top$, $C \perp (T,A,Z)$.}\label{scenario1}
\small
\begin{center}
\renewcommand{\arraystretch}{1.2}
\begin{tabular}{@{\extracolsep{5pt}} ccl} 
\hline\hline
  \\[-3ex] 
Scenario & Data-generating mechanism & Fitted models \\ 
  \hline 
   \hline
      \multirow{4}{*}{1} &    $Z_1,Z_2 \sim U(0,0.5)$  &    \\ 
 & $\logit \pi(Z)=Z_1-Z_2$ &  $\mathcal{A}_{log}$: CORRECT and $\mathcal{A}_{tw}$ \\
& $\lambda_{j}(t)=0.1A+1+Z_1+Z_2$ & $\mathcal{B}$: CORRECT \\
& $C\sim U(0,3)$ & \\\hline

       \multirow{4}{*}{2} &    $Z_1,Z_2 \sim U(0,0.5)$  &    \\
 & $\logit \pi(Z)=0.25(Z_1-Z_2)-0.5Z_1Z_2$ &  $\mathcal{A}_{log}$: WRONG and $\mathcal{A}_{tw}$  \\
& $\lambda_{j}(t)=0.1A+0.3+Z_1+Z_2$ & $\mathcal{B}$: CORRECT \\
& $C\sim U(0,3)$ &\\ \hline

      \multirow{5}{*}{3} &    $Z_1 \sim \mathcal{N}(0,1)$, $Z_2 \sim \mathcal{N}(Z_1,1)$  &    \\
 & $\logit \pi(Z)=0.25(Z_1-Z_2)+0.5Z_1Z_2-1$ &  $\mathcal{A}^{*}_{log}$: CORRECT and $\mathcal{A}_{tw}$  \\
& $\lambda_{j}(t)=0.1A+0.3+|Z_1|+\log(1+|Z_2|)$ & $\mathcal{B}$: WRONG \\
& $C\sim U(0,3)$ & \\\hline

      \multirow{5}{*}{4} &    $Z_1 \sim \mathcal{N}(0,1)$, $Z_2 \sim \mathcal{N}(Z_1,1)$  &    \\
 & $\logit \pi(Z)=0.25(Z_1-Z_2)+0.5Z_1Z_2-1$ & $\mathcal{A}^{*}_{log}$: CORRECT and 
 $\mathcal{A}_{tw}$ \\
& $\lambda_{j}(t)=0.1A+\exp{(Z_1+Z_2)}$ & $\mathcal{B}$: WRONG \\
& $C\sim U(0,3)$ & \\\hline
\end{tabular}
\end{center}
\vskip .2in
\hskip 1in $\mathcal{A}_{log}: 
\pi(z ; \alpha)=\expit(\alpha^\top z)$ 

\hskip 1in  $\mathcal{A}_{tw}$: twang 

\hskip 1in $\mathcal{A}^{*}_{log}: 
\pi(z ; \alpha)=\expit(\alpha^\top z + \alpha^{*}z_1z_2)$ 

\hskip 1in  $\mathcal{B}: 
 \Lambda_j(t, z; G_j, \gamma_j)=G_j(t) + \gamma_j^\top z$, $j=1, 2$
\end{table}

\begin{table}[h]
\footnotesize
\caption{Results of simulations from Scenarios 1-4, independent censoring; true $\beta_1= \beta_2 =0.1$. 
Column PS indicates the working model for the propensity score.
For Scenario 4, the first row of SE and CP
 are model-based, and the second row from bootstrap. SD, standard deviation; SE, standard error; CP, coverage of the $95\%$ confidence interval.}\label{res1}
\begin{center}
\setlength{\tabcolsep}{4.5pt}
\begin{tabular}{cccrccccrccccrccc}
\hline\hline
\multicolumn{2}{l}{} && \multicolumn{4}{c}{Score 1} &&  \multicolumn{4}{c}{Score 2} & &  \multicolumn{4}{c}{Regression}\tabularnewline 
\cline{4-7}\cline{9-12}\cline{14-17}
\\[-1.5ex] 
Scenario && PS &\multicolumn{1}{c}{Bias}&\multicolumn{1}{c}{SD}&\multicolumn{1}{c}{SE}&\multicolumn{1}{c}{CP}&&\multicolumn{1}{c}{Bias}&\multicolumn{1}{c}{SD}&\multicolumn{1}{c}{SE}&\multicolumn{1}{c}{CP} & &\multicolumn{1}{c}{Bias}&\multicolumn{1}{c}{SD}&\multicolumn{1}{c}{SE}&\multicolumn{1}{c}{CP}

\tabularnewline \hline\hline
\multirow{4}{*}{1}&\multirow{2}{*}{$\beta_1$}& logistic 
& $-0.012$&$0.156$&$0.146$&$0.93$&&$-0.006$&$0.157$&$0.146$&$0.93$
& \multirow{2}{*}{} & \multirow{2}{*}{-0.006} & \multirow{2}{*}{0.157} &  \multirow{2}{*}{0.147} & \multirow{2}{*}{0.93}\tabularnewline

& & twang 
&$-0.012$&$0.159$&$0.150$&$0.93$&&$-0.006$&$0.161$&$0.149$&$0.93$
&&&&& \tabularnewline
 \\[-1.5ex] 

&\multirow{2}{*}{$\beta_2$} & logistic 
&$0.0006$&$0.144$&$0.147$&$0.95$&&$0.006$&$0.146$&$0.146$&$0.95$ 
&\multirow{2}{*}{} & \multirow{2}{*}{0.006} & \multirow{2}{*}{0.146} & \multirow{2}{*}{0.147} & \multirow{2}{*}{0.95}\tabularnewline

& &twang 
&$-0.0003$&$0.147$&$0.150$&$0.96$&&$0.005$&$0.148$&$0.149$&$0.95$ &&&&\tabularnewline 
\hline
\multirow{4}{*}{2}&\multirow{2}{*}{$\beta_1$}& logstic &$-0.010$&$0.108$&$0.120$&$0.97$&&$-0.007$&$0.108$&$0.120$&$0.97$
& \multirow{2}{*}{} & \multirow{2}{*}{-0.007} & \multirow{2}{*}{0.108} &  \multirow{2}{*}{0.121} & \multirow{2}{*}{0.97}\tabularnewline

&&twang 
&$-0.011$&$0.110$&$0.124$&$0.97$&&$-0.008$&$0.110$&$0.124$&$0.97$
&&&&&\tabularnewline
 \\[-1.5ex] 

&\multirow{2}{*}{$\beta_2$} & logistic 
&$ 0.001$&$0.127$&$0.121$&$0.95$&&$ 0.005$&$0.129$&$0.121$&$0.95$ 
&  \multirow{2}{*}{} & \multirow{2}{*}{0.005} & \multirow{2}{*}{0.128} &  \multirow{2}{*}{0.121} & \multirow{2}{*}{0.95}\tabularnewline

&&twang 
&$ 0.001$&$0.131$&$0.124$&$0.95$&&$ 0.004$&$0.133$&$0.124$&$0.94$ &&&&&\tabularnewline 
\hline
\multirow{4}{*}{3}& \multirow{2}{*}{$\beta_1$} & logistic 
&$-0.009$&$0.160$&$0.163$&$0.96$&&$ 0.001$&$0.162$&$0.163$&$0.95$
&  \multirow{2}{*}{} & \multirow{2}{*}{0.336} & \multirow{2}{*}{0.170} &  \multirow{2}{*}{0.163} & \multirow{2}{*}{0.48}\tabularnewline

& &twang 
&$-0.006$&$0.158$&$0.162$&$0.97$&&$ 0.006$&$0.161$&$0.162$&$0.96$
&&&&&\tabularnewline
 \\[-1.5ex] 

& \multirow{2}{*}{$\beta_2$} & logistic 
&$ 0.000$&$0.157$&$0.163$&$0.97$&&$ 0.011$&$0.158$&$0.163$&$0.96$ 
& \multirow{2}{*}{} & \multirow{2}{*}{0.350} & \multirow{2}{*}{0.163} &  \multirow{2}{*}{0.163} & \multirow{2}{*}{0.42}\tabularnewline

& & twang 
&$ 0.006$&$0.153$&$0.162$&$0.97$&&$ 0.018$&$0.155$&$0.163$&$0.96$ &&&&&\tabularnewline
\hline
\multirow{4}{*}{4}& \multirow{2}{*}{$\beta_1$} & logistic
&$0.004$ & $0.091$  & \Centerstack{0.077 \\ 0.099} & \Centerstack{0.90 \\ 0.98} &&$0.006$ & $0.091$  & \Centerstack{0.080\\ 0.099}  & \Centerstack{0.91\\0.98} 
& \multirow{2}{*}{} & \multirow{2}{*}{$0.570$} & \multirow{2}{*}{$0.127$} & \multirow{2}{*}{$0.095$} & \multirow{2}{*}{$0$}  \tabularnewline
 \\[-1.5ex] 
 
& & twang 
&$0.044$&$0.089$  & \Centerstack{0.075 \\ 0.095} & \Centerstack{0.84\\ 0.93} &&$0.047$ & $0.092$  & \Centerstack{0.079\\ 0.100}  & \Centerstack{0.86\\0.95} 
& &&  &&\tabularnewline 
\\[-1.5ex] 

& \multirow{2}{*}{$\beta_2$} & logistic
&$0.002$ & $0.094$   &\Centerstack{0.077 \\ 0.099} & \Centerstack{0.89 \\ 0.96}& &$0.003$ & $0.095$  & \Centerstack{0.080\\ 0.099}  & \Centerstack{0.90\\0.96} 
& \multirow{2}{*}{} &  \multirow{2}{*}{$0.566$} & \multirow{2}{*}{$0.128$} & \multirow{2}{*}{$0.095$} & \multirow{2}{*}{$0$}  \tabularnewline
\\[-1.5ex] 

& & twang 
&$0.041$&$0.092$   &\Centerstack{0.075 \\ 0.095} & \Centerstack{0.89 \\ 0.90}&& $0.044$ & $0.095$  & \Centerstack{0.079\\ 0.101}  & \Centerstack{0.86\\ 0.93} &  & &  & & \tabularnewline 
\hline
\end{tabular}
\end{center}
\end{table}

\subsection{Dependent censoring}\label{cens}

For dependent censoring we consider the four scenarios described in Table \ref{scenario2}.
For estimating the censoring distribution, we consider the proportional hazards working model for all four scenarios; 
in Scenario 7 we also estimate the censoring distribution using the random survival forest \citep{ishwaran2008random}.  For the latter we use the R package `randomForestSRC' and its default hyperparameters.
We report the results using both the simplified scores, \eqref{score1nc} and \eqref{score2nc}, which assume independent censoring, 
and Score 1 in \eqref{score1s} with the estimated $S_c$. 
Model-based standard errors 
are used to construct $95\%$ confidence intervals in all cases.




The results of the simulations are reported in Table \ref{res2}. 
It is interesting  to note that although censoring depends on $A$ and $Z$, 
 the simplified scores using \eqref{score1nc} and \eqref{score2nc} in general perform better than \eqref{score1s} with the estimated $S_c$, which has generally over 10\% bias except when the random survival forest is used to estimate $S_c$ in Scenario 7. 
In Scenario 7,  when  the  proportional hazards  model, which is wrong, is used to estimate $S_c$,   the model-based SE underestimates SD, leading to substantial under  coverage  of the CI's. On the other hand, when the random survival forest is used to estimate $S_c$, the bias becomes small and the coverage is relatively accurate, so that the performance of  \eqref{score1s} is similar to those of \eqref{score1nc} and \eqref{score2nc}.


\begin{table}[h]
\small
\caption{Data-generating mechanisms of Scenarios 5-8; $Z=[Z_1,Z_2]^\top$, $C\perp T| A,Z$.}\label{scenario2}
\renewcommand{\arraystretch}{1.2}
\begin{center}
\begin{tabular}{@{\extracolsep{5pt}} ccl} 
\hline\hline
  \\[-3ex] 
Scenario & Data-generating mechanism & Fitted models \\ 
  \hline 
   \hline
      \multirow{4}{*}{5} &    $Z_1,Z_2 \sim U(0,0.5)$  &    \\ [0.2cm]
 & $\logit \pi(Z)=Z_1-Z_2$ &  $\mathcal{A}_{log}$: CORRECT and $\mathcal{A}_{tw}$ \\
& $C\sim Exp(\exp(-1+A+Z_1+Z_2))$ & $\mathcal{C}$: CORRECT  \\ [0.2cm]
& $\lambda_{j}(t)=0.1A+1+Z_1+Z_2$ & $\mathcal{B}$: CORRECT\\\hline

       \multirow{4}{*}{6} &    $Z_1,Z_2 \sim U(0,0.5)$  &    \\ [0.2cm]
 & $\logit \pi(Z)=0.25(Z_1-Z_2)-0.5Z_1Z_2$ &  $\mathcal{A}_{log}$: WRONG and $\mathcal{A}_{tw}$   \\
& $C\sim Exp(\exp(-1+A+Z_1+Z_2))$ & $\mathcal{C}$: CORRECT\\ [0.2cm]
& $\lambda_{j}(t)=0.1A+0.3+Z_1+Z_2$ & $\mathcal{B}$: CORRECT \\ \hline

    \multirow{4}{*}{7} &    $Z_1,Z_2 \sim U(0,0.5)$  &    \\  [0.2cm]
 & $\logit \pi(Z)=0.25(Z_1-Z_2)-0.5Z_1Z_2$ &  $\mathcal{A}_{log}$: WRONG and $\mathcal{A}_{tw}$   \\
& $\lambda_c(t | A,Z)=2t+A-Z_1-Z_2$ & $\mathcal{C}$: WRONG and RSF \\ [0.2cm]
& $\lambda_{j}(t)=0.1*A+0.3+Z_1+Z_2$ & $\mathcal{B}$: CORRECT\\\hline

      \multirow{5}{*}{8} &    $Z_1 \sim \mathcal{N}(0,1)$, $Z_2 \sim \mathcal{N}(Z_1,1)$  &    \\
       [0.2cm]
 & $\logit \pi(Z)=0.25(Z_1-Z_2)+0.5Z_1Z_2-1$ &  $\mathcal{A}^{*}_{log}$: CORRECT and $\mathcal{A}_{tw}$   \\
& $C\sim Exp(\exp(-A+Z_1-Z_2))$ & $\mathcal{C}$: CORRECT  \\ [0.2cm]
& $\lambda_{j}(t)=0.1A+0.3+|Z_1|+\log(1+|Z_2|)$ & $\mathcal{B}$: WRONG \\\hline
\end{tabular}
\end{center}
\vskip .2in
\hskip 1in $\mathcal{A}_{log}: 
\pi(z ; \alpha)=\expit(\alpha^\top z)$ 

\hskip 1in  $\mathcal{A}_{tw}$: twang 

\hskip 1in $\mathcal{A}^{*}_{log}: 
\pi(z ; \alpha)=\expit(\alpha^\top z + \alpha^{*}z_1z_2)$ 

\hskip 1in  $\mathcal{B}: 
 \Lambda_j(t, z; G_j, \gamma_j)=G_j(t) + \gamma_j^\top z$, $j=1, 2$

\hskip 1in  $\mathcal{C}: 
S_c(t| a, z ; \eta, \Lambda_c)=\exp\left\{-\Lambda_c(t)e^{\eta^\top d}\right\}$
where $d= (a, z)^\top$

\hskip 1in  RSF: random survival forest 

\end{table}

\begin{table}[h]
\caption{Results of simulations from Scenarios 5-8, dependent censoring; true $\beta_1= \beta_2 =0.1$. 
Column PS indicates the working model for the propensity score. For Scenario 7, the first row of `Score 1 with $\hat{S_c}$' uses the Cox model to estimate $S_c$ while the second row uses the random survival forest. SD, standard deviation; SE, standard error; CP, coverage of the $95\%$ confidence interval. $^*$median is reported when the distribution of SE is left skewed, i.e.~with heavy right tail.}\label{res2}
\footnotesize
\begin{center}
\setlength{\tabcolsep}{4.5pt}
\begin{tabular}{ccc rrrr c rrlc c rrrr}

\hline\hline
\multicolumn{2}{l}{} & & \multicolumn{4}{c}{Score 1 - Simplified} &&  \multicolumn{4}{c}{Score 1 with $\hat{S_c}$} &&  \multicolumn{4}{c}{Score 2 - Simplified}\tabularnewline 
\cline{4-7}\cline{9-12}\cline{14-17}
\\[-1.5ex] 
Scenario && PS&\multicolumn{1}{c}{Bias}&\multicolumn{1}{c}{SD}&\multicolumn{1}{c}{SE}&\multicolumn{1}{c}{CP}&&\multicolumn{1}{c}{Bias}&\multicolumn{1}{c}{SD}&\multicolumn{1}{c}{SE}&\multicolumn{1}{c}{CP}&&\multicolumn{1}{c}{Bias}&\multicolumn{1}{c}{SD}&\multicolumn{1}{c}{SE}&\multicolumn{1}{c}{CP}
\tabularnewline\hline
\hline
\multirow{4}{*}{5}&\multirow{2}{*}{$\beta_1$}& logistic  &$-0.021$&$0.157$&$0.181$&$0.97$&&$-0.032$&$0.163$&$0.180$&$0.97$&&$-0.006$&$0.154$&$0.180$&$0.98$\tabularnewline

&&twang 
&$-0.021$&$0.159$&$0.186$&$0.97$&&$-0.030$&$0.166$&$0.184$&$0.97$&&$-0.007$&$0.157$&$0.184$&$0.97$\tabularnewline
 \\[-1.5ex] 

&\multirow{2}{*}{$\beta_2$}& logistic 
&$-0.018$&$0.160$&$0.182$&$0.97$&&$-0.026$&$0.168$&$0.181$&$0.96$&&$-0.003$&$0.158$&$0.180$&$0.97$\tabularnewline

&&twang &$-0.018$&$0.163$&$0.186$&$0.97$&&$-0.026$&$0.171$&$0.185$&$0.95$&&$-0.003$&$0.162$&$0.184$&$0.97$\tabularnewline
\hline
\multirow{4}{*}{6}&\multirow{2}{*}{$\beta_1$}&logistic &$-0.009$&$0.115$&$0.113$&$0.95$&&$-0.012$&$0.115$&$0.119$&$0.96$&&$ -0.006$&$0.115$&$0.113$&$0.95$\tabularnewline

&&twang 
&$-0.009$&$0.120$&$0.117$&$0.95$&&$-0.013$&$0.118$&$0.123$&$0.96$&&$-0.005$&$0.119$&$0.117$&$0.95$\tabularnewline
 \\[-1.5ex] 

&\multirow{2}{*}{$\beta_2$}&logistic  &$-0.011$&$0.127$&$0.113$&$0.92$&&$-0.013$&$0.124$&$0.119$&$0.94$&&$-0.008$&$0.127$&$0.113$&$0.93$\tabularnewline

&&twang 
&$-0.013$&$0.131$&$0.116$&$0.91$&&$-0.014$&$0.128$&$0.123$&$0.93$&&$ -0.009$&$0.131$&$0.116$&$0.91$\tabularnewline
\hline
\multirow{4}{*}{7}&\multirow{2}{*}{$\beta_1$} &logistic  &$-0.007$&$0.126$&$0.136$&$0.96$&& \Centerstack{$-0.012$\\$-0.008$}&\Centerstack{$0.128$\\$0.126$}&
 \Centerstack{$0.120$\\$0.128$}$^*$ &\Centerstack{$0.83$\\$0.94$} &&$ 0.000$&$0.127$&$0.135$&$0.95$\tabularnewline

&&twang  &$-0.007$&$0.128$&$0.140$&$0.96$&& \Centerstack{$-0.012$\\$-0.008$}& \Centerstack{$0.129$\\$0.128$}& 
 \Centerstack{$0.122$\\$0.131$}$^*$& \Centerstack{$0.83$\\$0.95$} &&$ 0.001$&$0.129$&$0.139$&$0.97$\tabularnewline
\\[-1.5ex] 

&\multirow{2}{*}{$\beta_2$}&logistic &$-0.008$&$0.128$&$0.136$&$0.97$&& \Centerstack{$-0.014$\\$-0.008$}& \Centerstack{$0.131$\\$0.129$}&
 \Centerstack{$0.124$\\$0.128$}$^*$&\Centerstack{$0.82$\\$0.95$}&&$-0.002$&$0.129$&$0.135$&$0.97$\tabularnewline

&&twang &$-0.009$&$0.135$&$0.140$&$0.96$&& \Centerstack{$-0.014$\\$-0.008$}& \Centerstack{$0.138$\\$0.136$}&
 \Centerstack{$0.126$\\$0.132$}$^*$& \Centerstack{$0.81$\\$0.94$}&&$-0.003$&$ 0.135$&$0.139$&$0.96$
\tabularnewline 
\hline
\multirow{4}{*}{8}&\multirow{2}{*}{$\beta_1$}&logistic  &$0.001$&$0.170$&$0.160$&$0.93$&&$-0.030$&$0.184$&$0.197$&$0.96$&&$ 0.012$&$0.168$&$0.161$&$0.94$\tabularnewline

&&twang &$0.011$&$0.168$&$0.159$&$0.94$&&$-0.024$&$0.185$&$0.197$&$0.96$&&$0.020$&$0.166$&$ 0.160$&$0.95$\tabularnewline
 \\[-1.5ex] 

&\multirow{2}{*}{$\beta_2$}&logistic  
&$0.002$&$0.164$&$0.160$&$0.95$&&$-0.034$&$0.177$&$0.197$&$0.97$&&$0.013$&$0.163$&$0.161$&$0.94$\tabularnewline

&&twang &$0.014$&$0.160$&$0.159$&$0.95$&&$-0.026$&$0.176$&$0.198$&$0.97$&&$0.022$&$0.160$&$0.159$&$0.94$\tabularnewline
\hline
\end{tabular}
\end{center}
\end{table}

\section{Application}\label{application}


Here we study the effect of mid-life alcohol exposure on late life development of 
cognitive impairment. Cognitive impairment is assessed using the  Cognitive Assessment and Screening Instrument (CASI), collected from the participants starting in 1991 during the HAAS period. A score below 74 is considered moderate impairment, which is the event of interest. 
The data set consist of 1881 observations with normal cognitive functions at the start of HAAS, which is considered baseline for this competing risks analysis. 

\begin{figure}
\caption{Cumulative incidence function curves for  the HHP-HAAS data; (a) Cognitive Impairment, (b) death without impairment.} \label{cif}
\centering
    \includegraphics[scale=1]{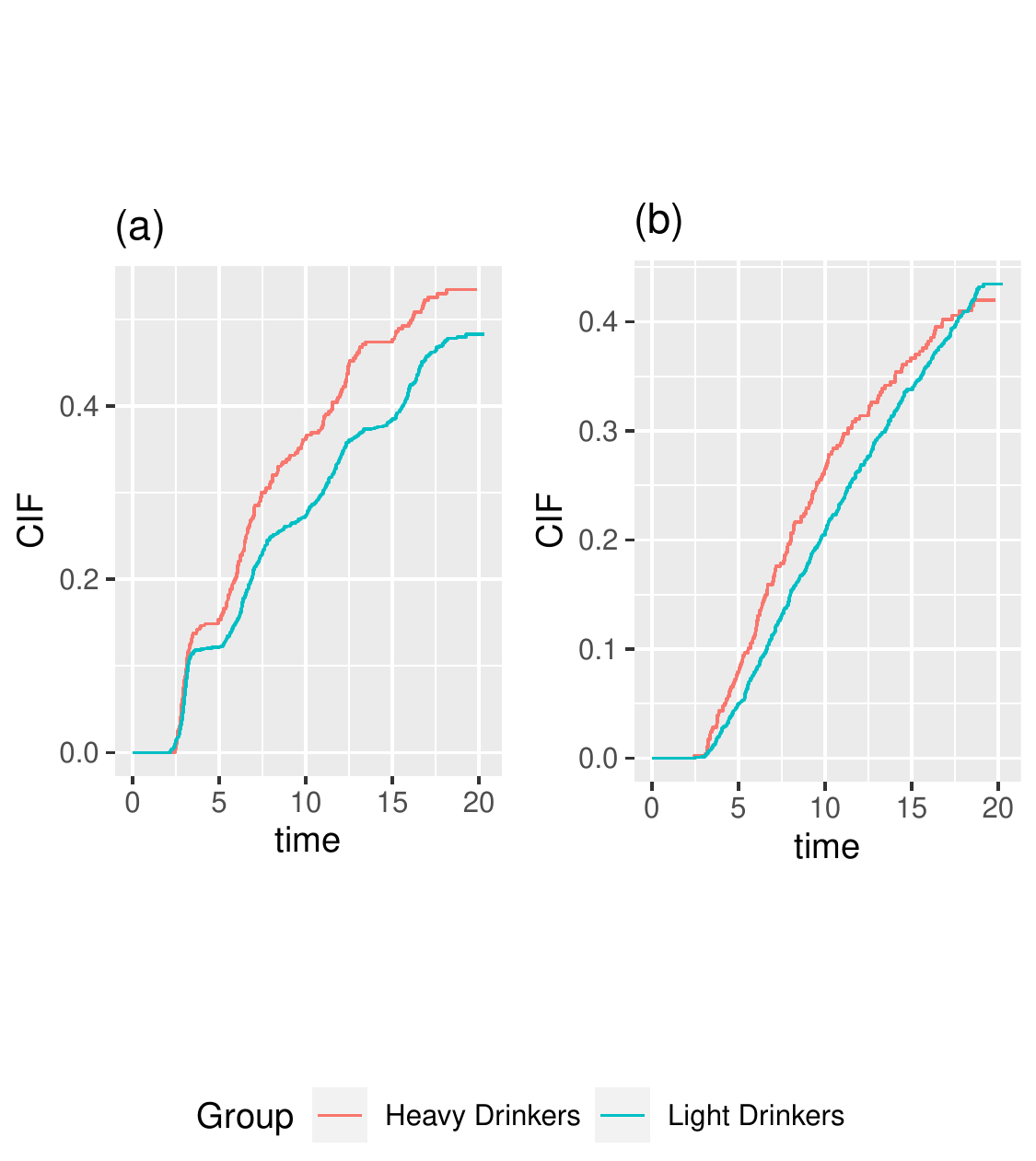}
 \end{figure}

Mid-life alcohol exposure was assessed during the HHP period between 1965 - 1974,  
and is divided into two groups of 1390  light drinkers, and 491  heavy drinkers at some point during mid-life.
At the end of follow-up, 
among light drinkers  557 (40\%) developed cognitive impairment and 474 (34\%) died without impairment, while among heavy drinkers 
216 (44\%) developed cognitive impairment and 163 (33\%) died without impairment. 
The cumulative incidence function curves for the two groups are presented in Figure \ref{cif}.

The  covariates used to adjust for confounding are maximum years of education,  age, systolic blood pressure and heart rate at the start of HHP,  and ApoE genotype. ApoE is known to be related to Alzheimer's disease (AD) and AD related dementia. 
In addition, since CASI at baseline (i.e.~start of HAAS) is post mid-life alcohol exposure, it might be considered as a mediator for the later development of cognitive impairment. 
Under the additive effect model \eqref{model}, similar to \cite{lange2011direct} and \cite{vanderweele2011causal}, 
 if  $M^a$ is the potential value of the mediator  under treatment $a=0,1$, we have
\eqnn
h_j(t | A=1,M^1,Z)- h_j(t | A=0,M^0,Z)&=&\beta_j +\lambda_j(t | M^1,Z)-\lambda_j(t | M^0,Z)\\
&=&\beta_j +h_j(t | A=1,M^1,Z)-h_j(t | A=1,M^0,Z),
\een
for $j=1,2$. 
The above gives  the usual decomposition of the total effect as the difference of the hazards in the left-hand side of the above, so that 
$\beta_j$ ($j=1,2$) may be seen as the {\it direct effect} of  mid-life alcohol exposure on the outcome (i.e.~competing risk) of interest, when we include CASI at baseline in the regression model \eqref{model}. 
For estimation of the {\it total effect}  on the left-hand side above, if we make the standard consistency assumption (i.e.~$T^{a,m}=T$ if $A=a,M=m$ and $T^a=T$ if $A=a$) and the composition assumption for mediators \cite[i.e.~$T^{a,M^a}=T^a$]{vanderweele2009conceptual}, where we again use the superscripts to indicate the potential values, 
it can be shown that 
\eqnn
h_j(t | A=a,M^a,Z) 
&=&\tilde h_j(t | A=a,Z) 
\een
for $j=1,2$ and $a=0, 1$, where $\tilde h_j $ denotes the conditional  cause-specific hazard when the mediator is not included in the regression model. Therefore the exposure effect from the latter, i.e.~when CASI at baseline is not included in model \eqref{model}, may be considered as the total effect of mid-life alcohol on the outcome.

\begin{figure}
\caption{Censoring distribution for the two exposure groups given different combinations of the covariates in the HHP-HAAS dataset. `p' indicates the log-rank test $p$-value.} \label{rdc}
    \includegraphics[width=1\textwidth]{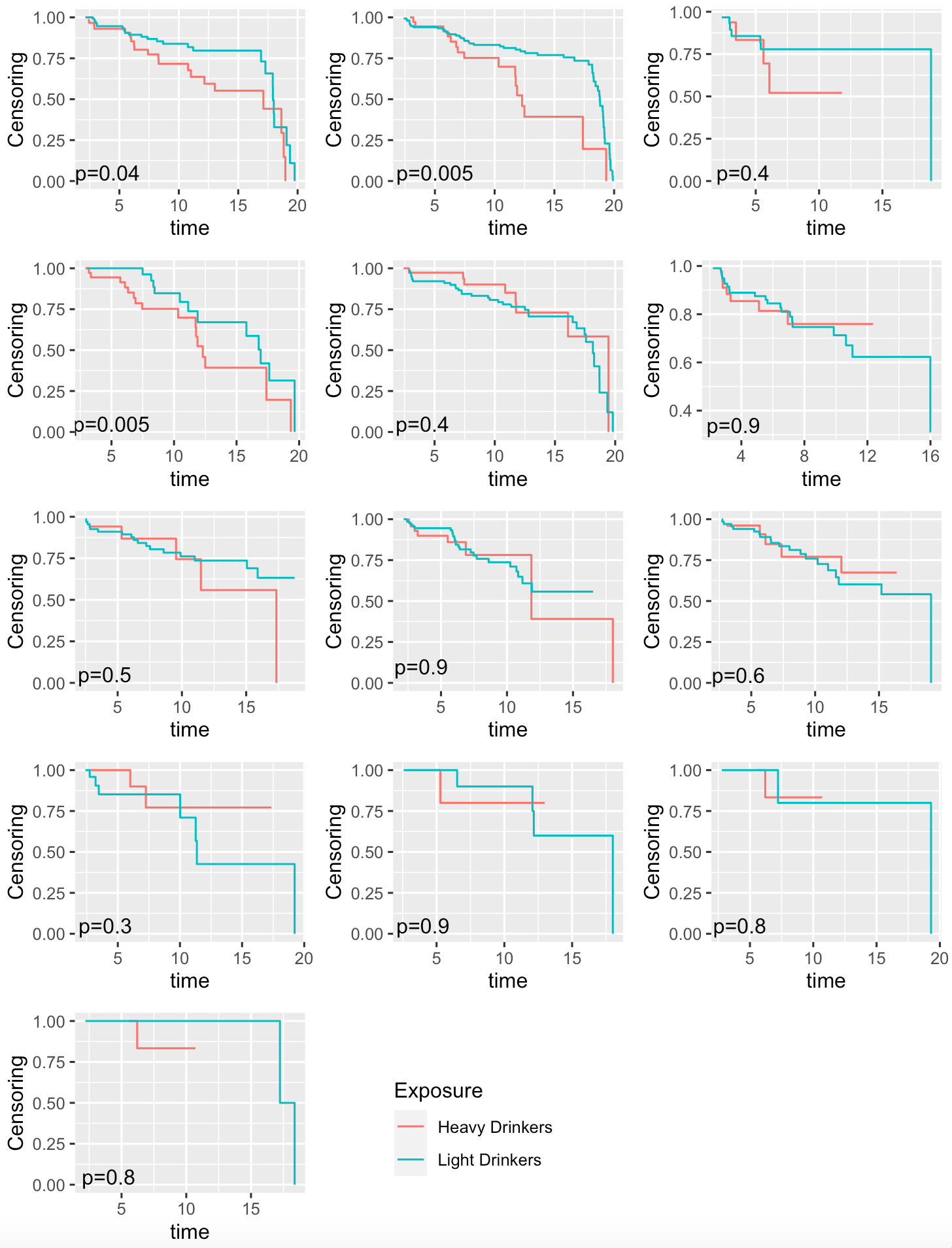}
\end{figure}

In order to estimate the above effects we use both  logistic regression without interaction and `twang' to estimate the propensity score. 
In Figure \ref{rdc} we plot the Kaplan-Meier curves of the censoring distribution for  different  groups defined by the exposure and the covariates, those that  have sufficient censoring events out of the 32 possible combinations of the 5 dichotomized  covariates (by their medians for the continuous).
 The plots seem to suggest that the stronger assumption of $C\perp (T,A) | Z$ may not hold here.
 
We  utilize the scores studied in the above simulations to estimate the effect of mid-life alcohol exposure on the development of moderate cognitive impairment and on the competing risk of death without cognitive impairment.
The censoring distribution is estimated using the proportional hazards model.
The results of the analysis are reported in Table \ref{rd1}. 
The results are similar quantitatively regardless of the estimation method, and seem to indicate that mid-life alcohol exposure has a significant effect on both the development of cognitive impairment and death without cognitive impairment, where the hazards are both increased (total effects). 
While there seems to be no obvious difference between the estimated total and direct effect  on death without cognitive impairment, the estimate direct effect is visibly less than the estimated total effect of alcohol on late life cognitive impairment, once the baseline CASI score has been accounted for. In other words, mid-life alcohol exposure conceivably contributed to late life cognitive impairment both through its earlier impact on cognitive function as well as through its sustained (i.e.~direct) impact later in life.

\begin{table}[h]
\caption{Estimated treatment effect for the HHP-HAAS data. 
Column PS indicates the working model for the propensity score.
Every first row of 95\% CI is model-based,  second row uses bootstrap. CI, confidence interval}  \label{rd1}
\small
\setlength{\tabcolsep}{3.5pt}
\begin{center}
\begin{tabular}{ccl cc c cr c cc} 
\hline\hline
\multicolumn{2}{l}{} & & \multicolumn{2}{c}{Score 1} &&  \multicolumn{2}{c}{Score 1-Cens} &&  \multicolumn{2}{c}{Score 2}\tabularnewline 
\cline{4-5}\cline{7-8}\cline{10-11}
\\[-1.5ex] 
Outcome &  PS & Effect &\multicolumn{1}{c}{$\hat \beta$}&\multicolumn{1}{c}{95\% CI} &&\multicolumn{1}{c}{$\hat \beta$}&\multicolumn{1}{c}{95\% CI}&&\multicolumn{1}{c}{$\hat \beta$}&\multicolumn{1}{c}{CI}\tabularnewline\hline
\hline
\multirow{8}{*}{Cog.~imp.}&\multirow{2}{*}{logistic}
& Total 
& $0.013$ & \Centerstack{$[0.004,0.022]$ \\ $[0.003,0.023]$} && 0.015 &  \Centerstack{$[0.005,0.025]$ \\ $[0.004,0.022]$} && 0.012 &  \Centerstack{$[0.004,0.021]$ \\ $[0.003,0.022]$}
\\
 \\[-1.5ex] 
 
&
& Direct
& $0.010$ & \Centerstack{$[0.001,0.019]$ \\ $[0.002,0.019]$} && 0.009 &  \Centerstack{$[0.000,0.019]$ \\ $[0.000,0.019]$} && 0.009 &  \Centerstack{$[0.001,0.018]$ \\ $[0.002,0.019]$}
\\
\\

&\multirow{2}{*}{twang}
& Total 
& $0.012$ & \Centerstack{$[0.003,0.020]$\\ $[0.001,0.022]$}&& 0.013  &  \Centerstack{$[0.004,0.023]$\\ $[0.002,0.025]$} && 0.011 &  \Centerstack{$[0.003,0.020]$ \\ $[0.002,0.021]$}
\tabularnewline
\\[-1.5ex] 

&
& Direct 
& $0.008$ & \Centerstack{$[0.000,0.017]$\\ $[0.005,0.011]$}& &0.008 &  \Centerstack{$[-0.002,0.017]$\\ $[-0.001,0.006]$} && 0.008 &  \Centerstack{$[0.000,0.016]$ \\ $[0.001,0.016]$}
\tabularnewline

\hline
\multirow{8}{*}{Death}
&\multirow{2}{*}{logistic}
& Total  
& $0.012$ & \Centerstack[c]{$[0.005,0.020]$\\ $[0.005,0.020]$}&& 0.013 &  \Centerstack[c]{$[0.004,0.022]$ \\ $[0.005,0.021]$} && 0.012 &  \Centerstack[c]{$[0.005,0.019]$ \\ $[0.006,0.018]$ }
\tabularnewline
\\[-1.5ex] 

&
&Direct  
& $0.012$ & \Centerstack[c]{$[0.005,0.020]$\\ $[0.005,0.022]$}&& 0.013 &  \Centerstack[c]{$[0.004,0.022]$ \\ $[0.004,0.023]$} && 0.012 &  \Centerstack[c]{$[0.005,0.019]$ \\ $[0.004,0.021]$}
\tabularnewline

\\

&\multirow{2}{*}{twang}
& Total 
&$0.012$ &  \Centerstack{$[0.004,0.020]$\\ $[0.002,0.021]$}&& 0.012 &  \Centerstack{$[0.003,0.022]$ \\ $[0.001,0.024]$} && 0.011 &  \Centerstack{$[0.004,0.019]$ \\ $[0.003,0.020]$}
\tabularnewline
\\[-1.5ex] 

&
& Direct 
&$0.010$ &  \Centerstack{$[0.003,0.018]$\\ $[0.007,0.014]$}&& 0.011 &  \Centerstack{$[0.002,0.020]$ \\ $[0.000,0.007]$} && 0.010 &  \Centerstack{$[0.003,0.018]$ \\ $[0.003,0.018]$}\\
\hline
\end{tabular}
\end{center}
\end{table}

\section{Discussion}




In this article we have proposed two doubly robust estimators for the conditional cause-specific hazard difference under competing risks.
We proposed two estimators that are model doubly robust: they are consistent and asymptotically normal if  both the propensity score  and the censoring distribution models are correctly specified,  or if the outcome models for the competing risks are correctly specified. In addition, they are rate doubly robust: they are consistent and asymptotically normal if both sets of models are correctly specified and the product of their convergence rates is $o(\sqrt{n})$. 

Rate double robustness 
gives the user the possibility
to use modern nonparametric methods, which are known to have rates of convergence slower than $\sqrt{n}$. In simulations we showed the performance of the proposed estimators when gradient boosted method is used for estimation of the propensity score, as well as 
 survival random forest for estimation of the censoring distribution. 
 In the absence of competing risks, \cite{hou2019estimating} proposed in their discussion to estimate nonparametrically the cumulative hazard function separately for the treated and the untreated, 
and then combine them using some weights to estimate what corresponds to $\Lambda_j(t,Z)$ under our model \eqref{model}. 
The procedure was not implemented or further investigated. 
For competing risks  \cite{ishwaran2014random} proposed survival random forest for estimation of both cumulative cause-specific hazard functions and cumulative incidence functions; these might be adapted in an approach similar to \cite{hou2019estimating}. This would be of interest for future work.

Recently further considerations of the competing event as a mediator to the event type of interest were described in  
 \cite{young:etal:2020} and \cite{stensrud2020separable}, in discrete time setting. 
 As with any such decomposition of total causal effects into direct and indirect effects, additional assumptions are needed. It is also not immediately clear how to extend the decomposition in the continuous time setting considered here as well as often encountered in practical applications. 

In simulations we have seen that the simple estimators using \eqref{score1nc} and \eqref{score2nc} appear  to be somehow robust  when the stronger independent censoring assumption is violated. On the other hand, 
when the censoring distribution is estimated, the model-based confidence intervals can have 
inaccurate coverage if the censoring model is misspecified. 
Using random survival forest to estimate the censoring distribution improves the performance of the treatment effect estimate. 
In practice 
bootstrap variance estimate might be used to construct confidence intervals in general.

The R codes developed in this work have been implemented in the R package 'HazardDiff' and are publicly available on CRAN (\url{https://CRAN.R-project.org/package=HazardDiff}).

\section*{Acknowledgements}
This research was partially supported by  NIH/NIA grant R03 AG062432. We thank Drs.~Steve Edland and Lon White for helpful discussion regarding the HHP-HAAS data, and Ms.~Yiran Zhang for preparation of the data.

\section{Supplementary materials}
\newtheorem{assumptionS}{Assumption}\setcounter{assumptionS}{+1}
\newtheorem{lemmas}{Lemma}\setcounter{lemmas}{+1}

\newtheorem{remark}{Remark}
\newcounter{assA}
\newenvironment{assumptionA}
  {\par\noindent
   \refstepcounter{assA}%
   A\textsc{ssumption }A\theassA.~\itshape\ignorespaces}
  {\par\ignorespacesafterend}
  
  \newcounter{assB}
\newenvironment{assumptionB}
  {\par\noindent
   \refstepcounter{assB}%
   A\textsc{ssumption }B\theassB.~\itshape\ignorespaces}
  {\par\ignorespacesafterend}
  
  \newcounter{assC}
\newenvironment{assumptionC}
  {\par\noindent
   \refstepcounter{assC}%
   A\textsc{ssumption }C\theassC.~\itshape\ignorespaces}
  {\par\ignorespacesafterend}
  
  \newcounter{assAn}
\newenvironment{assumptionA*}
  {\par\noindent
   \refstepcounter{assAn}%
   A\textsc{ssumption }A*\theassAn.~\itshape\ignorespaces}
  {\par\ignorespacesafterend}
  
    \newcounter{assBn}
\newenvironment{assumptionB*}
  {\par\noindent
   \refstepcounter{assBn}%
   A\textsc{ssumption }B*\theassBn.~\itshape\ignorespaces}
  {\par\ignorespacesafterend}
  
    \newcounter{assAnn}
\newenvironment{assumptionA'}
  {\par\noindent
   \refstepcounter{assAnn}%
   A\textsc{ssumption }A'\theassAnn.~\itshape\ignorespaces}
  {\par\ignorespacesafterend}
  
      \newcounter{assBnn}
\newenvironment{assumptionB'}
  {\par\noindent
   \refstepcounter{assBnn}%
   A\textsc{ssumption }B'\theassBnn.~\itshape\ignorespaces}
  {\par\ignorespacesafterend}
  
        \newcounter{assCnn}
\newenvironment{assumptionC'}
  {\par\noindent
   \refstepcounter{assCnn}%
   A\textsc{ssumption }C'\theassCnn.~\itshape\ignorespaces}
  {\par\ignorespacesafterend}

\newtheorem{innercustomgeneric}{\customgenericname}
\providecommand{\customgenericname}{}
\newcommand{\newcustomtheorem}[2]{%
  \newenvironment{#1}[1]
  {%
   \renewcommand\customgenericname{#2}%
   \renewcommand\theinnercustomgeneric{##1}%
   \innercustomgeneric
  }
  {\endinnercustomgeneric}
}


%

\subsection{Derivation of the semiparametrically efficient score}\label{derivation}

\subsubsection{Score for $\beta$}

As shorthand we denote $ \partial_{\beta_j} = \partial/ \partial{\beta_j}$, and similarly for other variables later. 

We first prove a generic result.
\begin{lemmas}\label{genericlemma}
For a generic cause-specific hazards model $h_j(t | W ; \theta)$, $j=1,\ldots,J$, where $\theta=(\beta, \eta)$ with finite dimensional  $\beta=[\beta_1^\top,\ldots,\beta_J^\top]^\top$  and covariates $W$, we have: 
\be
S_\beta=\left\{\int_0^\tau \left.\partial_{\beta_j}h_j(t | W ; \theta)\right|_{\beta=\beta_0}\frac{dM_j(t)}{h_j(t | W ; \theta)}\right\}_{j=1}^J.
\ee
\end{lemmas}
\newenvironment{newproof50}{\begin{proof}\textsc{\emph{of Lemma \ref{genericlemma}.}}}{\end{proof}}
\begin{newproof50}
The log likelihood for an individual is
\eqnn
\log L(\theta)&=&\sum_{j=1}^J\left[\mathbbm{1}\{\delta=1, \epsilon=j\}\log\left\{h_j(X | W ; \theta)\right\}-H_j(X | W; \theta)\right]\\&&+(1-\delta)\log\left\{\lambda_c(X | A,Z)\right\}-\Lambda_c(X | A,Z)+\log P(A| Z)+ \log f(Z),\nonumber
\een
and the associated martingales are:
\be
M_j(t)=N_j(t)- \int_0^t Y(u) h_j(u | W ; \theta_0) du, \ \ j=1,\ldots,J.
\ee
Therefore, 
\eqnn
\left\{S_{\beta}\right\}_j&=&\left.\frac{\partial\log L(\theta)}{\beta_j}\right|_{\theta=\theta_{0}}\\
&=&\sum_{j=1}^J{\mathbbm{1}\{\delta=1, \epsilon=j\}}\left[\left.\frac{\partial_{\beta_j}h_j(X | W ; \theta)}{h_j(X | W ; \theta)}\right|_{\theta=\theta_{0}}-\left.\partial_{\beta_j}H_j(X | W; \theta)\right|_{\theta=\theta_{0}}\right]\\
&=&\int_0^\tau\left.\frac{\partial_{\beta_j}h_j(t | W ; \theta)}{h_j(t | W ; \theta)}\right|_{\theta=\theta_{0}}dN_j(t)-\int_0^\tau\left.\partial_{\beta_j}h_j(t | W; \theta)\right|_{\theta=\theta_{0}}Y(t)dt\\
&=&\int_0^\tau \left.\partial_{\beta_j}h_j(t | W ; \theta)\right|_{\theta=\theta_0}\frac{dM_j(t)}{h_j(t | W ; \theta)}.
\een
\end{newproof50}

Application of the above Lemma to model \eqref{model} leads to:
\be\label{eq:S_beta}
S_\beta=\left\{\int_0^\tau A \frac{dM_j(t)}{h_j(t | A,Z)}\right\}_{j=1}^J.
\ee

\subsubsection{Proof of Lemma \ref{ortlemma}} 

Under model \eqref{model} we have $J+2$ nuisance parameters: 
 $\lambda_1(t, z),\ldots,\lambda_J(t, z),\lambda_{c}(t | a,z),P(a| z)f(z)$. We call their tangent spaces $\Lambda_{1s}^1,\ldots,\Lambda_{1s}^J,\Lambda_{2s},\Lambda_{3s}$, respectively.
Lemma 5.1 of \citet{tsiatis2007semiparametric} proved that:
\be
\Lambda_{2s}=\left\{\int_0^\tau g(t,A,Z) dM_c(t)\;\;\;\;:\;\;for\;all\;g(t,A,Z)\right\},
\ee
where $M_c(t)$ is the martingale associated with the censoring distribution. 
Pag. 117 of \citet{tsiatis2007semiparametric} proved that:
\be
\Lambda_{3s}=\left\{g(A,Z) \;\;:\;\;E\left\{g(A,Z)\right\}=0\right\}.
\ee

\underline{Step 1:}

We show that $\mathbf\Lambda$ is a direct sum of following orthogonal spaces
\be\label{decspecn}
\mathbf\Lambda=\Lambda_{1s}^1\oplus\ldots\oplus\Lambda_{1s}^J\oplus \Lambda_{2s} \oplus \Lambda_{3s},
\ee
and
\be\label{con}
\Lambda_{1s}^j=\left\{\int_0^\tau g(t,Z) \frac{dM_j(t)}{h_j(t | A,Z)}\;\;:\;\;for\;all\;g(t,Z)\right\}.
\ee



The nuisance tangent space, when the nuisance parameter has finite dimension, is defined as the space spanned by the nuisance score.
The nuisance tangent space for a semiparametric model is the mean-square closure of all parametric submodel nuisance tangent spaces.
We therefore start by considering parametric submodels.
Let's assume that $j$ is fixed and
 consider a parametric submodel:
\eqnn
h_j(t | A,Z ; \eta)=\lambda_j(t,Z; \eta) + \beta_jA,
\een
and $\eta_0$ indicates the true value of the parameter.
For this parametric submodel, by Lemma \ref{genericlemma}, we have 
\eqnn
S_{\eta}=\int_0^\tau \partial_{\eta}\lambda_j(t,Z ; \eta)|_{\eta=\eta_{0}} \frac{dM_{j}(t)}{h_j(t | A,Z)}.\een

We hence conjecture \eqref{con}.
By the above calculations, we know that, the nuisance tangent space of any parametric submodel belongs to $\Lambda^j_{1s}$. 
To complete our proof we need to prove that for any element of the conjectured \eqref{con}, indexed by $g(t,Z)$, there exists a parametric submodel such that, such element belongs to its nuisance tangent space. 
Given $g(t,Z)$, straightforward algebra proves that the score of the following parametric submodel:
\eqnn
h_j(t | A,Z ; \eta)=\lambda_{j}(t,Z; \eta_0) +\eta g(t,Z) + \beta_j A, \;\;\;\;\;\;\;\;
\een
corresponds to the element of $\Lambda^j_{1s}$ indexed by the chosen $g(t,Z)$.
Our conjecture is therefore proven.

We now focus on proving the orthogonality of these spaces.
For each $g_l(t,Z),g_j(t,Z)$ with $l\neq j$, we have:
\eqnn
&&E\left\{\int_0^\tau g_l(t,Z) \frac{dM_l(t)}{h_l(t \mid A,Z)} \times \int_0^\tau g_j(t,Z) \frac{dM_j(t)}{h_j(t \mid A,Z)}\right\}\\
&=&E\left\{\int_0^\tau g_l(t,Z)g_j(t,Z) \frac{1}{h_l(t \mid A,Z)h_j(t \mid A,Z)} <dM_l(t),dM_j(t)>\right\}=0,
\een
where the last equality comes from the fact that with absolutely continuous distributions of the event times, the probability that the competing events happen at the same time equals zero.
Therefore $\Lambda_{1s}^l \perp \Lambda_{1s}^j$ for $l\neq j$.

Finally the spaces $\Lambda_{1s}^j$, $\Lambda_{2s}$ and $\Lambda_{3s}$ are orthogonal to each other because the corresponding nuisance parameters are variationally independent and the likelihood factors into the likelihoods for each of them \citep{tsiatis2007semiparametric}. 


\underline{Step 2:}

If we don't put any restrictions on the density that generates the data,  it follows from Theorem 4.4 of \cite{tsiatis2007semiparametric} that the corresponding tangent space is the entire Hilbert space $\mathcal{H}=\{g(X,\delta,A,Z)\;:\;E\{g\}=0,\;E\{g^\top g\}<\infty\}$.
That is, 
\be\label{decgen}
\mathcal{H}=\Lambda_{1s}^{1*}\oplus\ldots,\oplus\Lambda_{1s}^{J*}\oplus \Lambda_{2s} \oplus \Lambda_{3s},
\ee
where $\Lambda_{1s}^{j*}$ is the tangent space associated with $h_j(t | A,Z)$, now left arbitrary.
Similarly to \eqref{con},
 it is easy to show that:
\be\label{genspace}
\Lambda_{1s}^{j*}=\left\{\int_0^\tau g_j(t,A,Z) \frac{dM_j(t)}{h_j(t | A,Z)}\;\;:\;\;for\;all\;g_j(t,A,Z)\right\},
\ee
and that, for any $l\neq j$:
\be\label{orth}
\Lambda_{1s}^{l*}\perp\Lambda_{1s}^{j*},\; \Lambda_{1s}^{l*}\perp\Lambda_{1s}^{j},\; \Lambda_{1s}^{l}\perp\Lambda_{1s}^{j*}.
\ee


Therefore, to find $\mathbf\Lambda^\perp$ it is sufficient to find the residual of the projection of an arbitrary element of $\Lambda_{1s}^{1*}\oplus\ldots,\oplus\Lambda_{1s}^{J*}$ onto $\Lambda_{1s}^{1}\oplus\ldots,\oplus\Lambda_{1s}^{J}$.
For any $g_1(t,A,Z), \ldots, g_J(t,A,Z)$:
\eqnn
&&\prod\left\{ \sum_{j=1}^J \left.\int_0^\tau g_j(t,A,Z) \frac{dM_j(t)}{h_j(t | A,Z)}\right| \Lambda_{1s}^{1}\oplus\ldots,\oplus,\Lambda_{1s}^{J}\right\}\\
&=&\sum_{l=1}^J\sum_{j=1}^J\prod\left\{\left.\int_0^\tau g_j(t,A,Z) \frac{dM_j(t)}{h_j(t | A,Z)}\right| \Lambda_{1s}^{l}\right\}\\&=&\sum_{j=1}^J\prod\left\{\left.\int_0^\tau g_j(t,A,Z) \frac{dM_j(t)}{h_j(t | A,Z)}\right| \Lambda_{1s}^{j}\right\}.\een

\underline{Step 3:}

Finally we show that 
for any $g_j(t,A,Z)$:
\eqnn
\prod\left\{\left.\int_0^\tau g_j(t,A,Z)\frac{dM_j(t)}{h_j(t | A,Z)} \right| \Lambda_{1s}^{j}\right\}=\int_0^\tau g^{*}_j(t,Z)\frac{dM_j(t)}{h_j(t | A,Z)},
\een
where
\eqnn
g^{*}_j(t, Z)&=&\frac{E\left[\left.g_j(t,A,Z)h_j^{-1}(t | A,Z)S_{c}(t | A,Z) e^{-\sum_{l=1}^J\beta_{l}At} \right| Z\right]}{E\left[\left.h_j^{-1}(t | A,Z)S_{c}(t | A,Z) e^{-\sum_{l=1}^J\beta_{l}At} \right| Z\right]}.
\een


By definition of projection, we need, for any $g(t, Z)$, that:
\eqnn
0&=&E\left[\int_0^\tau \left\{g_j(t,A,Z)-g^{*}_j(t,Z)\right\}\frac{dM_j(t)}{h_j(t | A,Z)}\times\int_0^\tau g(t,Z)\frac{dM_j(t)}{h_j(t | A,Z)}\right]\\
&=&E\left[\int_0^\tau \left\{g_j(t,A,Z)-g^{*}_j(t,Z)\right\}g(t,Z)h^{-2}_j(t | A,Z) <dM_j(t)>\right]\\
&=&E\left[\int_0^\tau\left\{g_j(t,A,Z)-g^{*}_j(t,Z)\right\}g(t,Z)h^{-1}_j(t | A,Z) Y(t)dt\right]\\
&=&\int_0^\tau E\left(E\left[\left\{g_j(t,A,Z)-g^{*}_j(t,Z)\right\}h^{-1}_j(t | A,Z)Y(t) | Z\right]g(t,Z)\right)dt,
\een
implying that, almost surely,
\eqnn
E\left[\left\{g_j(t,A,Z)-g^{*}_j(t,Z)\right\}h^{-1}_j(t | A,Z)Y(t) | Z\right]=0.
\een
By contradiction, let's assume that the above expectation is not zero on an interval with positive measure. If we take 
\eqnn g(t,Z)=E\left[\left\{g_j(t,A,Z)-g^{*}_j(t,Z)\right\}h^{-1}_j(t | A,Z)Y(t) | Z\right],\een
then 
\eqnn \int_0^\tau E\left(E\left[\left\{g_j(t,A,Z)-g^{*}_j(t,Z)\right\}h^{-1}_j(t | A,Z)Y(t) | Z\right]g(t,Z)\right)dt\neq0. \een
 and so the contradiction.

Therefore:
\eqnn
g^{*}_j(t, Z)&=&\frac{E\left\{g_j(t,A,Z)h_j^{-1}(t | A,Z)Y(t) | Z\right\}}{E\left\{h_j^{-1}(t | A,Z)Y(t) | Z\right\}}\\
&=&\frac{E\left[g_j(t,A,Z)h_j^{-1}(t | A,Z)E\left\{Y(t)| A,Z\right\} | Z\right]}{E\left[h_j^{-1}(t | A,Z)E\left\{Y(t)| A,Z\right\} | Z\right]}\\
&=&\frac{E\left\{g_j(t,A,Z)h_j^{-1}(t | A,Z)S_{c}(t | A,Z)  e^{-\sum_{l=1}^J\beta_{l}At} | Z\right\}}{E\left\{h_j^{-1}(t | A,Z)S_{c}(t | A,Z)  e^{-\sum_{l=1}^J\beta_{l}At} | Z\right\}}.
\een





\subsubsection{Lemma \ref{lemmaorth}} 

\begin{lemmas}\label{lemmaorth}
Let's consider a generic probability model $p(x ; \beta_0, \eta_0)$ for which $\beta_0$ is the true parameter of interest and $\eta_0$ is the nuisance parameter.
Let $\phi(x ,\beta, \eta)$ be such that $E_{\beta,\eta}\left\{\phi(x, \beta, \eta)\right\}=0$ and let $\Lambda^\perp$ be the space orthogonal to the nuisance tangent space.
Then, $\phi \in \Lambda^\perp$ if and only if the score is orthogonal, that is
\be
\frac{\partial}{\partial r} \left. E\left\{\phi(x, \beta_0, \eta^r)\right\}\right |_{r=0}=0,
\ee
where
$
 \eta^r=\eta_0+r\Delta \eta.
$
\end{lemmas}

\newenvironment{newproof10}{\begin{proof}\textsc{\emph{of Lemma \ref{lemmaorth}.}}}{\end{proof}}

\begin{newproof10}

We have:
\eqnn
\int \phi(x, \beta_0, \eta^r) p(x ; \beta_0, \eta^r)dx=0,
\een
and so
\eqnn
0&=&\left.\frac{\partial}{\partial r} \int \phi(x, \beta_0, \eta^r) p(x ; \beta_0, \eta^r)dx \right|_{r=0}\\
&=&\int \left.\partial_r\phi(x,\beta_0, \eta^r)\right|_{r=0} \left.p(x ; \beta_0, \eta^r)\right|_{r=0}dx+\int \left.\phi(x, \beta, \eta^r)\right|_{r=0} \left.\partial_r p(x ;\beta_0, \eta^r)\right|_{r=0}dx\\
&=&\int \left.\partial_r\phi(x, \beta_0, \eta^r)\right|_{r=0} p(x ; \beta_0, \eta_0)dx+\int\phi(x, \beta_0, \eta_0) \partial_r\left.\log p(x ; \beta_0, \eta^r)\right|_{r=0} p(x ; \beta_0, \eta_0)dx\\
&=&\frac{\partial}{\partial r} E\left.\left\{\phi(x, \beta_0, \eta^r)\right\}\right|_{r=0}+E\left\{\phi(x, \beta_0, \eta_0) S_\eta \right\}.
\een
Therefore, if $\phi \in \Lambda^\perp$, and therefore $E \left\{\phi(x, \beta_0, \eta_0) S_\eta \right\}=0$, we obtain $\frac{\partial}{\partial r}E\left.\left\{\phi(x, \beta_0, \eta^r)\right\}\right|_{r=0}=0$.
On the other hand, if $\frac{\partial}{\partial r} E\left.\left\{\phi(x, \beta_0, \eta^r)\right\}\right|_{r=0}=0$, we have $E \left\{\phi(x, \beta_0, \eta_0) S_\eta \right\}=0$ and so $\phi \in \Lambda^\perp$.
\end{newproof10}

\subsubsection{Proof of Theorem \ref{dr2}} 

For $j=1,\ldots,J$, we have:
\eqnn
&&E\left[\{{S}_{1}\}_j(\beta_0;S_c, \pi, \Lambda)\right]\\
&=&\int_0^\tau E \left[ e^{\sum_{j=1}^J \beta_{j0} At} S_c^{-1}(t | A,Z)\left\{A - \pi(Z)\right\}\left\{dN_{j}(t)-Y(t)d \Lambda_{j}(t,Z )-Y(t)\beta_{j0}Adt\right\}\right]\\
&=&\int_0^\tau  E [e^{\sum_{j=1}^J \beta_{j0} A t} S_c^{-1}(t | A,Z)\left\{A - \pi(Z)\right\}dM_{j}(t)]
\\
&& +\int_0^\tau  E [e^{\sum_{j=1}^J \beta_j A t} S_c^{-1}(t | A,Z)\left\{A - \pi(Z)\right\}Y(t)d\left\{\Lambda_{j0}(t,Z)- \Lambda_{j}(t,Z)\right\}].
\een
Note that under model \eqref{model} from the main paper, ${E}\left\{Y(t)| A, Z\right\}=e^{-\sum_{j=1}^J \beta_{j0} A t}S_{c0}(t | A,Z)e^{-\sum_{j=1}^J \Lambda_{j0}(t,Z)}$.
Therefore 
\eqnn
&&E\left[\{{S}_{1}\}_j(\beta_0; S_c, \pi, \Lambda)\right]\\
&=&\int_0^\tau  E( E [e^{\sum_{j=1}^J \beta_{j0} A t} S_c^{-1}(t | A,Z)\left\{A - \pi(Z)\right\} E \left\{Y(t)| A,Z\right\}| Z]d\left\{\Lambda_{j0}(t,Z)- \Lambda_{j}(t,Z)\right\})\\
&=&\int_0^\tau  E( E [ S_c^{-1}(t | A,Z)S_{c0}(t | A,Z)\left\{A - \pi(Z)\right\}| Z]e^{-\sum_{l=1}^J\Lambda_{l0}(t,Z)}d\left\{\Lambda_{j0}(t,Z)- \Lambda_{j}(t,Z)\right\})\\
&=&\int_0^\tau  E\left(\left[ S_c^{-1}(t | 1,Z)S_{c0}(t | 1,Z)\left\{1 - \pi(Z)\right\}\pi_0(Z)- S_c^{-1}(t | 0,Z)S_{c0}(t | 0,Z)\pi(Z)\left\{1-\pi_0(Z)\right\}\right]\right.\nonumber
\\
&&\times \left.e^{-\sum_{l=1}^J\Lambda_{l0}(t,Z)}d\left\{\Lambda_{j0}(t,Z)- \Lambda_{j}(t,Z)\right\}\right).
\een
The above is zero if either 
$\left\{S_c\left(\cdot | \cdot,\cdot \right)= S_{c0}(\cdot| \cdot,\cdot)\;and\;\pi(\cdot)= \pi_0(\cdot)\right\}$, or
$\Lambda_{j}(\cdot,\cdot)=\Lambda_{j0}(\cdot,\cdot )$.

For $S_2$ we have, for $j=1,\ldots,J$:
\eqnn
&&E\left[\{{S}_{2}\}_j(\beta_0;  S_c, \pi, \Lambda)\right]\\
&=&\int_0^\tau  E \left[\left\{A -{\cal E}(t; \beta_0,  S_c,  \pi)\right\}\left\{dN_{j}(t)-Y(t)d  \Lambda_{j0}(t,Z )-Y(t)\beta_{j0}Adt\right\}\right]\\
&=&\int_0^\tau  E \left[\left\{A -{\cal E}(t; \beta_0,  S_c,  \pi)\right\}dM_{j}(t)\right]\\
&&+\int_0^\tau  E \left[\left\{A -{\cal E}(t; \beta_0,  S_c,  \pi)\right\}Y(t)d\left\{\Lambda_{j0}(t,Z)-  \Lambda_{j}(t,Z )\right\}\right]\\
&=&\int_0^\tau  E\left( E \left[\left\{A -{\cal E}(t; \beta_0,  S_c,  \pi)\right\}E \left\{Y(t)| A,Z\right\}| Z\right]d\left\{\Lambda_{j0}(t,Z)- \Lambda_{j}(t,Z)\right\}\right).
\een
Therefore
\eqnn
&&E\left[\{{S}_{2}\}_j(\beta_0;  S_c, \pi, \Lambda)\right]\\
&=&\int_0^\tau  E\left( E\left[\left\{A -{\cal E}(t; \beta_0,  S_c,  \pi)\right\}e^{-\sum_{l=1}^J\beta_{l0}At}S_{c0}(t | A,Z)| Z\right]\right.
\\
&&\left. \times e^{-\sum_{l=1}^J\Lambda_{l0}(t,Z)}d\left\{\Lambda_{j0}(t,Z)-  \Lambda_{j}(t,Z )\right\}\right)\\
&=&\int_0^\tau  E\left\{\left(e^{-\sum_{l=1}^J\beta_{l0}At}S_{c0}(t | A=1,Z)\pi_0(Z) \right.\right.
\\
&&\left.\left.-{\cal E}(t; \beta_0,  S_c,  \pi)\left[e^{-\sum_{l=1}^J\beta_{l0}At}S_{c0}(t | A=1,Z )\pi_0(Z)+S_{c0}(t | A=0,Z)\left\{1-\pi_0(Z)\right\}\right]\right)\right.
\\
&&\times\left.e^{-\sum_{l=1}^J\Lambda_{l0}(t,Z)}d\left\{\Lambda_{j0}(t,Z)- \Lambda_{j}(t,Z )\right\}\right\}.
\een
The above is zero if either 
$\left\{S_c\left(\cdot | \cdot,\cdot \right)= S_{c0}(\cdot| \cdot,\cdot)\;and\;\pi(\cdot)= \pi_0(\cdot)\right\}$, or
$\Lambda_{j}(\cdot,\cdot)=\Lambda_{j0}(\cdot,\cdot )$.

\newpage
\subsection{Technical assumptions for the asymptotic properties}\label{ass}

\newcustomtheorem{customass}{Assumption}

\subsubsection{General Assumptions} 

We note below that Assumptions \ref{ass0n} - \ref{boundZn} and \ref{variationn} are standard regularity assumptions, Assumptions \ref{lowerboundn} and \ref{ass4n} are positivity assumptions often used in causal inference, and
Assumptions \ref{ass4n1} and \ref{lowerboundsc1} are  technical assumptions.

\begin{assumptionS}\label{ass0n} 
$\beta_0$ is contained in the interior of a compact set.
\end{assumptionS}

\begin{assumptionS}\label{boundln}
There exists an upper bound of time $\tau<\infty$ and, for $j=1,\ldots,J$, there exist $L_j<\infty$ such that $\sup_{z \in \mathcal{Z}}\Lambda_j(\tau, z)\leq L_j$ where $\mathcal{Z}$ is the sample space of the random variable $Z$.
\end{assumptionS}


\begin{assumptionS}\label{boundZn}
There exists $C_1$, such that 
$
P\left(\norma{Z}_{\infty}\leq C_1\right)=1.
$
\end{assumptionS}

\begin{assumptionS}\label{lowerboundn}
There exist  $C_2$ such that 
\eqnn \inf_{z \in \mathcal{Z}, a=0,1}S_{c0}(\tau | a,z)>C_2>0,\een
and $C_3$, $C_4$ such that:
\eqnn
0<C_3<\inf_{z \in \mathcal{Z}} \pi_0(z) <\sup_{z \in \mathcal{Z}} \pi_0(z)<C_4<1.
\een
Also there exists 
$\epsilon>0$ such that
$Var(A | Z)>\epsilon$, 
$E\left\{N(\tau) | A=0, Z\right\}<1-\epsilon$, 
$E\left\{Y(\tau) | A, Z\right\}>\epsilon$.
\end{assumptionS}

\begin{assumptionS}\label{variationn}
If the estimator $\hat \Lambda$ depends on the unknown $\beta$ for $j=1,2$ in a neighborhood of $\beta_0$, 
\eqnn
\bigvee_{t=0}^\tau\sup_{z \in \mathcal{Z}}\left\{\hat \Lambda_j(t,z ; \beta_j)-\hat \Lambda_j(t,z ; \beta_{j0})\right\}=O_p(\left|\beta_j-\beta_{j0}\right|),\nonumber
\een
where
$
\bigvee_{t=0}^\tau g(t)=\sup_{0<t_0<...<t_N=\tau, N \in \mathbb{N}}\sum_{j=1}^N \left|g(t_{j-1})-g(t_{j})\right|.
$
\end{assumptionS}

\begin{assumptionS}\label{ass4n}
There exist a positive $C_5$ such that 
\eqnn \inf_{z \in \mathcal{Z}, a=0,1}S^{*}_c(\tau | a,z)>C_5>0,\een
and $C_6$, $C_7$ such that:
\eqnn
0<C_6<\inf_{z \in \mathcal{Z}} \pi^{*}(z) <\sup_{z \in \mathcal{Z}} \pi^{*}(z)<C_7.
\een
\end{assumptionS}

\begin{assumptionS}\label{ass4n1}
There exists $\epsilon>0$ such that: 
\eqnn
E\left|\int_0^{\tau}\left[1+td\left\{\Lambda_1^{*}(t,Z)-\Lambda_{10}(t,Z)\right\}+t d\left\{\Lambda_2^{*}(t,Z)-\Lambda_{20}(t,Z)\right\}\right]dt\right|>\epsilon.
\een
Moreover, If the estimator $\hat \Lambda$ depends on the unknown $\beta$,
\eqnn
E\left[\left\{A -\pi^{*}( Z)\right\}\left\{A-E(q_j(t))\right\} | Z\right]>\epsilon,
\een
where we call $q_{j}(t)$ a function, such that:
\eqnn
\hat \Lambda_{j}(t, Z ; \beta)-\hat \Lambda_{j}(t, Z ; \beta_0)&=&(\beta_j-\beta_{j0})\times\frac{1}{n}\sum_{i=1}^n\int_0^\tau q_{ji}(t).
\een
\end{assumptionS}

\begin{assumptionS}\label{lowerboundsc1}
There exists $\epsilon>0$ such that
\eqnn
&&\int_0^\tau E\left[A-te^{-\sum_{j=1}^J\beta_{j0}At}S_c(t | A,Z)d\left\{\Lambda^{*}_1(t,Z)-\Lambda_{10}(t,Z)+\Lambda^{*}_2(t,Z)-\Lambda_{20}(t,Z)\right\}| Z\right]>\epsilon,\\
&&\int_0^\tau E\left[\left\{A+d\partial_{\beta_j}\Lambda^{*}_j(t,Z)\right\} \left\{A-{\cal E}(t; \beta, S_c, \pi)\right\}| Z\right]>\epsilon.
\een
\end{assumptionS}


\subsubsection{Specific assumptions for $\beta^{(1)}$}

In the following, for generic infinite-dimensional parameter $\eta$, a specific value $\eta_0$ and a generic function $f$ we use:
$$
\partial_{\eta}f(\eta_0, \cdot)=\partial_r \left.f\left\{\eta_0 +r(\eta + \eta_0) , \cdot\right\}\right|_{r=0}.
$$
We also use $\hat G_j(t)$ to denote a generic estimate of $G_j(t)$, unless noted otherwise.
\begin{assumptionA}\label{convn}
For $j=1,2$ let
\eqnn
P^{(a)}_{1j}(t) &:=& \frac{1}{n}\sum_{i=1}^n e^{(\beta_{10}+\beta_{20})A_it}\left\{A_i-\pi(Z_i ; \alpha_0)\right\}\partial_{\eta}S_c^{-1}(t | A,Z ; \eta_0, \Lambda_{c0})dM_{ji}(t; \beta_{j0}, \Lambda^{*}_j),\\
P^{(a)}_{2j}(t) &:=&\frac{1}{n}\sum_{i=1}^n e^{(\beta_{10}+\beta_{20})A_it}\left\{A_i-\pi(Z_i ; \alpha_0)\right\}\partial_{\Lambda_c}S_c^{-1}(t | A,Z ; \eta_0, \Lambda_{c0})dM_{ji}(t; \beta_{j0}, \Lambda^{*}_j),\\
P^{(a)}_{3j}(t) &:=& \frac{1}{n}\sum_{i=1}^n e^{(\beta_{10}+\beta_{20})A_it}\partial_\alpha \pi(Z_i ; \alpha_0)S_c^{-1}(t | A,Z ; \eta_0, \Lambda_{c0})dM_{ji}(t; \beta_{j0}, \Lambda^{*}_j).
\een
For $l=1,2,3$, there exist some bounded $p^{(a)}_{lj}(t)$ and a neighborhood $\mathcal{B}$ of  \\$\{\beta_0,S_{c0}(\cdot | \cdot,\cdot),\pi_0(\cdot), \Lambda^{*}_j(\cdot,\cdot)\}$
 such that:
\eqnn
\sup_{t \in [0,\tau], \left\{\beta,S_c, \pi, \Lambda_j\right\} \in \mathcal{B}} \left\Vert P^{(a)}_{lj}(t)-p^{(a)}_{lj}(t)\right\Vert\overset{p}{\rightarrow} 0,
\een
where  $\Vert \cdot \Vert$ denotes the $L^2$ norm.
\end{assumptionA}

\begin{assumptionA}\label{ifc}
There exist influence functions $\sigma_{1},\sigma_2(\cdot),\sigma_3$ such that, for any $t \in [0,\tau]$:
\eqnn
\hat\eta-\eta_0 &=& \frac{1}{n}\sum_{i=1}^n \sigma_{1i},\\
\hat\Lambda_c(t)-\Lambda_{c0}(t) &=& \frac{1}{n}\sum_{i=1}^n \sigma_{2i}(t),\\
\hat\alpha-\alpha_0 &=& \frac{1}{n}\sum_{i=1}^n \sigma_{3i}.
\een
\end{assumptionA}

\begin{assumptionB}\label{if1nn}
Let
\eqnn
P^{(b)}_{1j}(t) &:=&  \frac{1}{n}\sum_{i=1}^n e^{ (\beta_{10}  + \beta_{20}) A_i t}\left\{S^{*}_c(t | A_i,Z_i)\right\}^{-1}\left\{A_i -\pi^{*}( Z_i)\right\}Y_i(t)\partial_{\gamma_j}dL_j(t, Z ; G_{j0}, \gamma_{j0}),\\
P^{(b)}_{2j}(t) &:=& \frac{1}{n}\sum_{i=1}^n e^{ (\beta_{10}  + \beta_{20}) A_i t}\left\{S^{*}_c(t | A_i,Z_i)\right\}^{-1}\left\{A_i -\pi^{*}( Z_i)\right\}Y_i(t)\partial_{G_j}L_j(t, Z ; G_{j0}, \gamma_{j0}).
\een
We assume that, there exist $p^{(b)}_{lj}(t)$, for $l=1,2$ and a neighborhood $\mathcal{B}$ of  \\$\{\beta_0,\Lambda_{j0}(\cdot,\cdot), S^{*}_{c}(\cdot| \cdot,\cdot), \pi^{*}(\cdot,\cdot)\}$ such that :
\eqnn
\sup_{t \in [0,\tau],  \left\{\beta,\Lambda_j, S_c, \pi \right\} \in \mathcal{B}} \left\Vert P^{(b)}_{lj}(t)-p^{(b)}_{lj}(t)\right\Vert \overset{p}{\rightarrow} 0.\een
\end{assumptionB}

\begin{assumptionB}\label{ifl}
There exists influence functions $\sigma_4,\sigma_5(\cdot)$ such that, for any $t \in [0,\tau]$, $j=1,\ldots,J$:
\eqnn
\hat \gamma_j-\gamma_{j0} &=& \frac{1}{n}\sum_{i=1}^n \sigma_{4i},\\
\hat G_j(t)-G_{j0}(t)& =& \frac{1}{n}\sum_{i=1}^n \sigma_{5i}(t).
\een
\end{assumptionB}

\begin{assumptionC}\label{ass2n}
Let
\eqnn
h(t; A,Z) &=& e^{ (\beta_{10}  + \beta_{20}) A_i t}\left\{S_{c0}(t | A,Z)\right\}^{-1}\left\{A -\pi_0( Z)\right\},\\
P(t) &=& \frac{1}{n}\sum_{i=1}^n h^2(t; A_i, Z_i) A_iY_i(t),\\
Q_j(t) &=& \frac{1}{n}\sum_{i=1}^n h^2(t; A_i, Z_i)\Lambda_{j0}(t, Z_i)Y_i(t).
\een
We assume that, there exists $p(t),q_j(t)$ and a neighborhood $\mathcal{B}$ of the true \\$\{\beta_0, S_{c0}(\cdot| \cdot,\cdot), \pi_0(\cdot),\Lambda_0(\cdot,\cdot)\}$ such that:
\eqnn
\sup_{t \in [0,\tau], \left\{\beta,S_c, \pi, \Lambda\right\} \in \mathcal{B}} \left|P(t)-p(t)\right|&\overset{p}{\rightarrow}& 0,\\
\sup_{t \in [0,\tau], \left\{\beta,S_c, \pi, \Lambda\right\} \in \mathcal{B}} \left|Q_j(t)-q_j(t)\right|&\overset{p}{\rightarrow}& 0.
\een
\end{assumptionC}

\begin{assumptionC}\label{ass5n}
Let $p(t),q_j(t)$ be as defined in Assumption C\ref{ass2n}. For $j=1,2$, we assume that
$\int_0^\tau  \left\{p(u)\beta_j + q_j(u)\right\}du>0$.
\end{assumptionC}

\subsubsection{Specific assumptions for $\beta^{(2)}$}

\begin{assumptionA'}\label{ifD}
There exist $\sigma_6$ such that: 
\eqnn
{S}_{2,n}({\beta}_0, \hat S_c, \hat \pi,\Lambda^{*})-{S}_{2,n}({\beta}_0, S_{c0}, \pi_0, \Lambda^{*})=\frac{1}{n}\sum_{i=1}^n \sigma_{6i}.
\een
\end{assumptionA'}

\begin{assumptionB'}\label{ifE}
There exist $\sigma_7$ such that: 
\eqnn
{S}_{2,n}({\beta}_0, S^{*}_c,  \pi^{*} ,\hat \Lambda)-{S}_{2,n}({\beta}_0, S^{*}_{c}, \pi^{*}, \Lambda_0)=\frac{1}{n}\sum_{i=1}^n \sigma_{7i}.
\een
\end{assumptionB'}

\begin{assumptionC'}\label{ass2nd}
Let
\eqnn
h(t; A,Z)= \left\{A -{\cal E}(t; \beta_0, S_{c0}, \pi_0)\right\},
\een
\eqnn
P'(t)=\frac{1}{n}\sum_{i=1}^n h^2(t; A_i, Z_i) A_iY_i(t),
\een
and
\eqnn
Q'_j(t)=\frac{1}{n}\sum_{i=1}^n h^2(t; A_i, Z_i)\Lambda_{j0}(t, Z_i)Y_i(t).
\een
We assume that, there exist $p'(t),q'_j(t)$ and a neighborhood $\mathcal{B}$ of the true \\$\{\beta_0, S_{c0}(\cdot | \cdot,\cdot), \pi_0(\cdot), \Lambda_{0}(\cdot,\cdot)\}$ such that:
\eqnn
\sup_{t \in [0,\tau], \left\{\beta, S_c, \pi, \Lambda\right\} \in \mathcal{B}} \left|P'(t)-p'(t)\right|\overset{p}{\rightarrow} 0,\een
and
\eqnn
\sup_{t \in [0,\tau], \left\{\beta,S_c, \pi, \Lambda\right\} \in \mathcal{B}} \left|Q'_j(t)-q'_j(t)\right|\overset{p}{\rightarrow} 0.\een
\end{assumptionC'}

\begin{assumptionC'}\label{ass5nd}
Let $p'(t),q'_j(t)$ as in assumption C'\ref{ass2nd}, then for $j=1,2$, we assume that: 
\eqnn
\int_0^\tau  \left\{p'(u)\beta_j + q'_j(u)\right\}du>0.
\een
\end{assumptionC'}

\begin{remark}
Similarly to score 1, Assumption A' \ref{ifD} can be proved assuming some regularity assumptions and assuming that there exist influence functions for  $\hat S_c(\cdot | \cdot,\cdot)-S_{c0}(\cdot | \cdot,\cdot)$  and $\hat \pi(\cdot)-\pi_0(\cdot)$.
In the same way Assumption B' \ref{ifE} can be proved assuming some regularity assumptions and assuming that there exist an influence function for  $\hat \Lambda(\cdot,\cdot)-\Lambda_0(\cdot,\cdot)$.
\end{remark}


\subsection{Quantities related to asymptotic properties}\label{tq}


\subsubsection{For $\beta^{(1)}$}

For ease of reading we introduce the following additional notation, for $j=1, 2$: 
\be\label{K1}
K^{(1)}(\beta, S_c, \pi)&=&\frac{1}{n}\sum_{i=1}^n\int_0^\tau e^{ (\beta_1  + \beta_2) A_i t}S^{-1}_c(t \mid A_i,Z_i)\left\{A_i -\pi( Z_i)\right\}Y_i(t)A_idt, \\
\label{K2}
K^{(2)}_j(\beta,S_c, \pi)&=&\frac{1}{n}\sum_{i=1}^n\int_0^\tau e^{ (\beta_{10}  + \beta_{20}) A_i t}S^{-1}_c(t \mid A_i,Z_i)\left\{A_i -\pi( Z_i)\right\}Y_i(t)
\nonumber\\
&&\times d\left\{\hat \Lambda_{j}(t, Z_i ; \beta)-\hat \Lambda_{j}(t, Z_i ; \beta_0)\right\},\\
\label{K3}
K^{(3)}_{ji}(t, \beta, S_c, \pi)&=& S^{-1}_c(t \mid A_i,Z_i)\left\{A_i -\pi( Z_i)\right\}dM_{ji}(t ; \beta_j, \hat\Lambda),\\
\label{K4}
K^{(4)}_j(\beta, S_c, \pi, \Lambda)&=&\frac{1}{n}\sum_{i=1}^n\int_0^\tau e^{ (\beta_{1}  + \beta_{2})A_i t}A_i t \left\{S_c(t \mid A_i,Z_i)\right\}^{-1}\left\{A_i -\pi( Z_i)\right\}Y_i(t)
\nonumber\\
&&\times d\left\{\Lambda_j(t,Z_i)-\Lambda_{j0}(t,Z_i)\right\}.
\ee
The introduction of the above quantities will become clear  in the proof of Lemma \ref{lm:deco}.


\underline{Part (a)}:

$\Sigma^{(a)}=\left\{{E}(K^{(a)})^{-1}\right\}^\top \mbox{Var}(\psi^{(a)}){E}(K^{(a)})^{-1})$, where
$\left\{K^{(a)}\right\}^{-1}\psi^{(a)}$ is the influence function of $\hat \beta^{(1)}$ with
\eqnn
\psi^{(a)} &=&\left\{\int_0^\tau e^{ (\beta_{10}  + \beta_{20}) A t} \left\{S_{c0}(t | A,Z)\right\}^{-1}\left\{A -\pi_{0}( Z)\right\}dM_{j}(t; \beta_{j0},\Lambda^{*}_j)\right.\\&&\left.+\int_0^\tau \left[\{p^{(a)}_{1j}(t)\}^\top \sigma_{1}dt+\int_0^\tau p^{(a)}_{2j}(t)  \sigma_{2}(t)dt-\int_0^\tau \{p^{(a)}_{3j}(t)\}^\top  \sigma_{3} dt \right]\right\}_{j=1,2},
\een
and ${K^{(a)}}$  a $2\times2$ matrix with the following elements:
\eqnn
{K}^{(a)}_{jj}&=&-K^{(1)}(\beta_0,S_{c0},\pi_0)-K^{(4)}_j(\beta_0,S_{c0},\pi_0,\Lambda^{*}),
\een
\eqnn
{K}^{(a)}_{12}=-K^{(4)}_{1}(\beta_0,S_{c0},\pi_0,\Lambda^{*}),\;\;\;{K}^{(a)}_{21}=-K^{(4)}_{2}(\beta_0,S_{c0},\pi_0,\Lambda^{*}).
\een

\underline{Part (b)}:

$\Sigma^{(b)}=\left\{{E}(K^{(b)})^{-1}\right\}^\top \mbox{Var}(\psi^{(b)}){E}(K^{(b)})^{-1})$, where
$\left\{K^{(b)}\right\}^{-1}\psi^{(b)}$ is the influence function of $\hat \beta^{(1)}$ with
\eqnn
\psi^{(b)}&=&\left[\int_0^\tau e^{ (\beta_{10}  + \beta_{20}) A t} \left\{S^{*}_{c}(t | A,Z)\right\}^{-1}\left\{A -\pi^{*}( Z)\right\}dM_{j}(t)\right.\\&&\left.+\int_0^\tau \left\{ \sigma_{4} p^{(b)}_{1j}(t)dt+p^{(b)}_{2j}(t)d\sigma_{5}(t)+dp^{(b)}_{2j}(t)\sigma_{5}(t)\right\}\right]_{j=1,2},\een
and ${K^{(b)}}$  a $2\times2$ diagonal matrix with:
\eqnn
{K}^{(b)}_{jj}&=&-K^{(1)}(\beta_0,S^{*},\pi^{*})-K^{(2)}_j(\beta,S^{*}_c, \pi^{*})/(\beta_j-\beta_{j0}).
\een

\underline{Part (c)}:

  $\Sigma^{(c)}=\left\{(W^{(c)})^{-1}\right\}^\top V^{(c)} (W^{(c)})^{-1})$, where $V^{(c)}$ and $W^{(c)}$ are diagonal matrices with elements $\int_0^\tau  \left\{p(u)\beta_{j0}+q_j(u)\right\}du$ and 
${E}\left(\int_0^\tau e^{ (\beta_{10}  + \beta_{20}) A t}A\left\{S_{c0}(t | A,Z)\right\}^{-1}\left\{A - \pi_{0}( Z)\right\}Y(t)dt\right)$, respectively.

$\Sigma^{(c)}$
can be consistently estimated by: 
\be\label{score1v}
\left\{\hat{W}^{(c)}\right\}^{-1}\hat V^{(c)}(\tau)\left\{\hat{W}^{(c)}\right\}^{-1},
\ee
where
\eqnn
\hat{W}^{(c)}_{jj}=\frac{1}{n}\sum_{i=1}^nA_i\left\{A_i-\hat\pi(Z_i)\right\}\int_0^{X_i}\left\{\hat S_c(t | A_i,Z_i)\right\}^{-1}e^{(\hat\beta^{(1)}_1+\hat \beta^{(1)}_2)t}dt,
\een
and
\eqnn
\hat V^{(c)}_{jj}(\tau)=\frac{1}{n}\sum_{i=1}^n \mathbbm{1}\{\delta_i=1,\epsilon_i=j\}  e^{2(\hat\beta^{(1)}_1+\hat \beta^{(1)}_2)A_i X_i} \hat S^{-2}_c(X_i | A_i,Z_i)\left\{A_i-\hat\pi(Z_i)\right\}^2,
\een
for $j=1,2$.


\subsubsection{For $\beta^{(2)}$}

Define
\eqnn
J^{(1)}_{jj} &=& \frac{1}{n}\sum_{i=1}^n\int_0^\tau  \left\{A_i-{\cal E}_i(t; {\beta}^{*}, S^{*}_c,\pi^{*})\right\}A_iY_i(t)dt,\\
J^{(2)}_{jj} &=& \frac{1}{n}\sum_{i=1}^n\int_0^\tau  \partial_{\beta_j}{\cal E}_i(t; {\beta}^{*}, S^{*}_c,\pi^{*})Y_i(t)d\left\{\Lambda_j^{*}(t,Z_i ; \beta_{j0})-\Lambda_{j0}(t,Z_i)\right\},\\
J^{(3)}_{jj} &=& \frac{1}{n}\sum_{i=1}^n\int_0^\tau  \left\{A_i- {\cal E}_i(t; {\beta}^{*}, S^{*}_c,\pi^{*})\right\}Y_i(t)\partial_{\beta_j}d\Lambda^{*}_j(t,Z_i),\\
J^{(1)}_{12} &=& \frac{1}{n}\sum_{i=1}^n\int_0^\tau  \partial_{\beta_2}{\cal E}_i(t; {\beta}^{*}, S^{*}_c,\pi^{*})Y_i(t)d\left\{\Lambda_1^{*}(t,Z_i ; \beta_{10})-\Lambda_{10}(t,Z_i)\right\},\\
J^{(1)}_{21} &=& \frac{1}{n}\sum_{i=1}^n\int_0^\tau  \partial_{\beta_1}{\cal E}_i(t; {\beta}^{*}, S^{*}_c,\pi^{*})Y_i(t)d\left\{\Lambda_2^{*}(t,Z_i ; \beta_{10})-\Lambda_{20}(t,Z_i)\right\}.
\een

\underline{Part (a)}:

$\Gamma^{(a)}=\left\{{E}(J^{(a)})^{-1}\right\}^\top {Var}(\phi^{(a)}){E}(J^{(a)})^{-1}$, where
$\left\{J^{(a)}\right\}^{-1}\phi^{(a)}$ is the influence function of $\hat \beta^{(2)} $ with 
\eqnn
\phi^{(a)}&=&\left[\int_0^\tau \left\{A-{\cal E}(t; \beta, S_c,\pi)\right\}dM_{j}(t; \beta_{j0},\Lambda^{*}_j)+\{\sigma_{6}\}_j\right]_{j=1,2},
\een
and $J^{(a)}$  a $2\times2$ matrix with  $J^{(a)}_{jj}=J^{(1)}_{jj}+J^{(2)}_{jj}$ and $J^{(a)}_{12}=J^{(3)}_{12}$, $J^{(a)}_{21}=J^{(3)}_{21}$. 

\underline{Part (b)}:

$\Gamma^{(b)}=\left\{{E}(J^{(b)})^{-1}\right\}^\top{Var}(\phi^{(b)}){E}(J^{(b)})^{-1}$, where
$\left\{J^{(b)}\right\}^{-1}\phi^{(b)}$ is the influence function of $\hat \beta^{(2)} $ with 
\eqnn\phi^{(b)}&=&\left[\int_0^\tau \left\{A -{\cal E}(t; \beta, S_c,\pi)\right\}dM_{j}(t; \beta_{j0},\Lambda^{*}_j)+\{\sigma_{7}\}_j\right]_{j=1,2},\een
and $J^{(b)}$  a diagonal matrix with
$
J^{(b)}_{jj}=J^{(1)}_{jj}+J^{(3)}_{jj}
$.

\underline{Part (c)}:
$\Gamma^{(c)}=(W^{(c')})^{-1}V^{(c')}(\tau)(W^{(c')})^{-1})$ where
$W^{(c')}$ is a $2\times2$ diagonal matrix with diagonal element
${E}\left[A\int_0^{X}\left\{A-{\cal E}(t; \beta_0, S_{c0}, \pi_0)\right\}dt\right]$
and $V^{(c')}(\tau)$ is a diagonal matrix with diagonal elements $\int_0^\tau  \left\{p'(u)\beta_{j0}+q'_j(u)\right\}du$.
$\Gamma^{(c)}$ can be consistently estimated by 
\be\label{score2v}
(\hat{W}^{(c')})^{-1}\hat V^{(c')}(\tau)(\hat{W}^{(c')})^{-1},
\ee
where:
\eqnn
\hat{W}^{(c')}_{jj}=\frac{1}{n}\sum_{i=1}^nA_i\int_0^{X_i}\left\{A_i-{\cal E}_i(t; \hat \beta, \hat S_c,\hat \pi )\right\}dt,
\een
and
\eqnn
\hat V^{(c')}_{jj}(\tau)=\frac{1}{n}\sum_{i=1}^n \mathbbm{1}\{\delta_i=1,\epsilon_i=j\} \left\{A_i-{\cal E}_i (X_i; \hat \beta, \hat S_c,\hat \pi )\right\}^2.
\een






\newpage
\subsection{Proofs of the asymptotic results}\label{mainproof}

\subsubsection{Asymptotic properties using Score 1}

We suppose that Assumptions \ref{ass0n}-\ref{ass4n1} hold.
In the following,  
for asymptotic normality we need $\sqrt{n}$-convergence under the correctly specified model if the other model is misspecified. 
For this we will assume that the correctly specified  working models for $S_c(\cdot | \cdot,\cdot)$ and $\Lambda(\cdot,\cdot)$ are semiparametric, with a parametric component encoded by $\eta$ and $\gamma$, and a nonparametric component encoded by $\Lambda_{c}(t)$ and $G(t)$, respectively. 
The correctly specified  working models for $\pi(\cdot)$ will be assumed to be parametric, with parameter $\alpha$. 

Recall that 
\eqnn
{S}_{1,n}(\beta; S_c,  \pi,  \Lambda)=\left\{\frac{1}{n}\sum_{i=1}^n\int_0^\tau e^{ (\beta_1  + \beta_2) A_i t} S^{-1}_c(t \mid A_i,Z_i)\left\{A_i -\pi( Z_i)\right\}dM_{ji}(t ; \beta_j, \Lambda_{j})\right\}_{j=1,2}.
\een
Consider the following decomposition of the score:
\eqnn
{S}_{1,n}({\beta}, \hat S_c,  \hat \pi, \hat \Lambda)
&=&  {S}_{1,n}({\beta}, \hat S_c,  \hat \pi,  \hat \Lambda)-{S}_{1,n}({\beta}_0, \hat S_c,  \hat \pi,  \hat \Lambda)\\
&&+{S}_{1,n}({\beta}_0, \hat S_c,  \hat \pi,  \hat \Lambda)-{S}_{1,n}({\beta}_0, \hat S_c,  \hat \pi,  \Lambda^{*})\\
&&+ {S}_{1,n}({\beta}_0, \hat S_c,  \hat \pi, \Lambda^{*})-{S}_{1,n}({\beta}_0, S^{*}_c,  \pi^{*}, \Lambda^{*})\\
&&+ {S}_{1,n}({\beta}_0, S^{*}_c,  \pi^{*}, \Lambda^{*}).
\een
The last term ${S}_{1,n}({\beta}_0, S^{*}_c,  \pi^{*}, \Lambda^{*})$ is sum of i.i.d mean zero terms by Theorem \ref{dr2}. 
In the following lemma, we will show that: 
${S}_{1,n}({\beta}, \hat S_c,  \hat \pi,  \hat \Lambda)-{S}_{1,n}({\beta}_0, \hat S_c,  \hat \pi,  \hat \Lambda)$ can be written as $(\beta-\beta_0)$ times a positive definite matrix;
${S}_{1,n}({\beta}_0, \hat S_c,  \hat \pi,  \hat \Lambda)-{S}_{1,n}({\beta}_0, \hat S_c,  \hat \pi,  \Lambda^{*})$ is negligible when the censoring model and the propensity score model are correctly specified, otherwise it is a sum of i.i.d mean zero terms plus a negligible term, as long as $\Lambda^{*}(\cdot,\cdot)=\Lambda_0(\cdot,\cdot)$ and the rate of convergence of $\hat \Lambda(\cdot,\cdot)$ is $\sqrt{n}$
${S}_{1,n}({\beta}_0, \hat S_c,  \hat \pi, \Lambda^{*})-{S}_{1,n}({\beta}_0, S^{*}_c,  \pi^{*}, \Lambda^{*})$ is negligible when $\Lambda(\cdot,\cdot)$ is correctly specified, otherwise it is a sum of i.i.d mean zero terms plus a negligible term, as long as $S_c^{*}(\cdot | \cdot,\cdot)=S_{c0}(\cdot | \cdot,\cdot), \pi^{*}(\cdot)=\pi_0(\cdot)$ and the rate of convergence of $\hat S_c(\cdot | \cdot,\cdot),\hat \pi(\cdot)$ is $\sqrt{n}$.

Therefore, in each of the three scenarios,
$\hat \beta^{(1)}-\beta_0$ can be written as a sum of i.i.d mean zero terms and hence the consistency and the asymptotic normality of $\hat \beta_1$.

\begin{lemmas}\label{lm:deco}
For ${\beta}$ in a compact neighborhood of ${\beta}_{0}$, under Assumptions \ref{assumption}-\ref{ass4n1} we have:
\eqnn
{S}_{1,n}(\beta, \hat S_c,  \hat \pi, \hat \Lambda)&=&{S}_{1,n}(\beta_0, S^{*}_c,  \pi^{*}, \Lambda^{*})+{Q}^{(21)}+{Q}^{(3)}+{K}({\beta}-{\beta}_0)\\
&&+O_p\left(n^{-1/2}\left|\beta_1+\beta_2-\beta_{10}-\beta_{20}\right|+\left|\beta_1+\beta_2-\beta_{10}-\beta_{20}\right|^2\right),
\een
where
\be
Q^{(21)}_j&=&\frac{1}{n}\sum_{i=1}^n\int_0^\tau e^{ (\beta_{10}  + \beta_{20}) A_i t}\left\{S^{*}_c(t | A_i,Z_i)\right\}^{-1}\left\{A_i -\pi^{*}( Z_i)\right\}Y_i(t)\nonumber
\\
&&\times d\left\{\hat \Lambda_{j}(t, Z_i ; \beta_0)-\Lambda^{*}_{j}(t, Z_i)\right\}=o_p(1),\\
Q^{(3)}_j &=& \left\{{S}_{1,n}\right\}_j(\beta_0, \hat S_c,  \hat \pi, \Lambda^{*})-\left\{{S}_{1,n}\right\}_j(\beta_0, S^{*}_c,  \pi^{*}, \Lambda^{*})=o_p(1),\ee 
 ${K}$ is a $2\times2$ matrix with the following elements:
\eqnn
{K}_{jj}&=&-K^{(1)}(\beta_0,S^{*}_c, \pi^{*})-K^{(2)}_j(\beta,S^{*}_c, \pi^{*})/(\beta_j-\beta_{j0})-K^{(4)}_j(\beta_0,S^{*}_c, \pi^{*},\Lambda^{*}),
\een
and
\eqnn
{K}_{12}=-K^{(4)}_{1}(\beta_0,S^{*}_c, \pi^{*},\Lambda^{*}),\;\;\;{K}_{21}=-K^{(4)}_{2}(\beta_0,S^{*}_c, \pi^{*},\Lambda^{*}).
\een
In addition, 

a) If $S^{*}_c(t | a,z)=S_{c}(t | a,z ; \eta_0, \Lambda_{c0})=S_{c0}(t | a,z)$ and $\pi^{*}(z)=\pi(z ; \alpha_0)=\pi_0(Z)$ 
for some known functions $S_c$ and $\pi$ with $a_n=n^{-1/2},b_n=n^{-1/2}$; specifically, under Assumptions A\ref{convnn}-\ref{ifc}: $\quad Q^{(21)}=o_p(n^{-1/2})$ and $Q^{(3)}=O_p(n^{-1/2})$.

b) If $\Lambda^{*}(t,z)=L(t,z ; G_{0} , \gamma_{0})=\Lambda_0(t, z)$
for some known function $L$ with $c_n=n^{-1/2}$, specifically under Assumptions B\ref{if1nn}-\ref{ifl}: $\quad Q^{(3)}=o_p(n^{-1/2})$ and $Q^{(21)}=O_p(n^{-1/2})$.

c) If $S_c^{*}(\cdot|\cdot,\cdot)=S_{c0}(\cdot|\cdot,\cdot)$, $\pi^{*}(\cdot)=\pi_0(\cdot)$ and $\Lambda^{*}(\cdot,\cdot)=\Lambda_0(\cdot,\cdot)$ with $a_nc_n=o(n^{-1/2})$ and $b_nc_n=o(n^{-1/2})$ under assumptions C\ref{ass2n}-\ref{ass5n} : $\quad Q^{(21)}=o_p(n^{-1/2})$ and $Q^{(3)}=o_p(n^{-1/2})$.
\end{lemmas}

The proof of the Lemma is given in Section \ref{lemmas}.

We prove separately consistency and asymptotic normality of $\hat \beta^{(1)}$.

\textbf{\emph{Consistency:}}
In Lemma \ref{lm:deco} we showed that for ${\beta}$ in a neighboorhood of ${\beta}_0$, under case a) or b) or c):
\eqnn
{S}_{1,n}(\beta, \hat S_c,  \hat \pi, \hat \Lambda)&=&{S}_{1,n}(\beta_0, S^{*}_c,  \pi^{*}, \Lambda^{*})+{K}({\beta}-{\beta}_0)\\
&&+O_p\left(n^{-1/2}\left|\beta_1+\beta_2-\beta_{10}-\beta_{20}\right|+\left|\beta_1+\beta_2-\beta_{10}-\beta_{20}\right|^2+n^{-1/2}\right).\een

By Theorem \ref{dr2} we have: 
\eqnn
E\left[{S}_{1,n}(\beta_0, S^{*}_c,  \pi^{*}, \Lambda^{*})\right]={0}.
\een
Therefore by Lemma \ref{ci}, and Assumptions \ref{ass0n}, \ref{boundln} and \ref{ass4n} we have $S_{1,n}(\beta_0, S^{*}_c,  \pi^{*}, \Lambda^{*})=O_p(n^{-1/2})$.
Hence
\be\label{decocs}
{S}_{1,n}(\beta, \hat S_c,  \hat \pi, \hat \Lambda)&=&({\beta}-{\beta}_0){K}\\
&&+O_p\left(n^{-1/2}\left|\beta_1+\beta_2-\beta_{10}-\beta_{20}\right|+\left|\beta_1+\beta_2-\beta_{10}-\beta_{20}\right|^2+n^{-1/2}\right).\nonumber\ee

By the above, we prove that, for $|\delta|<1/2$:
\eqnn
{S}_{1,n}(\beta_0\pm n^{-\delta}, \hat S_c,  \hat \pi, \hat \Lambda)&=&{S}_{1,n}(\beta_0, S^{*}_c,  \pi^{*}, \Lambda^{*})+n^{-\delta}{K}+O_p(n^{-1/2}).
\een

If ${K}$ is invertible we can conclude that component-wise, either: 
\eqnn
{S}_{1,n}(\beta_0- n^{-\delta}, \hat S_c,  \hat \pi, \hat \Lambda)<{0}<{S}_{1,n}(\beta_0+n^{-\delta}, \hat S_c,  \hat \pi, \hat \Lambda),
\een
or
\eqnn
{S}_{1,n}(\beta_0+ n^{-\delta}, \hat S_c,  \hat \pi, \hat \Lambda)<{0}<{S}_{1,n}(\beta_0-n^{-\delta}, \hat S_c,  \hat \pi, \hat \Lambda).
\een

Therefore by definition of $\hat \beta^{(1)}$, we can conclude that $\hat \beta^{(1)}-{\beta}_0=O_p(n^{-\delta})=o_p(1)$.

We are now left to prove that ${K}$ is invertible. 
This is done according to 
the following two cases.


\begin{itemize}
\item{Case a): $S^{*}_c(\cdot | \cdot,\cdot)=S_{c0}(\cdot | \cdot,\cdot)$ and $\pi^{*}(\cdot)=\pi_0(\cdot)$, or $\hat \Lambda(\cdot,\cdot)$ does not depend on $\beta$ beyond an initial estimator of it.
}
\end{itemize}
By Lemma \ref{marquis}, 
we have:
\eqnn
\sup_{t \in [0,\tau]} \left|\frac{1}{n}\sum_{i=1}^n \left\{A_i-\pi_0(Z_i)\right\}\left\{S_{c0}(t \mid A_i,Z_i)\right\}^{-1}Y_i(t)e^{(\beta_{10}+\beta_{20})A_i t}\right|=O_p\left(n^{-1/2}\right).
\een
Together with Assumption\ref{variationn}, we have:
\eqnn
K^{(2)}_j(\beta, S_{c0}, \pi_0)=O_p(n^{-1/2}|\beta_j-\beta_{j0}|)=o_p(1).
\een
Therefore ${K}$ simplifies and it has the following determinant:
\eqnn
|{K}|&=&\left\{K^{(1)}(\beta_0,S_{c0},\pi_0)\right\}^2+K^{(1)}(\beta_0,S_{c0},\pi_0)K^{(4)}_{2}(\beta_0,S_{c0},\pi_0, \Lambda^{*})\\
&&+K^{(4)}_{1}(\beta_0,S_{c0},\pi_0,\Lambda^{*})K^{(4)}_{2}(\beta_0,S_{c0},\pi_0,\Lambda^{*})
+K^{(1)}(\beta_0,S_{c0},\pi_0)K^{(4)}_{1}(\beta_0,S_{c0},\pi_0,\Lambda^{*})\\
&&-K^{(4)}_{1}(\beta_0,S_{c0},\pi_0,\Lambda^{*})K^{(4)}_{2}(\beta_0,S_{c0},\pi_0,\Lambda^{*})\\
&=&K^{(1)}(\beta_0,S_{c0},\pi_0)\left\{K^{(1)}(\beta_0,S_{c0},\pi_0)+K^{(4)}_{2}(\beta_0,S_{c0},\pi_0,\Lambda^{*})+K^{(4)}_{1}(\beta_0,S_{c0},\pi_0,\Lambda^{*})\right\}.
\een
We prove now that both $K^{(1)}(\beta_0,S_{c0},\pi_{0})$ and\\ $\left\{K^{(1)}(\beta_0,\pi_0,S_{c0})+K^{(4)}_{2}(\beta_0,\pi_0,S_{c0},\Lambda^{*})+K^{(4)}_{1}(\beta_0,\pi_0,S_{c0},\Lambda^{*})\right\}$ are different from zero.

We first focus on $K^{(1)}(\beta_0,S_{c0},\pi_{0})$.
By Assumptions \ref{ass0n} and \ref{ass4n}, we have for some finite constant $C$:
\be\label{boundedn}
\left|A_i\left\{A_i-\pi_{0}(Z_i)\right\}\int_0^{\tau}\left\{S_{c0}(t \mid A_i,Z_i)\right\}^{-1}e^{(\beta_{10}+\beta_{20})t}Y_i(t)dt\right|\leq C^{-1}e^{(\beta_{10}+\beta_{20})\tau}\tau<\infty.
\ee
Under model \eqref{model}, 
$
E\left\{\left.Y(t)\right| A,Z\right\}=S_{c0}(t \mid A,Z)e^{-(\beta_{10}+\beta_{20})At}e^{-\Lambda_{10}(t,Z)-\Lambda_{20}(t,Z)},
$ we have:
\eqnn
&&E\left[A\left\{A-\pi_{0}(Z)\right\}\int_0^{\tau}\left\{S_{c0}(t \mid A,Z)\right\}^{-1}e^{(\beta_{10}+\beta_{20})t}Y(t)dt\right]\\
&=&E\left(E\left[\left.A\left\{A-\pi_{0}(Z)\right\}\int_0^{\tau}\left\{S_{c0}(t \mid A,Z)\right\}^{-1}e^{(\beta_{10}+\beta_{20})At}E\left\{Y(t)| A,Z\right\}dt \right| Z\right]\right)\\
&=&E\left(E\left[\left. A\left\{A-\pi_{0}(Z)\right\}\int_0^{\tau}P\left\{\left.T\geq t\right| A=0,Z\right\} dt\right| Z\right]\right)\\
&\geq&E\left(E\left[\left. A\left\{A-\pi_0(Z)\right\}\right| Z\right] \int_0^{\tau}P\left\{\left.T\geq t\right| A=0,Z\right\} dt\right)\\&\geq&E\left[\mbox{Var}\left(\left. A\right| Z\right) \tau (1-E\left\{\left.N(\tau) \right| A=0,Z\right\})\right].
\een
Therefore, by  Assumption \ref{ass4n} and \ref{lowerboundn}, we have, for some positive $\epsilon$:
\be\label{eqposn}
E\left[A\left\{A-\pi_{0}(Z)\right\}\int_0^{\tau}\left\{S_{c0}(t \mid A,Z)\right\}^{-1}e^{(\beta_{10}+\beta_{20})t}Y(t)dt\right]&>&\epsilon>0.
\ee
Hence, by Assumptions \ref{ass0n}, \ref{boundln}, \ref{ass4n}, by Hoeffding's inequality: 
\eqnn
K^{(1)}(\beta_0,S_{c0}, \pi_{0})=E\left\{K^{(1)}(\beta_0,S_{c0}, \pi_{0})\right\}+O_p(n^{-1/2})>\epsilon>0.
\een

We now focus on $K^{(1)}(\beta_0,\pi_{0},S_{c0})+K^{(2)}_{4}(\beta_0,S_{c0},\pi_{0},\Lambda^{*})+K^{(1)}_{4}(\beta_0,S_{c0},\pi_{0},\Lambda^{*})$.
Similarly, by Assumptions \ref{lowerboundn} and \ref{ass4n1} we have:
\eqnn
&&\left|K^{(1)}(\beta_0,S_{c0},\pi_{0})+K^{(2)}_{4}(\beta_0,S_{c0},\pi_{0},\Lambda^{*})+K^{(1)}_{4}(\beta_0,S_{c0},\pi_{0},\Lambda^{*})\right|\\
&=&\left|\frac{1}{n}\sum_{i=1}^n A_i\left\{A_i-\pi_0(Z_i)\right\}\int_0^\tau\left\{S_{c0}(t \mid A_i,Z_i)\right\}^{-1}Y_i(t)e^{(\beta_{10}+\beta_{20})A_i t}\right.\\&&\left.\times \left[1 +td\left\{\Lambda_1^{*}(t,Z_i)-\Lambda_{10}(t,Z_i)\right\}+t d\left\{\Lambda_2^{*}(t,Z_i)-\Lambda_{20}(t,Z_i)\right\}\right]\right|\\
&\geq&\left|E\left(\left\{A-\pi_0(Z)\right\}\int_0^\tau\exp\left\{-\Lambda_{10}(t,Z)-\Lambda_{20}(t,Z)\right\}\right.\right.\\&&\left.\left.\times\left[A +Atd\left\{\Lambda_1^{*}(t,Z)-\Lambda_{10}(t,Z)\right\}+A t d\left\{\Lambda_2^{*}(t,Z)-\Lambda_{20}(t,Z)\right\}\right]\right)\right|+O_p(n^{-1/2})\\
&\geq& \left|E\left(E\left[A\left\{A-\pi_0(Z)\right\} | Z\right]\int_0^\tau\left[1 +td\left\{\Lambda_1^{*}(t,Z)-\Lambda_{10}(t,Z)\right\}+t d\left\{\Lambda_2^{*}(t,Z)-\Lambda_{20}(t,Z)\right\}\right]\right)\right|\nonumber\\&&\cdot(1-E\left\{N(\tau) | A=0,Z\right\})+O_p(n^{-1/2})\\
&=&E\left(\mbox{Var}\left(A | Z\right) \int_0^{\tau}\left|1+td\left\{\Lambda_1^{*}(t,Z)-\Lambda_{10}(t,Z)\right\}+t d\left\{\Lambda_2^{*}(t,Z)-\Lambda_{20}(t,Z)\right\}\right|\right.\\&&\left.  \times(1-E\left\{\left.N(\tau) \right| A=0,Z\right\})\right)+O_p(n^{-1/2})>\epsilon.
\een
We can therefore conclude that ${K}$ is invertible.

\begin{itemize}
\item{Case b)}: $\hat \Lambda(\cdot,\cdot)$ depends on the unknown $\beta$ and $\Lambda^{*}(\cdot,\cdot)=\Lambda_0(\cdot,\cdot)$.
\end{itemize}

By definition $K^{(4)}_j(\beta_0, S^{*}_c, \pi^{*}, \Lambda_0)=0$.
Again, we want to prove that ${K}$ is invertible by proving that the determinant is different from zero.
Since $K^{(4)}_j(\beta_0, S^{*}_c, \pi^{*}, \Lambda_0)=0$, ${K}$ is a diagonal matrix so we just need to verify that the diagonal elements are not null.
We have:
\eqnn
\left|{K}_{jj}\right|&=&\left|K^{(1)}(\beta_0,S^{*}_c, \pi^{*})+K^{(2)}_j(\beta,S^{*}_c, \pi^{*})/(\beta_j-\beta_{j0})\right|\\
&=&\left|\frac{1}{n}\sum_{i=1}^n\int_0^\tau e^{ (\beta_{10}  + \beta_{20}) A_i t}\left\{S^{*}_c(t \mid A_i,Z_i)\right\}^{-1}\left\{A_i -\pi^{*}( Z_i)\right\}Y_i(t)\left\{A_i+\frac{1}{n}\sum_{l=1}^nq_{jl}(t)\right\}dt\right|,\nonumber\een
where we call $q_{ji}(t)$ a function, such that:
\eqnn
\hat \Lambda_{j}(t, Z ; \beta)-\hat \Lambda_{j}(t, Z ; \beta_0)&=&(\beta_j-\beta_{j0})*\frac{1}{n}\sum_{i=1}^n\int_0^\tau q_{ji}(t).
\een

Similarly to before, by Assumptions \ref{ass0n}, \ref{ass4n}, and Hoeffding's inequality, we have:
\eqnn
&&\frac{1}{n}\sum_{i=1}^n\int_0^\tau e^{ (\beta_{10}  + \beta_{20}) A_i t}\left\{S^{*}_c(t \mid A_i,Z_i)\right\}^{-1}\left\{A_i -\pi^{*}( Z_i)\right\}Y_i(t)\left[A_i-\frac{1}{n}\sum_{l=1}^nq_{jl}(t)\right]dt\\
&&=E\left(\int_0^{\tau} \left\{S^{*}_c(t \mid A,Z)\right\}^{-1}S_{c0}(t \mid A,Z)e^{-\sum_{l=1}^2\Lambda_{l0}(t,Z)}\left\{A -\pi^{*}( Z)\right\}\left[A-E(q_j(t))\right]dt\right)\\&&+O_p(n^{-1/2})\nonumber\\
&\geq&CE\left(\int_0^{\tau}e^{-\sum_{l=1}^2\Lambda_{l0}(t,Z)}E\left[\left\{A -\pi^{*}( Z)\right\}\left\{A-E(q_j(t))\right\} | Z\right]dt\right)+O_p(n^{-1/2}).\nonumber
\een

Therefore, by Assumptions \ref{ass4n} and \ref{ass4n1}, we have $\left|{K}_{jj}\right|>\epsilon+O_p(n^{-1/2})$ and hence ${K}$ is invertible.


\textbf{\emph{Asymptotic normality:}}
In Lemma \ref{lm:deco} we proved that for $\beta$ in a neighboorhood of $\beta_0$:
\eqnn
{S}_{1,n}(\beta, \hat S_c,  \hat \pi, \hat \Lambda)&=&{S}_{1,n}(\beta_0, S^{*}_c,  \pi^{*}, \Lambda^{*})+Q^{(21)}+Q^{(3)}+{K}({\beta}-{\beta}_0)\\
&&+O_p\left(n^{-1/2}\left|\beta_1+\beta_2-\beta_{10}-\beta_{20}\right|+\left|\beta_1+\beta_2-\beta_{10}-\beta_{20}\right|^2\right),
\een
where
\eqnn
Q^{(21)}_j&=&\frac{1}{n}\sum_{i=1}^n\int_0^\tau e^{ (\beta_{10}  + \beta_{20}) A_i t}\left\{S^{*}_c(t \mid A_i,Z_i)\right\}^{-1}\left\{A_i -\pi^{*}( Z_i)\right\}Y_i(t)\nonumber
\\
&&\times d\left\{\hat \Lambda_{j}(t, Z_i ; \beta_0)-\Lambda^{*}_{j}(t, Z_i)\right\}=o_p(1),\een
\eqnn
Q^{(3)}_j&=&\left\{{S}_{1,n}\right\}_j(\beta_0, \hat S_c,  \hat \pi, \Lambda^{*})-\left\{{S}_{1,n}\right\}_j(\beta_0, S^{*}_c,  \pi^{*}, \Lambda^{*})=o_p(1).\een

In the previous part of the proof we proved that $\hat{{\beta}} -{\beta}_0=O_p(n^{-\delta})$ for any $|\delta|<1/2$ and that ${K}$ is invertible. Therefore we have:
\be\label{decoan}
\sqrt{n}(\hat \beta^{(1)}-{\beta}_{0})&=&{K}^{-1}\left\{\sqrt{n}{S}_{1,n}(\beta_0, S^{*}_c,  \pi^{*}, \Lambda^{*})+\sqrt{n}Q^{(21)}+\sqrt{n}Q^{(3)}\right\}+o_p(1).\ee

We remind the reader that if the censoring model and the propensity score model are correctly specified, $Q^{(21)}=o_p(n^{-1/2})$. If $\Lambda(\cdot)$ is correctly specified, $Q^{(3)}=o_p(n^{-1/2})$.
Hence, if every model is correctly specified, \eqref{decoan} simplifies and the asymptotic normality of $\hat \beta^{(1)}$ is obtained by the normality of $\sqrt{n}{S}_{1,n}(\beta_0, S^{*}_c,  \pi^{*}, \Lambda^{*})$, that is a sum of i.i.d multivariate martingale integral.

If only the censoring model and the propensity score model are correctly specified, under Assumption A\ref{ifc}, $Q^{(3)}$ is asymptotically linear. Asymptotic normality of $\hat \beta^{(1)}$ is therefore obtained by the normality of $\sqrt{n}S_{2,n}(\beta_0, S^{*}_c,  \pi^{*}, \Lambda^{*})+\sqrt{n}Q^{(3)}$ that is a sum of i.i.d mean zero random variables.

If only the baseline hazard model is correctly specified, under Assumptions B\ref{ifl}, $Q^{(21)}$ is asymptotically linear. Asymptotic normality of $\hat \beta^{(1)}$ is therefore obtained by the normality of $\sqrt{n}{S}_{1,n}(\beta_0, S^{*}_c,  \pi^{*}, \Lambda^{*})+\sqrt{n}Q^{(21)}$that is a sum of i.i.d mean zero random variables.

In the following we prove the above statements in details.

\begin{itemize}
\item{Case (a):}
\end{itemize}
We remind the reader that, if $S_c^{*}(\cdot | \cdot,\cdot)=S_{c0}(\cdot | \cdot,\cdot)$ and $\pi^{*}(\cdot)=\pi_{0}(\cdot)$, $Q^{(21)}=o_p(n^{-1/2})$ and $K^{(2)}_j(\beta, S^{*}_c,\pi^{*})=o_p(1)$. 
In the previous part of the proof we have proved that ${K}$ simplifies to a $2\times2$ matrix with
${K}_{jj}=-K^{(1)}(\beta_0,S^{*}_c,  \pi^{*})-K^{(4)}_j(\beta_0,S^{*}_c,  \pi^{*},\Lambda^{*}),$
and ${K}_{12}=-K^{(4)}_{1}(\beta_0,S^{*}_c,  \pi^{*},\Lambda^{*}),\;\;\;{K}_{21}=-K^{(4)}_2(\beta_0,S^{*}_c,  \pi^{*},\Lambda^{*}).$

Therefore, in \eqref{decoan} we are left with 
\be\label{decoana}
\sqrt{n}(\hat{{\beta}}-{\beta}_{0})={K}^{-1}\sqrt{n}\left\{S_{2,n}(\beta_0, S^{*}_c,  \pi^{*}, \Lambda^{*})+{Q}^{(3)}\right\}+o_p(1).
\ee

Since $\sqrt{n}{S}_{1,n}(\beta_0, S^{*}_c,  \pi^{*}, \Lambda^{*})$ is already a sum of i.i.d mean zero terms, with the help of Assumption A \ref{ifc}, we now prove that also term ${Q}^{(3)}$ can be written as the sum of i.i.d mean zero terms.
We can then apply the multivariate central limit theorem to $\sqrt{n}{S}_{1,n}(\beta_0, S^{*}_c,  \pi^{*}, \Lambda^{*})+\sqrt{n}{Q}^{(3)}$ and reach our conclusion.

We now look at the details.
Using the fact that $\hat\pi(z)=\pi(z ; \hat\alpha)$, $\hat S_c(t | a,z)=S_c(t | a,z ; \hat \eta, \hat \Lambda_c)$ we have, by Taylor expansion:
\be
Q^{(3)}_j&=&\sqrt{n}\left[\left\{{S}_{1,n}\right\}_j(\beta_0, \hat S_c,  \hat \pi,   \Lambda^{*})-\left\{{S}_{1,n}\right\}_j(\beta_0, S_{c0}, \pi_0,\Lambda^{*})\right] \nonumber\\
&=&\sqrt{n}\left[\left\{{S}_{1,n}\right\}_j(\beta_0, \hat\eta, \hat \Lambda_{c},  \hat \alpha,  \Lambda^{*})-\left\{{S}_{1,n}\right\}_j(\beta_0, \eta_0, \Lambda_{c0}, \alpha_0, \Lambda^{*})\right]\nonumber\\
&=&\sqrt{n}\frac{1}{n}\sum_{i=1}^n\int_0^\tau e^{(\beta_{10}+\beta_{20})A_it}\left\{{D}_{ij}^n(t,\eta_{0}, \Lambda_{c0}, \alpha_0, \Lambda^{*})\right\}^\top{\Delta} dM_{ji}(t; \beta_{j0}, \Lambda^{*}_j)+o_p(n^{-1/2}),\nonumber
\ee
where
\eqnn
{D}_{ij}^n(\eta,\Lambda_{c},\alpha):=[\partial_{\eta} f (A_i,Z_i;\eta,\Lambda_{c},\alpha),
\partial_{\Lambda_{c}} f(A_i,Z_i;\eta,\Lambda_{c},\alpha),
\partial_\alpha f(A_i,Z_i;\eta,\Lambda_{c},\alpha)]^\top,
\een
\eqnn
f (A_i,Z_i; \eta,\Lambda_{c},\alpha):=\left\{A_i-\pi(Z ; \hat\alpha)\right\}S_c^{-1}(t | A,Z ; \hat \eta, \hat \Lambda_c),\een
and 
\eqnn
{\Delta}:=[ \hat\eta-\eta_0,
\hat\Lambda_{c}(t)-\Lambda_{c0}(t),
\hat\alpha-\alpha_0]^\top.
\een

Standard algebra gives us:
\eqnn
{D}_{ij}^n(\eta,\Lambda_{c},\alpha)&=&[\left\{A_i-\pi(Z_i ; \alpha)\right\}\partial_{\eta}S_c^{-1}(t | A,Z ; \eta, \Lambda_c),
\left\{A_i-\pi(Z_i ; \alpha)\right\}\partial_{\Lambda_c}S_c^{-1}(t | A,Z ; \eta, \Lambda_c)\\&&,
-\partial_\alpha \pi(Z_i ; \alpha)S_c^{-1}(t | A,Z ; \eta, \Lambda_c)
]^\top.
\een

Moreover, by Assumption A\ref{ifc}, we have:
\eqnn
\hat \alpha-\alpha_0=O_p(n^{-1/2}),\;\;\;\;\;\;\hat \eta-\eta_0=O_p(n^{-1/2}),\;\;\;\;\;\;\;\sup_{t \in [0,\tau]} \left\{\hat\Lambda_{c}(t)-\Lambda_{c0}(t)\right\}=O_p(n^{-1/2}).
\een

Therefore, by the above and by Assumptions A\ref{convn} and A\ref{ifc} we have:
\be
Q^{(3)}_j&=&\int_0^\tau \left[\left\{P^{(a)}_{1j}(t)\right\}^\top (\hat\eta-\eta_0)+ P^{(a)}_{2j}(t) \left\{\hat\Lambda_{c}(t;  \hat \eta)-\Lambda_{c0}(t)\right\}- \left\{P^{(a)}_{3j}(t)\right\}^\top (\hat\alpha-\alpha_0)\right]dt\nonumber\nonumber\\
&=&\int_0^\tau \left[\left\{p^{(a)}_{1j}(t)\right\}^\top (\hat\eta-\eta_0)+ p^{(a)}_{2j}(t) \left\{\hat\Lambda_{c}(t;  \hat \eta)-\Lambda_{c0}(t)\right\}- \left\{p^{(a)}_{3j}(t)\right\}^\top (\hat\alpha-\alpha_0)\right]dt\nonumber\nonumber\\
&&+o_p(n^{-1/2})\nonumber\\
&=& \frac{1}{n}\sum_{i=1}^n\int_0^\tau \left[\left\{p^{(a)}_{1j}(t)\right\}^\top \sigma_{1i}+\int_0^\tau p^{(a)}_{2j}(t)  \sigma_{2i}(t)-\int_0^\tau  \left\{p^{(a)}_{3j}(t)\right\}^\top  \sigma_{3i}  \right]dt+o_p(n^{-1/2}).\nonumber
\ee

Therefore we have:
\be\label{anpswna}
\sqrt{n}{S}_{1,n}(\beta_0, S_{c0},  \pi_0,  \Lambda^{*})+\sqrt{n}{Q}^{(3)}=\frac{1}{\sqrt{n}}\sum_{i=1}^n \psi^{(a)}_{i}(t)+o_p(1),
\ee
where 
\eqnn
\psi^{(a)}_{i,j}&=&\int_0^\tau e^{ (\beta_{10}  + \beta_{20}) A_i t} \left\{S_{c0}(t \mid A_i,Z_i)\right\}^{-1}\left\{A_i -\pi_{0}( Z_i)\right\}dM_{ji}(t; \beta_{j0},\Lambda^{*}_j)\\&&+\int_0^\tau \left[\left\{p^{(a)}_{1j}(t)\right\}^\top \sigma_{1}+ p^{(a)}_{2j}(t)  \sigma_{2}(t)-\int_0^\tau \left\{p^{(a)}_{3j}(t)\right\}^\top  \sigma_{3} \right]dt.
\een

By Theorem \ref{dr2} and by construction of ${Q}^{(3)}$ the right hand side of \eqref{decoana} is a sum of i.i.d mean zero and the multivariate central limit theorem can be applied.
Therefore, case a) of the asymptotic normality is proven. 

\begin{itemize}
\item{Case (b):}
\end{itemize}
We remind the reader that, if $\Lambda^{*}(\cdot,\cdot)=\Lambda_{0}(\cdot,\cdot)$, ${Q}^{(3)}=o_p(n^{-1/2})$.
In the previous part of the proof we have proved that ${K}$ simplifies to a $2\times2$ diagonal matrix with:
\eqnn
{K}_{jj}&=&-K^{(1)}(\beta_0,S^{*}_c,  \pi^{*})-K^{(2)}_j(\beta,S^{*}_c,  \pi^{*})/(\beta_j-\beta_{j0}).
\een

Therefore, in \eqref{decoan} we are left with 
\be\label{decoanb}
\sqrt{n}(\hat{{\beta}}-{\beta}_{0})={K}^{-1}\sqrt{n}\left\{{S}_{1,n}(\beta_0, S^{*}_c,  \pi^{*}, \Lambda^{*})+{Q}^{(21)}\right\}+o_p(1).
\ee

Since $\sqrt{n}{S}_{1,n}(\beta_0, S^{*}_c,  \pi^{*}, \Lambda^{*})$ is already a sum of i.i.d mean zero terms, with the help of Assumption B\ref{ifl}, we now prove that also term ${Q}^{(21)}$ can be written as a sum of i.i.d mean zero terms.
We can then apply the multivariate central limit theorem to $\sqrt{n}{S}_{1,n}(\beta_0, S^{*}_c,  \pi^{*}, \Lambda^{*})+\sqrt{n}{Q}^{(21)}$ and reach our conclusion.

We now look at the details.

By Assumption B\ref{ifl}, we know that 
\eqnn
\hat \gamma_{j}-\gamma_{j0}=O_p(n^{-1/2}),\;\;\;\hat G_j(t)-G_{j0}(t)=O_p(n^{-1/2}).
\een

Therefore by Taylor expansion we have:
\eqnn
\sqrt{n}Q^{(21)}_j&=&-\frac{1}{\sqrt{n}}\sum_{i=1}^n\int_0^\tau e^{ (\beta_{10}  + \beta_{20}) A_i t} \left\{S^{*}_c(t \mid A_i,Z_i)\right\}^{-1}\left\{A_i -\pi^{*}( Z_i)\right\}Y_i(t)\\&&\times d\left\{L(t,Z; \hat G_{j}, \hat \gamma_j)-L(t,Z; G_{j0}, \gamma_{j0})\right\}+o_p(n^{-1/2})\nonumber\\
&=&-\frac{1}{\sqrt{n}}\sum_{i=1}^n\int_0^\tau e^{ (\beta_{10}  + \beta_{20}) A_i t} \left\{S^{*}_c(t \mid A_i,Z_i)\right\}^{-1}\left\{A_i -\pi^{*}( Z_i)\right\}Y_i(t)\\&&\times \left[(\hat\gamma_j-\gamma_{j0})^\top\partial_{\gamma_j}dL(t, Z ; G_{j0}, \gamma_{j0})  +d\left\{\hat G_j(t)-G_{j0}(t)\right\}\partial_{\gamma_j}L(t, Z ; G_{j0}, \gamma_{j0})\right.\\&&\left.+\left\{\hat G_j(t)-G_{j0}(t)\right\}\partial_{\gamma_j}dL(t, Z ; G_{j0}, \gamma_{j0})\right]+o_p(n^{-1/2}).
\een

Hence, by the above and by Assumption B\ref{if1nn} and B\ref{ifl}, we have:
\be
Q^{(21)}_j&=&\int_0^\tau \left[(\hat\gamma_j-\gamma_{j0})^\top P^{(b)}_{1j}(t)dt +P^{(b)}_{2j}(t)d\left\{\hat G_j(t)-G_{j0}(t)\right\}\right.
\\
&&\left.\quad
+dP^{(b)}_{2j}(t)\left\{\hat G_j(t)-G_{j0}(t)\right\}\right]\nonumber\\
&=&\int_0^\tau \left[(\hat\gamma_j-\gamma_{j0})^\top p^{(b)}_{1j}(t)dt +p^{(b)}_{2j}(t)d\left\{\hat G_j(t)-G_{j0}(t)\right\}\right.
\nonumber\\
&&\left.\quad +dp^{(b)}_{2j}(t)\left\{\hat G_j(t)-G_{j0}(t)\right\}\right]+o_p(n^{-1/2})\nonumber\\
&=&\frac{1}{\sqrt{n}}\sum_{i=1}^n\int_0^\tau \left[ \sigma_{4i} p^{(b)}_{1j}(t)dt +p^{(b)}_{2j}(t)d\sigma_{5i}(t)+dp^{(b)}_{2j}(t)\sigma_{5i}(t)\right]+o_p(n^{-1/2}).\nonumber
\ee

Therefore we have:
\be\label{anpswnb}
\sqrt{n}{S}_{1,n}(\beta_0, S_{c0},  \pi_0,  \Lambda^{*})+\sqrt{n}{Q}^{(21)}=\frac{1}{\sqrt{n}}\sum_{i=1}^n \psi^{(b)}_i(t)+o_p(1)
\ee
where
\eqnn
\psi^{(b)}_{i,j}&=&\int_0^\tau e^{ (\beta_{10}  + \beta_{20}) A_i t} \left\{S^{*}_{c}(t \mid A_i,Z_i)\right\}^{-1}\left\{A_i -\pi^{*}(Z_i)\right\}dM_{ji}(t; \beta_{j0},\Lambda^{*}_j)\\&&+\int_0^\tau \left[ \sigma_{4i} p^{(b)}_{1j}(t)dt +p^{(b)}_{2j}(t)d\sigma_{5i}(t)dt+p^{(b)}_{2j}(t)\sigma_{5i}(t)\right].\een

By Theorem \ref{dr2} and by construction of $Q^{(21)}$ the right hand side of \eqref{decoanb} is a sum of i.i.d mean zero random variable and the multivariate central limit theorem can be applied.
Therefore, by the above together with \eqref{decoanb}, we can prove part b) of the asymptotic normality. 

\begin{itemize}
\item{Case (c):}
\end{itemize}
We remind the reader that, if $S_c^{*}(\cdot | \cdot,\cdot)=S_{c0}(\cdot | \cdot,\cdot)$, $\pi^{*}(\cdot)=\pi_{0}(\cdot)$ and $\Lambda^{*}(\cdot,\cdot)=\Lambda_{0}(\cdot,\cdot)$, we have ${Q}^{(21)}=o_p(n^{-1/2})$, ${Q}^{(3)}=o_p(n^{-1/2})$ and therefore the influence function in this case simplifies.
In the previous part of the proof we have proved that ${K}$ simplifies to a $2\times2$ diagonal matrix with:
$
{K}_{jj}=-K^{(1)}(\beta_0,S_{c0},\pi_0, \Lambda_{0})$.

Indeed by this, by consistency of $\hat \beta^{(1)}$ proved in the above  and by \eqref{decoan}, we have:
\be\label{anpswnc}
\sqrt{n}({\beta}-{\beta}_0)&=&{K}^{-1}\sqrt{n}{S}_{1,n}(\beta_0, S_{c0},  \pi_0,  \Lambda_{0})+o_p(1).\nonumber
\ee

We prove that $\sqrt{n}{S}_{1,n}(\beta_0, S_{c0},  \pi_0,  \Lambda_{0})$ is normal by martingale central limit theorem.
Since here we assume that we plug in the true parameters, for ease of notation, in the following we will suppress the dependency of the martingale on $\beta,\Lambda_{0}$. 
We consider the following multivariate martingale: 
${M}_i(t)=[M_{1i}(t), M_{2i}(t)]^\top$
with respect to the filtration $\mathcal{F}_t=\sigma\left\{N_{j}(s), Y(s+), A, Z\;:j=1,2,\;0<s<t\right\}$.
We consider the following two-dimensional vector:
$
{M}^n(t)=\frac{1}{\sqrt{n}} \sum_{i=1}^n\int_0^t h(u; A_i,Z_i) d{M}_i(u),
$
where 
\eqnn h(t; A,Z)= e^{ (\beta_{10}  + \beta_{20}) A_i t}\left\{S_{c0}(t \mid A,Z)\right\}^{-1}\left\{A -\pi_0( Z)\right\}.\een
Since $h(t; A,Z)$ is predictable with respect to the filtration, then ${M}^n(t)$ is a multivariate martingale too.
We have \eqnn<M_{1i}(t),M_{2i}(t)>=<M_{1i}(t),M_{1j}(t)>=<M_{2i}(t),M_{2j}(t)>=<M_{1i}(t),M_{2j}(t)>=0,\een for each $i\neq j$ therefore:
\eqnn
<{M}_1^n(t),{M}_2^n(t)>&=&<\frac{1}{\sqrt{n}} \sum_{i=1}^n\int_0^t h(u; A_i,Z_i) d{M}_{1i}(u),\frac{1}{\sqrt{n}} \sum_{i=1}^n\int_0^t h(u; A_i,Z_i) d{M}_{2i}(u)>\\&=&\frac{1}{n} \sum_{i,j=1}^n\int_0^t h^2(u; A_i,Z_i) d<M_{1i}(u),M_{2j}(u)>=0,
\een
and so the two components of the multidimensional martingale ${M}^n(t)$ are orthogonal to each other.
Therefore, we can apply the multidimensional version of the martingale central limit theorem of Rebolledo (Theorem 5 of \cite{rebolledo1978applications}). 

First we verify Assumption 2 about the convergence of the variance.
We have, by Assumption\ref{ass2n}, for $j=1,2$:
\eqnn
<{M}_j^n(t),{M}_j^n(t)>&=&\frac{1}{n}\sum_{i=1}^n\int_0^t h^2(u; A_i,Z_i) d\Lambda_{j}(u \mid A_i, Z_i)Y_i(u)\\
&=&\frac{1}{n}\sum_{i=1}^n\int_0^t h^2(u; A_i,Z_i) \left\{d\Lambda_{j0}(u,Z_i) + \beta_{j0}Adu\right\}Y_i(u)\\
&=&\frac{1}{n}\sum_{i=1}^n\int_0^t e^{2(\beta_{10}  + \beta_{20} )A_i u}\left\{S_{c0}(u \mid A_i,Z_i)\right\}^{-2} \left\{A_i-\pi_0(Z_i)\right\}^2\\&&\times\left\{d\Lambda_{j0}(u,Z_i) + \beta_{j0}A_idu\right\}Y_i(u)\nonumber\\
&=&\int_0^t  \left\{P(u)\beta_j+Q_j(u)\right\}du\overset{p}{\rightarrow} \int_0^t  \left\{p(u)\beta_j+q_j(u)\right\}du=V_j(t),
\een
and so Assumption 2 of the MCLT is verified.

We now look at Assumption 1 about the jumps of each component of the martingale. \cite{rebolledo1978applications} at pag. 39 claims that if the Lindeberg condition is verified, then Assumption 1 of its theorem holds.
We therefore needs to prove that, for any $\epsilon$ and any $j$:
\eqnn
\int_0^\tau \frac{1}{n}\sum_{i=1}^n h^2(u ; A_i,Z_i) \mathbbm{1}\left\{\left|h(u ; A_i,Z_i)\right|>\sqrt{n}\epsilon\right\}Y_i(t)\left\{d\Lambda_{j0}(t, Z_i) + \beta_{j0} A_idt\right\} \overset{p}{\rightarrow} 0,
\een
by Assumption\ref{ass0n} and \ref{ass4n}, we know that: 
\eqnn
\left|h(t; A,Z)\right|\leq C^{-1}_ce^{ (\beta_{10}  + \beta_{20}) \tau}<\infty,
\een
so, we have:
\eqnn
&&\int_0^\tau \frac{1}{n}\sum_{i=1}^n h^2(u ; A,Z) \mathbbm{1}\left\{\left|h(u ; A,Z)\right|>\sqrt{n}\epsilon\right\}Y_i(t)\left\{d\Lambda_{j0}(t, Z_i) + \beta_{j0} A_idt\right\} \\
&\leq& \int_0^\tau \frac{1}{n}\sum_{i=1}^n h^2(u ; A,Z) \mathbbm{1}\left\{C^{-1}_c\exp(\beta_{10}\tau +\beta_{20} \tau)>\sqrt{n}\epsilon\right\}Y_i(t)
\\
&&\times\left\{d\Lambda_{j0}(t, Z_i) + \beta_{j0} A_idt\right\}.
\een
Moreover, by Assumption\ref{boundln}, we also know that: 
\eqnn
&&\left| \int_0^\tau \frac{1}{n}\sum_{i=1}^n h^2(u ; A,Z) \mathbbm{1}\left\{C^{-1}_ce^{ (\beta_{10}  + \beta_{20}) \tau}>\sqrt{n}\epsilon\right\}Y_i(t)\left\{d\Lambda_{j0}(t, Z_i) + \beta_{j0} A_idt\right\}\right| \\
 &\leq & C^{-2}_ce^{2(\beta_{10}  +\beta_{20})\tau} \mathbbm{1}\left\{C^{-1}_ce^{ (\beta_{10}  + \beta_{20})\tau}>\sqrt{n}\epsilon\right\}\tau \left|L_j + \beta_{j0} \right|  \overset{p}{\rightarrow} 0,
\een
and so Assumption 1 of the martingale central limit theorem holds.
 
Therefore, we can conclude that 
\eqnn
\sqrt{n}S_{1,n}(\beta_0, S_{c0}, \pi_0, \Lambda_{0})={M}^n(t)\overset{D}{\rightarrow}\mathcal{N}(0,V(\tau)).
\een


\textbf{\emph{Consistent variance estimator:}}
We now prove  that 
$\hat{W}^{(c)}_{jj}-{W}^{(c)}_{jj}=o_p(1)$
and that 
$
\hat V^{(c)}_{jj}(\tau)-V^{(c)}_{jj}(\tau)=o_p(1).
$

We have:
\eqnn
&&\hat{W}^{(c)}_{jj}-{W}^{(c)}_{jj}\\&&=\frac{1}{n}\sum_{i=1}^nA_i\left\{A_i-\hat\pi(Z_i)\right\}\int_0^{X_i}\left\{\hat S_c(t \mid A_i,Z_i)\right\}^{-1}e^{(\hat\beta_1+\hat \beta_2)t}dt\\&&-E\left[A\left\{A-\pi_0(Z)\right\}\int_0^{X}\left\{S_{c0}(t \mid A,Z)\right\}^{-1}e^{(\beta_{10}+\beta_{20})t}dt\right]\nonumber\\
&&=\frac{1}{n}\sum_{i=1}^nA_i\left\{A_i-\pi_0(Z_i)\right\}\int_0^{X_i}\left\{ S_{c0}(t \mid A_i,Z_i)\right\}^{-1}e^{(\beta_{10}+ \beta_{20})t}dt\\
&&-E\left[A\left\{A-\pi_0(Z)\right\}\int_0^{X}\left\{S_{c0}(t \mid A,Z)\right\}^{-1}e^{(\beta_{10}+\beta_{20})t}dt\right]\\
&&+\frac{1}{n}\sum_{i=1}^n\int_0^{X_i}A_i\left\{A_i-\pi_0(Z_i)\right\}\left\{S_{c0}(t \mid A_i,Z_i)\right\}^{-1}\left\{e^{(\hat\beta_1+\hat \beta_2)t}-e^{(\beta_{10}+ \beta_{20})t}\right\}dt\\
&&+\frac{1}{n}\sum_{i=1}^n\int_0^{X_i}A_i\left[\left\{A_i-\hat\pi(Z_i)\right\}\left\{\hat S_c(t \mid A_i,Z_i)\right\}^{-1}-\left\{A_i-\pi_0(Z_i)\right\}\left\{S_{c0}(t \mid A_i,Z_i)\right\}^{-1}\right]\\&&\times\left\{e^{(\hat\beta_1+\hat \beta_2)t}-e^{(\beta_{10}+ \beta_{20})t}\right\}dt\nonumber
\\&&+\frac{1}{n}\sum_{i=1}^n\int_0^{X_i}A_i\left[\left\{A_i-\hat\pi(Z_i)\right\}\left\{\hat S_c(t \mid A_i,Z_i)\right\}^{-1}-\left\{A_i-\pi_0(Z_i)\right\}\left\{ S_{c0}(t \mid A_i,Z_i)\right\}^{-1}\right]\nonumber\\
&&\times e^{(\beta_{10}+ \beta_{20})t}dt\nonumber\\
&&=Q_1-Q_2+Q_3+Q_4+Q_5.
\nonumber\een
As before, by Assumptions \ref{ass0n} and \ref{ass4n}, by Hoeffding's inequality we have:
$
Q_1-Q_2=O_p(n^{-1/2}).
$
by Assumptions \ref{ass0n} and \ref{ass4n} and by Lemma \ref{mvt}, we get:
\eqnn
|Q_3|\leq \tau C_c^{-1}\left\{e^{(\hat\beta_1+\hat \beta_2)\tau}-e^{(\beta_{10}+ \beta_{20})\tau}\right\}=\tau C_c^{-1}e^{(\beta^*_1+\beta^*_2)\tau}\tau\left(\hat \beta_1+\hat \beta_2-\beta_{10}-\beta_{20}\right).
\een
where $\beta^{*}_j$ are points between $\hat \beta_j$ and $\beta_{j0}$.
Therefore, by consistency of the estimator $\hat \beta$, we have $|Q_3|=o_p(1)$.\\
By Assumption \ref{assumption} and by consistency of $\beta$ we have $Q_4=o_p(1)$.
By this and by Assumptions \ref{assumption}, \ref{ass0n}, \ref{ass4n}, we have:
\be
\left|Q_5\right|&\leq&\sup_{t \in [0,\tau], Z \in \mathcal{Z}, A \in {0,1}}\left\{\left|\hat S^{-1}_c(t \mid A,Z)- \left\{S_{c0}(t \mid A,Z)\right\}^{-1}\right|+\left|\hat S^{-1}_c(t \mid A,Z)\right|\left|\hat \pi( Z) - \pi_0( Z)\right|\right\}\nonumber\\&&\times \tau e^{(\beta_{10}+\beta_{20})\tau}\nonumber\\ &=&o_p(1).
\ee
We can therefore conclude that $\hat{W}-W=o_p(1)$.

We have:
\eqnn
&&\hat V^{(c)}_{jj}(\tau)-V^{(c)}_{jj}(\tau)
\\&=&\frac{1}{n}\sum_{i=1}^n \int_0^\tau e^{2(\hat\beta_1+\hat \beta_2)A_i X_i} \hat S^{-2}_c(X_i \mid A_i,Z_i)\left\{A_i-\hat\pi(Z_i)\right\}^2dN_{ji}(t)-\int_0^\tau \left\{p(t)\beta_j+q_j(t)\right\}dt\nonumber\\
&=&\frac{1}{n}\sum_{i=1}^n \int_0^\tau e^{2(\beta_{10}+ \beta_{20})A_i t} \left\{S_{c0}(X_i \mid A_i,Z_i)\right\}^{-2}\left\{A_i-\pi_0(Z_i)\right\}^2dM_{ji}(t)\\
&& +\int_0^\tau \left\{P(t)\beta_j+Q_j(t)-p(t)\beta_j-q_j(t)\right\}dt\\
&&+\frac{1}{n}\sum_{i=1}^n\int_0^{\tau} \left\{S_{c0}(X_i \mid A_i,Z_i)\right\}^{-2}\left\{A_i-\pi_0(Z_i)\right\}^2\left\{e^{2(\hat\beta_1+\hat \beta_2)t}-e^{2(\beta_{10}+ \beta_{20})t}\right\}dN_{ji}(t)\nonumber\\
&&+\frac{1}{n}\sum_{i=1}^n\int_0^{\tau}\left[\hat S^{-2}_c(X_i \mid A_i,Z_i)\left\{A_i-\hat\pi(Z_i)\right\}^2-\left\{S_{c0}(X_i \mid A_i,Z_i)\right\}^{-2}\left\{A_i-\pi_0(Z_i)\right\}^2\right]\nonumber\\&&\times \left\{e^{2(\hat\beta_1+\hat \beta_2)t}-e^{2(\beta_{10}+ \beta_{20})t}\right\}dN_{ji}(t)\\
&&+\frac{1}{n}\sum_{i=1}^n\int_0^{\tau}\left[\hat S^{-2}_c(X_i \mid A_i,Z_i)\left\{A_i-\hat\pi(Z_i)\right\}^2-\left\{ S_{c0}(t \mid A_i,Z_i)\right\}^{-2}\left\{A_i-\pi_0(Z_i)\right\}^2\right]\nonumber\\&&\times e^{2(\beta_{10}+ \beta_{20})t}dN_{ji}(t)\\
&=&E_1+E_2+E_3+E_4+E_5.
\een

For $E_1$ we notice that, by Assumptions \ref{ass0n} and \ref{ass4n}, we have:
\eqnn
e^{2(\beta_{10}+ \beta_{20})A_i X_i} \left\{S_{c0}(X_i \mid A_i,Z_i)\right\}^{-2}\left\{A_i-\pi_0(Z_i)\right\}^2\leq C_c^{-2} e^{2(\beta_{10}+ \beta_{20})A_i \tau}<\infty,
\een
and so, by Lemma \ref{ci}, we have $E_1=o_p(1)$.\\
By Assumption\ref{ass2n}, we can prove that $E_2=o_p(1)$.\\
Similarly to what we have done for $Q_3$, $Q_4$ and $Q_5$, we can prove that $E_3=o_p(1)$, $E_4=o_p(1)$ and $E_5=o_p(1)$.\\
Therefore $\hat V_j-V_j=o_p(1)$.


\newpage
\subsubsection{Asymptotic variance for specific working models} 

Under the  assumptions required for  Score 1, 

(a) if $S^{*}_c(t | a,z)=S_{c0}(t | a,z)=\exp\left(-\Lambda_{c0}(t)e^{\eta^\top_0 d}\right)$ where $d=[a, z]'$, and $\pi^{*}(z)=\pi_0(z)=\left\{1+\exp(-\alpha^\top z)\right\}^{-1}$, 
then $ \Sigma^{(a)} =  \Sigma^{(a')}$ where the explicit expression for $ \Sigma^{(a')}$ is given below under additional regularity Assumptions  A*\ref{convnn}-\ref{posdefbisn}, also given below;

(b) if  $\Lambda^{*}(t,z)=\Lambda_0(t,z)=G_{0}(t) + \gamma_{0}^{\top} z t$, where a quantity without subscript $j$ indicates a vector of both components for $j=1,2$, 
and $G_{j}(t)$ is estimated using \eqref{estbas1} in the main paper, then $ \Sigma^{(b)} =  \Sigma^{(b')}$ where the explicit expression for $ \Sigma^{(b')}$ is given below under additional regularity Assumptions  B*\ref{if1nnn} ,\ref{if2n}, \ref{posdefn};
if $G_j(t)$ is estimated using \eqref{estbas2},
 then $ \Sigma^{(b)} =  \Sigma^{(b'')}$ where the explicit expression for $ \Sigma^{(b'')}$ is given below under additional regularity Assumptions  B*\ref{if1nnn} ,\ref{if3n}, \ref{posdefn}.



\eqnn
\Sigma^{(a')} &=&\left\{{E}(K^{(a)})^{-1}\right\}^\top {Var}(\psi^{(a')}){E}(K^{(a)})^{-1},\\
\Sigma^{(b')} &=& \left\{{E}(K^{(b')})^{-1}\right\}^\top{Var}(\psi^{(b')}){E}(K^{(b')})^{-1},\\
\Sigma^{(b'')} &=& \left\{{E}(K^{(b'')})^{-1}\right\}^\top{Var}(\psi^{(b'')}){E}(K^{(b'')})^{-1},
\een
where $K^{(a)},K^{(b')},K^{(b'')},\psi^{(a')},\psi^{(b')},\psi^{(b'')}$ are given below.
\eqnn
\psi^{(a')}&=&\left[\int_0^\tau e^{ (\beta_{10}  + \beta_{20}) At} \left\{S_{c0}(t | A,Z)\right\}^{-1}\left\{A -\pi_0( Z)\right\}dM_{j}(t; \beta_{j0},\Lambda^{*}_j)\right.\\
&&\left. +\left(\int_0^\tau\left[\frac{s_d^{(2)}(t)}{s_d^{(0)}(t)}-\left\{\frac{s_d^{(1)}(t)}{s_d^{(0)}(t)}\right\}^2\right]s_d^{(0)}(t)d\Lambda_{c0}(t)\right)^{-1}\int_0^\tau \left\{D-\frac{s_d^{(1)}(t)}{s_d^{(0)}(t)}\right\}dM^c(t)
\right.\\
&&\left. 
\times\left[\int_0^\tau \{p^{(a')}_{1j}\}^\top(t)dt-\int_0^\tau p^{(a')}_{2j}(t)\int_0^t d\Lambda_{c0}(u; \eta_0)\frac{s_d^{(1)}(u)}{s_d^{(0)}(u)}du dt\right]
\right.\\
&&\left. +\int_0^\tau p^{(a')}_{2j}(t)\int_0^t \left\{s_d^{(0)}(u)\right\}^{-1}dM^c(u)dt
\right.\\
&&\left.- \int_0^\tau\{p^{(a')}_{3j}\}^\top(t) \left(E\left[Z^\top Z \pi_0(Z)\left\{1-\pi_0(Z)\right\}\right]\right)^{-1}Z\left\{A-\pi_0(Z)\right\}dt\right]_{j=1,2}.
\een
\eqnn
\psi^{(b')}&=&\left[\int_0^\tau e^{ (\beta_{10}  + \beta_{20}) A t} \left\{S^{*}_{c}(t | A,Z)\right\}^{-1}\left\{A -\pi^{*}( Z)\right\}dM_{j}(t)\right.\\&&\left.
+\int_0^\tau \left(\left[\frac{1}{n}\sum_{i=1}^n\int_0^{\tau}\left\{D_i-\frac{s_d^{(1)}(t)}{s_d^{(0)}(t)}\right\}^{\otimes 2}Y_i(t)dt\right]^{-1}\int_0^\tau \left\{Z-\frac{s_z^{(1)}(t)}{s_z^{(0)}(t)}\right\}dM_{j}(t)\right)^\top 
\right.\\
&&\left.
\times
\left\{p^{(b')}_{1}(t)dt -p^{(b')}_{0}(t)\frac{s_{z}^{(1)}(t)}{s_{z}^{(0)}(t)}dt\right\}
+\int_0^\tau p^{(b')}_{0}(t)\left\{s_{z}^{(0)}(t)\right\}^{-1}dM_{j}(t)
\right]_{j=1,2}.
\een
${K^{(b')}}$ is a $2\times2$ diagonal matrix with:
\eqnn
{K}^{(b')}_{jj}&=&E\left[\int_0^\tau e^{ (\beta_{10}  + \beta_{20}) A t}\left\{S^{*}_c(t | A, Z)\right\}^{-1}\left\{A - \pi^{*}( Z)\right\}Y(t)\left[A-\frac{s_a^{(1)}(t)}{s^{(0)}(t)}\right]dt\right].
\een
\eqnn
\psi^{(b'')}&=&\left[\int_0^\tau e^{ (\beta_{10}  + \beta_{20}) A t} \left\{S^{*}_{c}(t | A,Z)\right\}^{-1}\left\{A -\pi^{*}( Z)\right\}dM_{j}(t)\right.\\&&\left.
+\int_0^\tau \left(\left[\frac{1}{n}\sum_{i=1}^n\int_0^{\tau}\left\{D_i-\frac{s_d^{(1)}(t)}{s_d^{(0)}(t)}\right\}^{\otimes 2}Y_i(t)dt\right]^{-1}\int_0^\tau \left\{Z-\frac{s_z^{(1)}(t)}{s_z^{(0)}(t)}\right\}dM_{j}(t)\right)^\top 
\right.\\
&&\left.
\cdot
\left\{p^{(b')}_{1}(t)dt -p^{(b')}_{0}(t)\frac{s_{wz}^{(1)}(t; S^{*}_c,\pi^{*})}{s_{wz}^{(0)}(t; S^{*}_c,\pi^{*})}dt\right\}
+\int_0^\tau p^{(b')}_{0}(t)\left\{s_{wz}^{(0)}(t; S^{*}_c,\pi^{*})\right\}^{-1}dM_{j}(t)
\right]_{j=1,2}.
\een
${K^{(b'')}}$ is a $2\times2$ diagonal matrix with:
\eqnn
{K}^{(b'')}_{jj}=E\left(\int_0^\tau e^{ (\beta_{10}  + \beta_{20}) A t}\left\{S^{*}_c(t | A,Z)\right\}^{-1}\left\{A - \pi^{*}( Z)\right\}Y(t)\left[A-\frac{s_{aw}^{(1)}(t; S^{*},\pi^{*})}{s_{aw}^{(0)}(t;  S^{*},\pi^{*})}\right]dt\right).
\een

\vskip .4in
\begin{assumptionA*}\label{convnn}
For $j=1,2$ let
\eqnn
P^{(a')}_{1j}(t)
&=&\frac{1}{n}\sum_{i=1}^n e^{(\beta_{10}+\beta_{20})A_it}\left\{A_i- \mbox{expit}(\alpha_0^\top Z_i)\right\}\exp\left(\Lambda_{c0}(t)e^{\eta_0^\top D_i}\right)\Lambda_{c0}(t)e^{\eta_0^\top D_i}D_idM_{ji}(t; \beta_{j0}, \Lambda^{*}_j),\\
P^{(a')}_{2j}(t)&=&\frac{1}{n}\sum_{i=1}^n e^{(\beta_{10}+\beta_{20})A_it}\left\{A_i-\expit(\alpha_0^\top Z_i)\right\}\exp\left(\Lambda_{c0}(t)e^{\eta_0^\top D_i}\right)e^{\eta_0^\top D_i}dM_{ji}(t; \beta_{j0}, \Lambda^{*}_j),\\
P^{(a')}_{3j}(t) &=& -\frac{1}{n}\sum_{i=1}^n e^{(\beta_{10}+\beta_{20})A_it}\exp\left(\Lambda_{c0}(t)e^{\eta_0^\top D_i}\right)\expit(\alpha_0^\top Z_i)e^{\alpha_0^\top Z_i}Z_idM_{ji}(t; \beta_{j0}, \Lambda^{*}_j),
\een
where $D=[A,Z]^\top$.
There exist, for $l=1,2,3$, some bounded $p^{(a)}_{lj}(t)$ and a neighborhood $\mathcal{B}$ of  $\{\beta_0, S_{c0}(\cdot | \cdot,\cdot), \pi_0(\cdot), \Lambda^{*}(\cdot,\cdot)\}$ such that:
\eqnn
\sup_{t \in [0,\tau], \left\{\beta,S_c, \pi, \Lambda_j\right\} \in \mathcal{B}} \left\Vert P^{(a)}_{lj}(t)-p^{(a)}_{lj}(t)\right\Vert \overset{p}{\rightarrow} 0,
\een
where we indicate with $\Vert \cdot \Vert$ the $L^2$ norm.
\end{assumptionA*}

\begin{assumptionA*}\label{boundbasn}
$
\sup_{Z \in \mathcal{Z}}|\Lambda^{*}_j(\tau,Z)|<\infty.
$
\end{assumptionA*}

\begin{assumptionA*}\label{ifn}
For $l=0,1,2$ let
\eqnn
S_d^{(l)}(t)=\frac{1}{n}\sum_{i=1}^nY_i(t)D^{l}_ie^{\eta_0^\top D_i}.
\een
There exist some bounded $s_d^{(l)}(t)$ such that:
\eqnn
\sup_{t \in [0,\tau]} \left\Vert S_d^{(l)}(t)-s_d^{(l)}(t)\right\Vert \overset{p}{\rightarrow} 0.
\een
\end{assumptionA*}

\begin{assumptionA*}\label{posdefbisn}
\eqnn\int_0^\tau \left\{\frac{s_d^{(2)}(t)}{s_d^{(0)}(t)}-\left(\frac{s_d^{(1)}(t)}{s_d^{(0)}(t)}\right)^2\right\}E\left\{Y(t)e^{\eta_0^\top D}\right\}d\Lambda_{c0}(t)\een and 
$ E\left[Z^\top Z \pi_0(Z_i)\left\{1-\pi_0(Z_i)\right\}\right]$ are positive definite.
\end{assumptionA*}

\begin{assumptionB*}\label{if1nnn}
For $l=0,1$ let
\eqnn
P^{(b')}_{1}(t) = \frac{1}{n}\sum_{i=1}^n e^{ (\beta_{10}  + \beta_{20}) A_i t} \left\{S^{*}_c(t | A_i,Z_i)\right\}^{-1}\left\{A_i -\pi^{*}( Z_i)\right\}Y_i(t)Z_i^l.
\een
We assume that, for $l=0,1$, there exist $p^{(b')}_{l}(t)$ and a neighborhood $\mathcal{B}$ of  \\$\{\beta_0,\Lambda_{0}(\cdot,\cdot), S^{*}_{c}(\cdot | \cdot,\cdot), \pi^{*}(\cdot)\}$ such that :
\eqnn
\sup_{t \in [0,\tau],  \left\{\beta,\Lambda, S_c, \pi\right\} \in \mathcal{B}} \left\Vert P^{(b')}_{l}(t)-p^{(b')}_{l}(t)\right\Vert \overset{p}{\rightarrow} 0.\een
\end{assumptionB*}

\begin{assumptionB*}\label{if2n}
For $l=0,1$ let
\eqnn
S_d^{(1)}(t) &=& \frac{1}{n}\sum_{i=1}^nY_i(t)D_i,\\
S_z^{(1)}(t) &=& \frac{1}{n}\sum_{i=1}^nY_i(t)Z_i,\\
S^{(0)}(t) &=& \frac{1}{n}\sum_{i=1}^nY_i(t),
\een
where $D=[A,Z]^\top$.
There exist $s_d^{(1)}(t),s_z^{(1)}(t),s^{(0)}(t)$ such that:
\eqnn
\sup_{t \in [0,\tau]} \left\Vert S_d^{(1)}(t)-s_d^{(1)}(t)\right\Vert &\overset{p}{\rightarrow} & 0,\\
\sup_{t \in [0,\tau]} \left\Vert S_z^{(1)}(t)-s_z^{(1)}(t)\right\Vert  &\overset{p}{\rightarrow} & 0,\\
\sup_{t \in [0,\tau]} \left|S^{(0)}(t)-s^{(0)}(t)\right| &\overset{p}{\rightarrow} & 0.\een
\end{assumptionB*}

\begin{assumptionB*}\label{if3n}
Let for $l=0,1$
\eqnn
S_{wd}^{(l)}(t; S^{*},\pi^{*}) &=& \frac{1}{n}\sum_{i=1}^nw_i(S^{*},\pi^{*})Y_i(t)D^l_i, \\
S_{wz}^{(l)}(t; S^{*},\pi^{*}) &=& \frac{1}{n}\sum_{i=1}^nw_i(S^{*},\pi^{*})Y_i(t)Z_i^l.
\een
We assume that, there exist $s_{wd}^{(l)}(t; S^{*},\pi^{*}),s_{wz}^{(l)}(t; S^{*},\pi^{*})$ such that:
\eqnn
\sup_{t \in [0,\tau]} \left\Vert S_{wd}^{(l)}(t; S^{*},\pi^{*})-s_{wd}^{(l)}(t; S^{*},\pi^{*})\right\Vert  &\overset{p}{\rightarrow} & 0,\\
\sup_{t \in [0,\tau]} \left\Vert S_{wz}^{(l)}(t; S^{*},\pi^{*})-s_{wz}^{(l)}(t; S^{*},\pi^{*})\right\Vert  &\overset{p}{\rightarrow} & 0.
\een
\end{assumptionB*}

\begin{assumptionB*}\label{posdefn}
$\int_0^{\tau}E\left[\left\{D-{s_d^{(1)}(t)}/ {s_d^{(0)}(t)}\right\}^{\otimes 2}Y(t)\right]dt$ is positive definite.
\end{assumptionB*}


\vskip .4in
The proof of the results 
 follows directly from finding the influence functions defined in Assumptions A\ref{ifc} and  B\ref{ifl} for the specific working models $S_c(t | a,z; \eta, \Lambda_c)=\exp\left(-\Lambda_{c}e^{\eta^\top d}\right)$, where $D=[A,Z]^\top$ and $\pi(z ; \alpha)=\left\{1+\exp(-\alpha^\top z)\right\}^{-1}$ for case a) and $\Lambda_j(t,z; G_j, \gamma_j)=G_{j}(t) + \gamma_{j}^{\top} z t$ for case b) respectively. 
We indeed remind the reader that under case a), we have: 
\eqnn
\sqrt{n}(\hat{{\beta}}-{\beta}_{0})={K}^{-1}\sqrt{n}\left\{S_{1,n}(\beta_0, S^{*}_c,  \pi^{*}, \Lambda^{*})+{Q}^{(3)}\right\}+o_p(1),
\een
where $Q^{(3)}$, defined in Lemma \ref{lm:deco}, directly depends on the form of the influence functions of estimators $\hat \alpha, \hat \Lambda_{c},\hat \eta$.
Under case b), we instead have:
\eqnn
\sqrt{n}(\hat{{\beta}}-{\beta}_{0})={K}^{-1}\sqrt{n}\left\{S_{1,n}(\beta_0, S^{*}_c,  \pi^{*}, \Lambda^{*})+{Q}^{(21)}\right\}+o_p(1),
\een
where $Q^{(21)}$, defined in Lemma \ref{lm:deco}, directly depends on the form of the influence functions of estimators $\hat G, \hat \gamma$.

The next Lemma defines the specific form of $Q^{(3)}$ when the logistic model and the Cox model are assumed on the propensity score and the censoring distribution, respectively.
\begin{lemmas}\label{ifQ4}
We assume, $\pi(Z ; \alpha)=\left\{1+\exp(-\alpha^\top Z)\right\}^{-1}$
and $S_c(t | A,Z; \eta, \Lambda_c)=\exp\left(-\Lambda_{c}e^{\eta^\top D}\right)$.
Under Assumptions \ref{boundln}, \ref{boundZn} and Assumption A*\ref{convnn}- A*\ref{posdefbisn} we have:
\eqnn
\sqrt{n}Q^{(3)}_j
&=& \left(\int_0^\tau\left[\frac{s_d^{(2)}(t)}{s_d^{(0)}(t)}-\left\{\frac{s_d^{(1)}(t)}{s_d^{(0)}(t)}\right\}^2\right]s_d^{(0)}(t)d\Lambda_{c0}(t)\right)^{-1}\frac{1}{\sqrt{n}}\sum_{i=1}^n\int_0^\tau \left\{D_i-\frac{s_d^{(1)}(t)}{s_d^{(0)}(t)}\right\}dM^c_i(t)
\\
&&\times\left[\int_0^\tau \left\{p^{(a')}_{1}\right\}^\top(t)dt-\int_0^\tau p^{(a')}_{2}(t)\int_0^t d\Lambda_{c0}(u; \eta_0)\frac{s_d^{(1)}(u)}{s_d^{(0)}(u)}dt\right]
\\
&& +\int_0^\tau p^{(a')}_{2}(t)\frac{1}{\sqrt{n}}\sum_{i=1}^n\int_0^t \left\{s_d^{(0)}(u)\right\}^{-1}dM^c_i(u)
\\
&&- \int_0^\tau\left\{p^{(a')}_{3}\right\}^\top(t) \left(E\left[Z^\top Z \pi_0(Z_i)\left\{1-\pi_0(Z_i)\right\}\right]\right)^{-1}\frac{1}{\sqrt{n}}\sum_{i=1}^n Z_i\left\{A_i-\pi_0(Z_i)\right\}dt
\\&&+o_p(1).\een
\end{lemmas}

The next Lemma defines the specific form of $Q^{(21)}$ when the traditional additive hazard model is assumed on the cause-specific hazards.
\begin{lemmas}\label{ifQ21a}
Let $\Lambda_j(t,Z ; G_j, \gamma_j)=G_j(t) + \gamma_j^{\top} Z t$ and let  $\gamma_j$ be estimated by \eqref{tradgamma} in the main document and $G_j(t)$ be estimated using \eqref{estbas1} in the main document.
Under Assumptions \ref{boundln}, \ref{boundZn} and Assumptions B*\ref{if2n} and B*\ref{posdefn} it holds:
\eqnn
\sqrt{n}Q^{(21)}_j&=&\int_0^\tau \left(\left[\frac{1}{n}\sum_{i=1}^n\int_0^{\tau}\left\{D_i-\frac{s_d^{(1)}(t)}{s_d^{(0)}(t)}\right\}^{\otimes 2}Y_i(t)dt\right]^{-1}\frac{1}{\sqrt{n}}\sum_{i=1}^n
\int_0^\tau \left\{Z_i-\frac{s_z^{(1)}(t)}{s_z^{(0)}(t)}\right\}dM_{ji}(t)\right)^\top 
\\
&&
\times
\left\{p^{(b')}_{1}(t)dt -p^{(b')}_{0}(t)\frac{s_z^{(1)}(t)}{s_z^{(0)}(t)}dt\right\}
\\
&&+\int_0^\tau p^{(b')}_{0}(t)\left\{s_z^{(0)}(t)\right\}^{-1}\frac{1}{\sqrt{n}}\sum_{i=1}^ndM_{ji}(t)+o_p(1).
\een
\end{lemmas}
\begin{lemmas}\label{ifQ21b}
Let $\Lambda_j(t,Z ; G_j, \gamma_j)=G_j(t) + \gamma_j^{\top} Z t$ and let  $\gamma_j$ be estimated by \eqref{tradgamma} in the main document and $G_j(t)$ be estimated using \eqref{estbas2} in the main document.
Under Assumptions \ref{boundln}, \ref{boundZn} and Assumptions B*\ref{if3n} and B*\ref{posdefn} it holds:
\eqnn \sqrt{n}Q^{(21)}_j&=&\int_0^\tau \left(\left[\frac{1}{n}\sum_{i=1}^n\int_0^{\tau}\left\{D_i-\frac{s_d^{(1)}(t)}{s_d^{(0)}(t)}\right\}^{\otimes 2}Y_i(t)dt\right]^{-1}\frac{1}{\sqrt{n}}\sum_{i=1}^n
\int_0^\tau \left\{Z_i-\frac{s_z^{(1)}(t)}{s_z^{(0)}(t)}\right\}dM_{ji}(t)\right)^\top 
\\
&&
\times
\left\{p^{(b')}_{1}(t)dt -p^{(b')}_{0}(t)\frac{s_{wz}^{(1)}(t; S^{*}_c,\pi^{*})}{s_{wz}^{(0)}(t; S^{*}_c,\pi^{*})}dt\right\}
\\
&&+\int_0^\tau p^{(b')}_{0}(t)\left\{s_{wz}^{(0)}(t; S^{*}_c,\pi^{*})\right\}^{-1}\frac{1}{\sqrt{n}}\sum_{i=1}^ndM_{ji}(t)+o_p(1).\een
\end{lemmas}

Finally, the asymptotic results for $\beta^{(1)}$ together with Lemma \ref{ifQ4} proves part a) of the specific results;
with Lemma \ref{ifQ21a} proves part b1) of the specific results; and
with Lemma \ref{ifQ21b} proves part b2) of the specific results.


\newpage
\subsubsection{Asymptotic properties using Score 2}
Proofs of asymptotic properties of $\beta^{(2)}$
uses similar ideas and techniques as for $\beta^{(1)}$, 
under Assumption \ref{ass0n}-\ref{ass4n} and \ref{lowerboundsc1}. 
We therefore report here a sketch of their proofs.


\textbf{\emph{Consistency:}}
Under Assumptions \ref{assumption}, it follows that:
$
{S}_{2,n}({\beta}, \hat S_c, \hat \pi, \hat \Lambda)=S_{2,n}({\beta}, S^{*}_{c},  \pi^{*}, \Lambda^{*})+O_p(n^{-1/2}).
$

By Taylor expansion we have:
\eqnn
S_{2,n}({\beta}, S^{*}_{c},  \pi^{*}, \Lambda^{*})=S_{2,n}({\beta}_0, S^{*}_c,  \pi^{*}, \Lambda^{*})+\nabla_{{\beta}}S_{2,n}({\beta}^{*}, S^{*}_c,  \pi^{*}, \Lambda^{*})\left({\beta}-{\beta}_0\right)^\top,
\een
where ${\beta}^{*}$ lies between ${\beta}$ and ${\beta}_0$.

By double robustness of the score (Theorem \ref{dr2}), and by application of Hoeffding's inequality under Assumptions  \ref{boundln} and \ref{ass4n}, we have:
$
S_{2,n}({\beta}_0, S^{*}_c,  \pi^{*}, \Lambda^{*})=E\left\{S_{2,n}({\beta}_0, S^{*}_c,  \pi^{*}, \Lambda^{*})\right\}+O_p(n^{-1/2})=O_p(n^{-1/2}).
$
Therefore we have:
\be\label{decosc1n}
S_{2,n}({\beta}, \hat S_c, \hat \pi, \hat \Lambda)=\nabla_{{\beta}}S_{2,n}({\beta}^{*}, S^{*}_c,  \pi^{*}, \Lambda^{*})\left({\beta}-{\beta}_0\right)^\top + O_p(n^{-1/2}).
\ee

We now focus on $\nabla:=\nabla_{{\beta}}S_{2,n}({\beta}^{*}, S^{*}_{c},  \pi^{*}, \Lambda^{*})$.
We have as diagonal element, for $j=1,2$:
\eqnn
\nabla_{jj}&=&\partial_{\beta_j}\left\{{S}_{2,n}\right\}_j({\beta}^{*}, S^{*}_{c},  \pi^{*}, \Lambda^{*})\\
&=&-\frac{1}{n}\sum_{i=1}^n\int_0^\tau  \partial_{\beta_j}{\cal E}_i(t; {\beta}^{*}, S^{*}_c,\pi^{*})dM_{ji}(t ; \beta^{*}_j, \Lambda_j^{*})\\
&&+\frac{1}{n}\sum_{i=1}^n\int_0^\tau  \left\{A_i-{\cal E}_i(t; {\beta}^{*}, S^{*}_c,\pi^{*})\right\}A_iY_i(t)\\
&&+\frac{1}{n}\sum_{i=1}^n\int_0^\tau  \left\{A_i- {\cal E}_i(t; {\beta}^{*}, S^{*}_c,\pi^{*})\right\}Y_i(t)\partial_{{\beta}}d\Lambda^{*}(t,Z_i; \beta^{*})\\
&=&Q_1+Q_2+Q_3.
\een
We have:
\eqnn
Q_1&=&-\frac{1}{n}\sum_{i=1}^n\int_0^\tau  \partial_{\beta_j}{\cal E}_i(t; {\beta}^{*}, S^{*}_c,\pi^{*})dM_{ji}(t)\\
&&+(\beta^{*}_j-\beta_{j0})\frac{1}{n}\sum_{i=1}^n\int_0^\tau  \partial_{\beta_j}{\cal E}_i(t; {\beta}^{*}, S^{*}_c,\pi^{*})A_iY_i(t)dt \\
&&+\frac{1}{n}\sum_{i=1}^n\int_0^\tau  \partial_{\beta_j}{\cal E}_i(t; {\beta}^{*}, S^{*}_c,\pi^{*})Y_i(t)d\left\{\Lambda_j^{*}(t,Z_i ; \beta^{*}_j)-\Lambda_j^{*}(t,Z_i ; \beta_{j0})\right\}\\
&&+\frac{1}{n}\sum_{i=1}^n\int_0^\tau  \partial_{\beta_j}{\cal E}_i(t; {\beta}^{*}, S^{*}_c,\pi^{*})Y_i(t)d\left\{\Lambda_j^{*}(t,Z_i ; \beta_{j0})-\Lambda_{j0}(t,Z_i)\right\}\\
&=&Q_{11}+Q_{12}+Q_{13}+Q_{14}.
\een
$Q_{11}$ is a martingale integral with bounded integrand by Assumption\ref{ass4n}. Therefore, by concentration inequality of martingale integral is $o_p(1)$.
Under Assumption\ref{variationn}, we have 
\be Q_{13}=(\beta^{*}_j-\beta_{j0})\frac{1}{n}\sum_{i=1}^n\int_0^\tau  \partial_{\beta_j}{\cal E}_i (t; {\beta}^{*}, S^{*}_c,\pi^{*} )Y_i(t)E(q_{j}(t))dt,\nonumber\ee
where we call $q_{ji}(t)$ a function, such that:
\eqnn
\hat \Lambda_{j}(t, Z ; \beta)-\hat \Lambda_{j}(t, Z ; \beta_0)&=&(\beta_j-\beta_{j0})*\frac{1}{n}\sum_{i=1}^n\int_0^\tau q_{ji}(t).
\een
Term $Q_{14}=0$ if $\Lambda^{*}=\Lambda_0$.
What about $Q_3$?
If $S^{*}_{c}=S_{c0}$ and $\pi^{*}=\pi_0$, because everything is bounded, by Assumption\ref{ass4n} we have something along the following line:
\eqnn
Q_3&=&\frac{1}{n}\sum_{i=1}^n\int_0^\tau  \left\{A_i-{\cal E}_i(t; \beta, S_{c0},\pi_0 )\right\}Y_i(t)\partial_{\beta}d\Lambda(t,Z_i; \beta^{*})\\
&=&\int_0^\tau E \left[\left\{A-\frac{E\left[A e^{-\sum_{j=1}^J\beta_jAt}S_{c0}(t | A,Z) | Z \right]}{E\left[e^{-\sum_{j=1}^J\beta_jAt}S_{c0}(t | A,Z ) | Z\right]}\right\}Y(t)\partial_{\beta}d\Lambda(t,Z; \beta^{*})\right]+o_p(1)\\
&=&\int_0^\tau E \left[\left\{A-\frac{E\left[A e^{-\sum_{j=1}^J\beta_jAt}S_{c0}(t | A,Z) | Z \right]}{E\left[e^{-\sum_{j=1}^J\beta_jAt}S_{c0}(t | A,Z ) | Z\right]}\right\}E\left\{Y(t)|A,Z\right\}\partial_{\beta}d\Lambda(t,Z; \beta^{*})\right]+o_p(1)\\
&=&\int_0^\tau E \left[\left\{A-\frac{E\left[A e^{-\sum_{j=1}^J\beta_jAt}S_{c0}(t | A,Z) | Z \right]}{E\left[e^{-\sum_{j=1}^J\beta_jAt}S_{c0}(t | A,Z ) | Z\right]}\right\}
\right.\\
&&\left.\times e^{-\sum_{j=1}^J\beta_jAt}e^{-\sum_{j=1}^J\Lambda_j(t,Z)}S_{c0}(t | A,Z)\partial_{\beta}d\Lambda(t,Z; \beta^{*})\right]+o_p(1)\nonumber\\
&=&o_p(1).
\een
Therefore we have:
\eqnn
\nabla_{jj}=\partial_{\beta_j}\left\{{S}_{2,n}\right\}_j({\beta}^{*}, S^{*}_{c},  \pi^{*}, \Lambda^{*})=(\beta^{*}_j-\beta_{j0})(J^{(1')}_{jj}+J^{(2')}_{jj})+J^{(1)}_{jj}+J^{(2)}_{jj}+J^{(3)}_{jj},
\een
where
\eqnn
J^{(1')}_{jj}=\frac{1}{n}\sum_{i=1}^n\int_0^\tau  \partial_{\beta_j}{\cal E}_i(t; {\beta}^{*}, S^{*}_c,\pi^{*})A_iY_i(t)dt,
\een
\eqnn
J^{(2')}_{jj}=\frac{1}{n}\sum_{i=1}^n\int_0^\tau  \partial_{\beta_j}{\cal E}_i(t; {\beta}^{*}, S^{*}_c,\pi^{*})Y_i(t)E(q_{j}(t))dt,
\een
\eqnn
J^{(1)}_{jj}=\frac{1}{n}\sum_{i=1}^n\int_0^\tau  \left\{A_i-{\cal E}_i(t; {\beta}^{*}, S^{*}_c,\pi^{*})\right\}A_iY_i(t),
\een
\eqnn
J^{(2)}_{jj}=\frac{1}{n}\sum_{i=1}^n\int_0^\tau  \partial_{\beta_j}{\cal E}_i(t; {\beta}^{*}, S^{*}_c,\pi^{*})Y_i(t)d\left\{\Lambda_j^{*}(t,Z_i ; \beta_{j0})-\Lambda_{j0}(t,Z_i)\right\},
\een
\eqnn
J^{(3)}_{jj}=\frac{1}{n}\sum_{i=1}^n\int_0^\tau  \left\{A_i- {\cal E}_i(t; {\beta}^{*}, S^{*}_c,\pi^{*})\right\}Y_i(t)\partial_{\beta_j}d\Lambda^{*}_j(t,Z_i, \beta^{*}_j).
\een

We notice that $J^{(2)}_{jj}=0$ if $\Lambda^{*}(\cdot,\cdot)=\Lambda_0(\cdot,\cdot)$ and $J^{(3)}_{jj}=o_p(1)$ if or $S^{*}_{c}(\cdot | \cdot,\cdot)=S_{c0}(\cdot | \cdot,\cdot)$ and $\pi^{*}(\cdot)=\pi_0(\cdot)$ or $\hat \Lambda(\cdot,\cdot)$ does not depend on the unknown $\beta$.

On the other hand, similarly, we have:
\eqnn
\nabla_{12}&=&\partial_{\beta_2}\left\{{S}_{2,n}\right\}_1({\beta}^{*}, S^{*}_{c},  \pi^{*}, \Lambda^{*})\\
&=&(\beta^{*}_1-\beta_{10})\frac{1}{n}\sum_{i=1}^n\int_0^\tau  \partial_{\beta_2}{\cal E}_i(t; {\beta}^{*}, S^{*}_c,\pi^{*})A_iY_i(t)dt \\
&&+(\beta^{*}_1-\beta_{10})\frac{1}{n}\sum_{i=1}^n\int_0^\tau  \partial_{\beta_2}{\cal E}_i(t; {\beta}^{*}, S^{*}_c,\pi^{*})Y_i(t)E(q_{j}(t))dt\\
&&+\frac{1}{n}\sum_{i=1}^n\int_0^\tau  \partial_{\beta_2}{\cal E}_i(t; {\beta}^{*}, S^{*}_c,\pi^{*})Y_i(t)d\left\{\Lambda_1^{*}(t,Z_i ; \beta_{10})-\Lambda_{10}(t,Z_i)\right\}+o_p(1)\\
&=&(\beta^{*}_1-\beta_{10})(J^{(1')}_{12}+J^{(2')}_{12})+J^{(1)}_{12},\een
and
\eqnn
\nabla_{21}&=&\partial_{\beta_1}\left\{{S}_{2,n}\right\}_2({\beta}^{*}, S^{*}_{c},  \pi^{*}, \Lambda^{*})\\
&=&(\beta^{*}_2-\beta_{20})\frac{1}{n}\sum_{i=1}^n\int_0^\tau  \partial_{\beta_1}{\cal E}_i(t; {\beta}^{*}, S^{*}_c,\pi^{*})A_iY_i(t)dt \\
&&+(\beta^{*}_2-\beta_{20})\frac{1}{n}\sum_{i=1}^n\int_0^\tau  \partial_{\beta_1}{\cal E}_i(t; {\beta}^{*}, S^{*}_c,\pi^{*})Y_i(t)E(q_{j}(t))dt\\
&&+\frac{1}{n}\sum_{i=1}^n\int_0^\tau  \partial_{\beta_1}{\cal E}_i(t; {\beta}^{*}, S^{*}_c,\pi^{*})Y_i(t)d\left\{\Lambda_2^{*}(t,Z_i ; \beta_{20})-\Lambda_{20}(t,Z_i)\right\}\\
&=&(\beta^{*}_2-\beta_{20})(J^{(1')}_{21}+J^{(2')}_{21})+J^{(1)}_{21},\een
where the last terms $J^{(1)}_{12}=J^{(1)}_{21}=0$ if $\Lambda^{*}(\cdot,\cdot)=\Lambda_0(\cdot,\cdot)$.

Therefore we have:
\eqnn
\nabla&=&J+\begin{bmatrix} \beta^{*}_1-\beta_{10} \\ \beta^{*}_2-\beta_{20}\end{bmatrix}J'
\een
where the above multiplication is intended componentwise and
\eqnn
J=\begin{bmatrix} J^{(1)}_{11}+J^{(2)}_{11}+J^{(3)}_{11}& J^{(1)}_{12}\\
J^{(1)}_{21} & J^{(1)}_{22}+J^{(2)}_{22}+J^{(3)}_{22}
\end{bmatrix},
\een
and
\eqnn
J'=\begin{bmatrix} J^{(1')}_{11}+J^{(2')}_{11} & J^{(1')}_{12}+J^{(2')}_{12}\\
J^{(1')}_{21}+J^{(2')}_{21}& J^{(1')}_{22}+J^{(2')}_{22}
\end{bmatrix}.
\een

We will prove that $J$ is invertible. If this is the case, for any $|\delta|<1/2$, by \eqref{decosc1n}  and the above we have:
\eqnn
S_{2,n}(\beta_0\pm n^{-\delta}, \hat S_c,  \hat \pi, \hat \Lambda)&=&n^{-\delta}J+n^{-2\delta}J'+O_p(n^{-1/2}).
\een
We can therefore conclude that either:
\eqnn
S_{2,n}(\beta_0- n^{-\delta}, \hat S_c,  \hat \pi, \hat \Lambda)<{0}<S_{2,n}(\beta_0+n^{-\delta}, \hat S_c,  \hat \pi, \hat \Lambda),
\een
or
\eqnn
S_{2,n}(\beta_0+ n^{-\delta}, \hat S_c,  \hat \pi, \hat \Lambda)<{0}<S_{2,n}(\beta_0-n^{-\delta}, \hat S_c,  \hat \pi, \hat \Lambda).
\een
Therefore by definition of $\hat{{\beta}}$, we can conclude that $\hat{{\beta}}-{\beta}_0=O_p(n^{-\delta})$.

We now prove that $J$ is invertible proving that its determinant is different from zero.
$J$ simplifies accordingly to which model is correct. We therefore divide the proof of its invertibility in two cases.

\begin{itemize}
\item{Case a): $J$ is invertible if $S^{*}_c(\cdot | \cdot,\cdot)=S_{c0}(\cdot | \cdot,\cdot)$ and $\pi^{*}(\cdot)=\pi_0(\cdot)$.}
\end{itemize} 
Noticing that $ \partial_{\beta_1}{\cal E}(t; {\beta}^{*}, S^{*}_c,\pi^{*})= \partial_{\beta_2}{\cal E}(t; {\beta}^{*}, S^{*}_c,\pi^{*})$ and $J^{(1)}_{11}=J^{(1)}_{22}$, after some algebra we have:
\eqnn
|J|&=&(J^{(1)}_{11}+J^{(2)}_{11})(J^{(1)}_{22}+J^{(2)}_{22})-J^{(1)}_{12}J^{(1)}_{21}+o_p(1)\\
&=&J^{(1)}_{11}(J^{(1)}_{11}+J^{(2)}_{11}+J^{(2)}_{22})+o_p(1).
\een

We now prove that both $J^{(1)}_{11}\neq 0$ and $J^{(1)}_{11}+J^{(2)}_{11}+J^{(2)}_{22} \neq 0$. Those would prove that $|J|\neq 0$ and so that $J$ is invertible.

Under model \eqref{model},
$
E\left\{\left.Y(t)\right| A,Z\right\}=S{c0}c(t | A,Z)e^{-(\beta_{10}+\beta_{20})At}e^{-\Lambda_{10}(t,Z)-\Lambda_{20}(t,Z)},
$ therefore we have:
\eqnn
&&E\left[\left\{A-\frac{E\left[A e^{-\sum_{j=1}^2\beta_{j0}At}S_{c0}(t | A,Z) | Z \right]}{E\left[e^{-\sum_{j=1}^2\beta_{j0}At}S_{c0}(t | A,Z ) | Z\right]}\right\}AY(t)dt\right]\\
&=&\int_0^\tau E \left[\left\{A-\frac{E\left[A e^{-\sum_{j=1}^2\beta_{j0}At}S_{c0}(t | A,Z) | Z \right]}{E\left[e^{-\sum_{j=1}^2\beta_{j0}At}S_{c0}(t | A,Z ) | Z\right]}\right\}E\left\{Y(t)|A,Z\right\}A dt\right],
\een
and so
\eqnn
&=&\int_0^\tau E \left[\left\{A-\frac{E\left[A e^{-\sum_{j=1}^2\beta_{j0}At}S_{c0}(t | A,Z) | Z \right]}{E\left[e^{-\sum_{j=1}^2\beta_{j0}At}S_{c0}(t | A,Z ) | Z\right]}\right\}e^{-\sum_{j=1}^2\beta_{j0}At}e^{-\sum_{j=1}^2\Lambda_{j0}(t,Z)}S_{c0}(t | A, Z)Adt\right]\nonumber\\
&=&\int_0^\tau E \left[E\left[A e^{-\sum_{j=1}^2\beta_{j0}At}S_{c0}(t | A,Z) | Z \right]\left\{1-\frac{E\left[A e^{-\sum_{j=1}^2\beta_{j0}At}S_{c0}(t | A,Z) | Z \right]}{E\left[e^{-\sum_{j=1}^2\beta_{j0}At}S_{c0}(t | A,Z ) | Z\right]}\right\}\nonumber\right.\\
&&\left.\times e^{-\sum_{j=1}^2\Lambda_{j0}(t,Z)}dt\right].\nonumber
\een
The above is strictly different from zero under the positivity Assumption\ref{lowerboundn}.
Therefore applying Hoeffding's inequality, for some positive $\epsilon$:
$
J^{(1)}_{jj}>\epsilon.
$

We now focus on $J^{(1)}_{11}+J^{(2)}_{11}+J^{(2)}_{22}$.
By algebra we have:
\eqnn
\partial_{\beta_j}{\cal E}(t; {\beta}^{*}, S^{*}_c,\pi^{*} )&=&-t \cdot {\cal E}(t; {\beta}^{*}, S^{*}_c,\pi^{*} )\left\{1-{\cal E}(t; {\beta}^{*}, S^{*}_c,\pi^{*} )\right\}.
\een We have:
\eqnn
J^{(1)}_{11}+J^{(2)}_{11}+J^{(2)}_{22}&=&+\frac{1}{n}\sum_{i=1}^n\int_0^\tau  \left\{A_i-{\cal E}_i(t; {\beta}^{*}, S^{*}_c,\pi^{*})\right\}Y_i(t)A_i\\
&&-\frac{1}{n}\sum_{i=1}^n\int_0^\tau t{\cal E}_i(t; {\beta}^{*}, S^{*}_c,\pi^{*})\left\{1-{\cal E}_i(t; {\beta}^{*}, S^{*}_c,\pi^{*})\right\}\\&&\times d\left\{\Lambda_1^{*}(t,Z_i ; \beta_{j0})-\Lambda_{10}(t,Z_i)+\Lambda_2^{*}(t,Z_i ; \beta_{j0})-\Lambda_{20}(t,Z_i)\right\}Y_i(t).
\een

Similarly to before, if we look at the expected value, we have:
\eqnn
&&E\left[J^{(1)}_{11}+J^{(2)}_{11}+J^{(2)}_{22}\right]\\&=&\int_0^\tau E \left[\left\{A-\frac{E\left[A e^{-\sum_{j=1}^2\beta_{j0}At}S_{c0}(t | A,Z) | Z \right]}{E\left[e^{-\sum_{j=1}^2\beta_{j0}At}S_{c0}(t | A,Z ) | Z\right]}\right\}AE\left\{ Y(t) | A, Z \right\}\right]\nonumber\\
&-&\int_0^\tau E\left[\frac{E\left[A e^{-\sum_{j=1}^2\beta_{j0}At}S_{c0}(t | A,Z) | Z \right]}{E\left[e^{-\sum_{j=1}^2\beta_{j0}At}S_{c0}(t | A,Z ) | Z\right]}\left\{1-\frac{E\left[A e^{-\sum_{j=1}^2\beta_{j0}At}S_{c0}(t | A,Z) | Z \right]}{E\left[e^{-\sum_{j=1}^2\beta_{j0}At}S_{c0}(t | A,Z ) | Z\right]}\right\}\right.\nonumber\\&&\times\left. td\left\{\Lambda_1^{*}(t,Z ; \beta_{j0})-\Lambda_{10}(t,Z)+\Lambda_2^{*}(t,Z; \beta_{j0})-\Lambda_{20}(t,Z)\right\}E\left\{ Y(t) | A, Z \right\}\right].\nonumber
\een
Therefore
\eqnn
&&E\left[J^{(1)}_{11}+J^{(2)}_{11}+J^{(2)}_{22}\right]\\
&=&\int_0^\tau E\left(\frac{E\left[A e^{-\sum_{j=1}^2\beta_{j0}At}S_{c0}(t | A,Z) | Z \right]}{E\left[e^{-\sum_{j=1}^2\beta_{j0}At}S_{c0}(t | A,Z ) | Z\right]}\left\{1-\frac{E\left[A e^{-\sum_{j=1}^2\beta_{j0}At}S_{c0}(t | A,Z) | Z \right]}{E\left[e^{-\sum_{j=1}^2\beta_{j0}At}S_{c0}(t | A,Z ) | Z\right]}\right\}\right.\nonumber\\&&\times \left.e^{-\sum_{j=1}^2\Lambda_{j0}(t,Z)}\left[A - te^{-\sum_{j=1}^2\beta_{j0}At}S_c(t | A,Z)\sum_{j=1}^2d\left\{\Lambda_j^{*}(t,Z ; \beta_{j0})-\Lambda_{j0}(t,Z)\right\}\right]\right).\nonumber
\een
Again, by Assumption\ref{lowerboundn} and \ref{lowerboundsc1}, we can conclude that $E\left[J^{(1)}_{11}+J^{(2)}_{11}+J^{(2)}_{22}\right]>\epsilon$ and therefore, by Hoeffding's inequality that $J^{(1)}_{11}+J^{(2)}_{11}+J^{(2)}_{22}>\epsilon+o_p(1)$.

\begin{itemize}
\item{Case b): $J$ is invertible if $\Lambda^{*}(\cdot,\cdot)=\Lambda_0(\cdot,\cdot)$.}
\end{itemize}  
We have:
\eqnn
|J^{(1)}|&=&(J^{(1)}_{11}+J^{(3)}_{11})(J^{(1)}_{22}+J^{(3)}_{22}),
\een

We are now left to prove that $J^{(1)}_{jj}+J^{(3)}_{jj}\neq 0$ when $\Lambda^{*}(\cdot)=\Lambda_0(\cdot)$.
We have:
\eqnn
J^{(1)}_{jj}+J^{(3)}_{jj}&=&\frac{1}{n}\sum_{i=1}^n\int_0^\tau  \left\{A_i-{\cal E}_i(t; {\beta}^{*}, S^{*}_c,\pi^{*})\right\}Y_i(t)\left\{A_i+\partial_{\beta_j}d\Lambda^{*}_j(t,Z_i)\right\},
\een
and similarly to before, by Assumption \ref{lowerboundsc1} we can prove that $J^{(1)}_{jj}+J^{(3)}_{jj}>\epsilon$.

\textbf{\emph{Asymptotic normality:}}
By Taylor expansion we have:
\be
S_{2,n}({\beta}, \hat S_{c},  \hat \pi, \hat \Lambda)=S_{2,n}({\beta}_0, \hat S_c,  \hat \pi, \hat \Lambda)+\nabla_{{\beta}}S_{2,n}({\beta}^{*}, \hat S_c,  \hat \pi, \hat \Lambda)\left({\beta}-{\beta}_0\right)^\top,
\ee
where ${\beta}^{*}$ lies between ${\beta}$ and ${\beta}_0$.

Under Assumptions \ref{assumption}-\ref{variationn}, it can be proved that:
\be
\nabla_{{\beta}}S_{2,n}({\beta}^{*}, \hat S_c,  \hat \pi, \hat \Lambda)=\nabla_{{\beta}}S_{2,n}({\beta}^{*}, S^{*}_c,  \pi^{*}, \Lambda^{*})+o_p(1).
\ee

In the previous part of the proof we moreover proved that:
\eqnn
\nabla_{{\beta}}S_{2,n}({\beta}^{*}, S^{*}_c,  \pi^{*}, \Lambda^{*})&=&J+\begin{bmatrix} \beta^{*}_1-\beta_{10} \\ \beta^{*}_2-\beta_{20}\end{bmatrix}J',
\een
where the above multiplication is intended componentwise and
\eqnn
J=\begin{bmatrix} J^{(1)}_{11}+J^{(2)}_{11}+J^{(3)}_{11}& J^{(1)}_{12}\\
J^{(1)}_{21} & J^{(1)}_{22}+J^{(2)}_{22}+J^{(3)}_{22}
\end{bmatrix},
\een
and
\eqnn
J'=\begin{bmatrix} J^{(1')}_{11}+J^{(2')}_{11} & J^{(1')}_{12}+J^{(2')}_{12}\\
J^{(1')}_{21}+J^{(2')}_{21}& J^{(1')}_{22}+J^{(2')}_{22}
\end{bmatrix}.
\een

We now focus on term $S_{2,n}(\beta_0, \hat S_c, \hat \pi, \hat \Lambda)$.
We have the following decomposition:
\eqnn
S_{2,n}(\beta_0, \hat S_c, \hat \pi, \hat \Lambda)&=&+S_{2,n}(\beta_0, \hat S_c, \hat \pi, \hat \Lambda)-S_{2,n}(\beta_0, S^{*}_{c},  \pi^{*}, \hat \Lambda)\\
&&+S_{2,n}(\beta_0, S^{*}_{c},  \pi^{*}, \hat \Lambda)-S_{2,n}(\beta_0, S^{*}_{c},  \pi^{*}, \Lambda^{*})\\
&&+S_{2,n}(\beta_0, S^{*}_{c},  \pi^{*}, \Lambda^{*})\\
&=&Q_1+Q_2+Q_3.
\een

We remind the reader that in the previous part of the proof we proved that $\hat \beta^{(2)}-\beta_0=O_p(n^{-\delta})$ for $|\delta|<1/2$.

Putting all the above together, by definition of $\hat \beta$ we have:
\be\label{decosc1}
\sqrt{n}(\hat \beta^{(2)}-\beta_0)&=&J^{-1}\sqrt{n}\left\{S_{2,n}(\beta_0, S^{*}_{c},  \pi^{*}, \Lambda^{*})+Q_1+Q_2\right\}+o_p(1).
\ee

$S_{2,n}(\beta_0, S^{*}_{c},  \pi^{*}, \Lambda^{*})$ is by double robustness of the score  (Theorem \ref{dr2}), a sum of i.i.d. mean zero terms.
Similarly to the proof for $\beta^{(1)}$
we can prove that $Q_2=o_p(n^{-1/2})$ if $\Lambda^{*}(\cdot,\cdot)=\Lambda_0(\cdot,\cdot)$ and $Q_1=o_p(n^{-1/2})$ if $S_c^{*}(\cdot | \cdot,\cdot)=S_{c0}(\cdot | \cdot,\cdot)$ and $\pi^{*}(\cdot)=\pi_0(\cdot)$.
We therefore now divide the proof in three different cases according to which model is correctly specified.

\begin{itemize}
\item{ Case a): $S^{*}_c(\cdot | \cdot,\cdot)=S_{c0}(\cdot | \cdot,\cdot)$, $\pi^{*}(\cdot)=\pi_0(\cdot)$ and $\Lambda^{*}(\cdot,\cdot)\neq \Lambda_0(\cdot,\cdot)$ with $a_n=n^{-1/2}$, $b_n=n^{-1/2}$.}
\end{itemize}
As said before, we can prove that $Q_1=o_p(n^{-1/2})$, therefore, by \eqref{decosc1} and by Assumption A'\ref{ifD}, we have:
\eqnn
\sqrt{n}(\hat \beta-\beta_0)&=&J^{-1}\sqrt{n}\left\{S_{2,n}(\beta_0, S^{*}_{c},  \pi^{*}, \Lambda^{*})+\frac{1}{n}\sum_{i=1}^n\sigma_{6i}\right\}+o_p(1).
\een

$\sqrt{n}(\hat \beta-\beta_0)$ can be therefore written as  sum of i.i.d mean zero terms, and therefore, by multivariate central limit theorem, it is asymptotically normal.

Part a) of the asymptotic normality follows directly.

\begin{itemize}
\item{Case b): $\Lambda^{*}(\cdot,\cdot)=\Lambda_{0}(\cdot,\cdot)$, $S_c^{*}(\cdot | \cdot,\cdot)\neq S_{c0}(\cdot|\cdot,\cdot)$ and $\pi^{*}(\cdot)\neq \pi_0(\cdot)$ with $c_n=n^{-1/2}$.}
\end{itemize}
As said before, we can prove that $Q_2=o_p(n^{-1/2})$, therefore, by \eqref{decosc1} and by Assumption B' \ref{ifE}, we have:
\eqnn
\sqrt{n}(\hat \beta-\beta_0)&=&J^{-1}\sqrt{n}\left\{S_{2,n}(\beta_0, S^{*}_{c},  \pi^{*}, \Lambda^{*})+\frac{1}{n}\sum_{i=1}^n\sigma_{7i}\right\}+o_p(1).
\een

$\sqrt{n}(\hat \beta-\beta_0)$ can be therefore written as sum of i.i.d mean zero terms, and therefore, by multivariate central limit theorem, it is asymptotically normal.

Part b) of the asymptotic normality follows directly.

\begin{itemize}
\item{Case c): $S^{*}_c(\cdot | \cdot,\cdot)=S_{c0}(\cdot | \cdot,\cdot)$ and $\pi^{*}(\cdot)=\pi_0(\cdot)$ and $\Lambda^{*}(\cdot,\cdot)=\Lambda_{0}(\cdot,\cdot)$ with $a_nc_n=o(n^{-1/2})$ and $b_nc_n=o(n^{-1/2})$.}
\end{itemize}
In this case we have both $Q_1=o_p(n^{-1/2})$ and $Q_2=o_p(n^{-1/2})$. Therefore:
\be\label{decosc1c}
\sqrt{n}(\hat \beta-\beta_0)&=&J^{-1}\sqrt{n}S_{2,n}(\beta_0, S^{*}_{c},  \pi^{*}, \Lambda^{*})+o_p(1).
\ee

Moreover, when both models are correct, $J$ simplifies to a diagonal matrix with diagonal element equals to $J^{(1)}_{jj}=\frac{1}{n}\sum_{i=1}^n\int_0^\tau  \left\{A_i-{\cal E}_i(t; {\beta}^{*}, S^{*}_c,\pi^{*})\right\}A_iY_i(t)$.

MCLT can be applied to $\sqrt{n}S_{2,n}({\beta}_0, S_{c0},  \pi_0, \Lambda_0)$ to prove asymptotic normality. Specifically we consider the following multivariate martingale 
${M}_i(t)=[
           M_{1i}(t) ,
           M_{2i}(t)
      ]^\top $
with respect to the filtration $\mathcal{F}_t=\sigma\left\{N_{ji}(s), Y_i(s+), A_i, Z_i\;:j=1,2,\;i=1,\ldots,n,\;0<s<t\right\}$.
We consider the following two-dimensional vector:
\eqnn
{M}^n(t)=\frac{1}{\sqrt{n}} \sum_{i=1}^n\int_0^t h(u; A_i,Z_i) d{M}_i(u),
\een
where $h(t; A,Z)= A -{\cal E}(t; \beta, S_{c0},\pi_0)$.
Since $h(t; A,Z)$ is predictable with respect to the filtration, then ${M}^n(t)$ is a multivariate martingale too.
We have $<M_{1i}(t),M_{2i}(t)>=<M_{1i}(t),M_{1j}(t)>=<M_{2i}(t),M_{2j}(t)>=<M_{1i}(t),M_{2j}(t)>=0$ for each $i\neq j$ therefore:
\eqnn
<{M}_1^n(t),{M}_2^n(t)>&=&<\frac{1}{\sqrt{n}} \sum_{i=1}^n\int_0^t h(u; A_i,Z_i) d{M}_{1i}(u),\frac{1}{\sqrt{n}} \sum_{i=1}^n\int_0^t h(u; A_i,Z_i) d{M}_{2i}(u)>\\&=&\frac{1}{n} \sum_{i,j=1}^n\int_0^t h^2(u; A_i,Z_i) d<M_{1i}(t),M_{2j}(t)>=0,
\een
and so the two components of the multidimensional martingale ${M}^n(t)$ are orthogonal to each other.
Therefore, we can apply the multidimensional version of the martingale central limit theorem of Rebolledo (Theorem 5 of \cite{rebolledo1978applications}). 

First we verify Assumption 2 about the convergence of the variance.
We have, by Assumption C'\ref{ass2nd}, for $j=1,2$:
\eqnn
<{M}_j^n(t),{M}_j^n(t)>&=&\frac{1}{n}\sum_{i=1}^n\int_0^t h^2(u; A_i,Z_i) d\Lambda_{j0}(u | A_i, Z_i)Y_i(u)du\\
&=&\frac{1}{n}\sum_{i=1}^n\int_0^t h^2(u; A_i,Z_i) \left\{d\Lambda_{j0}(u,Z_i) + \beta_{j0}Adu\right\}Y_i(u)\\
&=&\int_0^t  \left\{P'(u)\beta_{j0}+Q'_j(u)\right\}du\overset{p}{\rightarrow} \int_0^t  \left\{p'(u)\beta_{j0}+q'_j(u)\right\}du=V'_j(t),
\een
and so Assumption 2 of the MCLT is verified.

We now look at Assumption 1 about the jumps of each component of the martingale. \cite{rebolledo1978applications} at pag. 39 claims that if the Lindeberg condition is verified, then Assumption 1 of its theorem holds.
We therefore needs to prove that, for any $\epsilon$ and any $j$:
\eqnn
\int_0^\tau \frac{1}{n}\sum_{i=1}^n h^2(u ; A_i,Z_i) \mathbbm{1}\left\{\left|h(u ; A_i,Z_i)\right|>\sqrt{n}\epsilon\right\}Y_i(t)\left\{d\Lambda_{j0}(t, Z_i) + \beta_{j0} A_idt\right\} \overset{p}{\rightarrow} 0,
\een

by Assumptions \ref{ass0n} and \ref{lowerboundn}, we know that: 
\eqnn
\left|h(t; A,Z)\right|\leq 1+\frac{\max\{1,e^{-(\beta_{10}+\beta_{20})\tau}\}C_3}{\min\{1,e^{-(\beta_{10}+\beta_{20})\tau}\}C_1C_2+1-C_3}<\infty,
\een
so, we have:
\eqnn
&&\int_0^\tau \frac{1}{n}\sum_{i=1}^n h^2(u ; A,Z) \mathbbm{1}\left\{\left|h(u ; A,Z)\right|>\sqrt{n}\epsilon\right\}Y_i(t)\left\{d\Lambda_{j0}(t, Z_i) + \beta_{j0} A_idt\right\} \\
&\leq& \int_0^\tau \frac{1}{n}\sum_{i=1}^n h^2(u ; A,Z) \mathbbm{1}\left\{1+\frac{\max\{1,e^{-(\beta_{10}+\beta_{20})\tau}\}C_3}{\min\{1,e^{-(\beta_{10}+\beta_{20})\tau}\}C_1C_2+1-C_3}>\sqrt{n}\epsilon\right\}Y_i(t)\nonumber
\\
&&\times\left\{d\Lambda_{j0}(t, Z_i) + \beta_{j0} A_idt\right\}.\nonumber
\een
Moreover, by Assumption \ref{boundln}, we also know that: 
\eqnn
&&\left| \int_0^\tau \frac{1}{n}\sum_{i=1}^n h^2(u ; A,Z) \mathbbm{1}\left\{1+\frac{\max\{1,e^{-(\beta_{10}+\beta_{20})\tau}\}C_3}{\min\{1,e^{-(\beta_{10}+\beta_{20})\tau}\}C_1C_2+1-C_3}>\sqrt{n}\epsilon\right\}\right.\\&&\cdot\left.Y_i(t)\left\{d\Lambda_{j0}(t, Z_i) + \beta_{j0} A_idt\right\}\right| \nonumber\\
 &\leq & \left\{1+\frac{\max\{1,e^{-(\beta_{10}+\beta_{20})\tau}\}C_3}{\min\{1,e^{-(\beta_{10}+\beta_{20})\tau}\}C_1C_2+1-C_3}\right\}^2  \nonumber \\&&\times\mathbbm{1}\left\{1+\frac{\max\{1,e^{-(\beta_{10}+\beta_{20})\tau}\}C_3}{\min\{1,e^{-(\beta_{10}+\beta_{20})\tau}\}C_1C_2+1-C_3}>\sqrt{n}\epsilon\right\}\tau\left|L_j + \beta_{j0} \right| \overset{p}{\rightarrow} 0,
\een
and so Assumption 1 of the martingale central limit theorem holds.
 
Therefore, we can conclude that 
\be\label{normsc1}
\sqrt{n}S_{2,n}(\beta_0, S_{c0}, \pi_0, \Lambda_{0})=\mathbbm{M}^n(t)\overset{D}{\rightarrow}\mathcal{N}(0,V'(\tau)).
\ee

The proof of the consistency of the variance estimator is similar to the proof of the consistency of the variance estimator for $\hat \beta^{(1)}$ and we leave it to the reader.

By the above and \eqref{decosc1c} part c) of the asymptotic normality follows.


\newpage
\subsection{Proofs of Lemmas}\label{lemmas}

\newenvironment{newproof7}{\begin{proof}\textsc{\emph{of Lemma \ref{lm:deco}.}}}{\end{proof}}

\begin{newproof7}
We remind the reader that:
\eqnn
{S}_{1,n}(\beta, S_c,  \pi, \Lambda)=\left\{\frac{1}{n}\sum_{i=1}^n\int_0^\tau e^{ (\beta_1  + \beta_2) A_i t}S^{-1}_c(t | A_i,Z_i)\left\{A_i -\pi( Z_i)\right\}dM_{ji}(t ; \beta_j, \Lambda_{j})\right\}_{j=1,2}.
\een
By algebra we have the following decomposition of the score:
\eqnn
&&{S}_{1,n}(\beta, \hat S_c,  \hat \pi, \hat \Lambda)\\&=&  S_{2,n}({\beta}, \hat S_c,  \hat \pi,  \hat \Lambda)-S_{2,n}({\beta}_0, \hat S_c,  \hat \pi,  \hat \Lambda)\\
&&+S_{2,n}({\beta}_0, \hat S_c,  \hat \pi,  \hat \Lambda)-S_{2,n}({\beta}_0, \hat S_c,  \hat \pi,  \Lambda^{*})\\
&&+ S_{2,n}({\beta}_0, \hat S_c,  \hat \pi, \Lambda^{*})-S_{2,n}({\beta}_0, S^{*}_c,  \pi^{*}, \Lambda^{*})\\
&&+ S_{2,n}({\beta}_0, S^{*}_c,  \pi^{*}, \Lambda^{*})\\
&=&Q^{(1)}+Q^{(2)}+Q^{(3)}+Q^{(4)}.
\een
We first of all notice that by Assumption \ref{assumption} and \ref{ass4n}, we have:
\be\label{nuisnew}
&&\left[\hat S^{-1}_c(t | A_i,Z_i)\left\{A_i -\hat \pi( Z_i)\right\}- \left\{S^{*}_c(t | A_i,Z_i)\right\}^{-1}\left\{A_i - \pi^{*}( Z_i)\right\}\right]\\&\leq&\sup_{t \in [0,\tau],Z \in \mathcal{Z}, A \in {0,1}}\left|\hat S^{-1}_c(t | A,Z)- \left\{S^{*}_c(t | A,Z)\right\}^{-1}\right|\\&&+\sup_{t \in [0,\tau], Z \in \mathcal{Z}, A \in {0,1}}\left| \left\{S^{*}_c(t | A,Z)\right\}^{-1}\right|\left|\hat \pi( Z) - \pi^{*}( Z)\right|=o_p(1).
\ee

Moreover, we notice that, by Assumption \ref{ass0n} and \ref{ass4n}
$
K^{(1)}(\beta_0,\pi^{*},S^{*}_c)=O_p(1).
$
By Assumptions \ref{ass0n}, \ref{ass4n}, \ref{variationn}, we have:
$
K^{(2)}_j(\beta,\pi^{*},S^{*}_c)=O_p(|\beta_j-\beta_{j0}|).
$

We now work on each term separately. 

\begin{itemize}
\item{Term $Q^{(1)}$}:
\end{itemize}
Algebra and the application of Lemma \ref{mvt} gives us:
\eqnn
Q^{(1)}_j&=&\frac{1}{n}\sum_{i=1}^n\int_0^\tau \left\{e^{ (\beta_1  + \beta_2)A_i t}-e^{ (\beta_{10}  + \beta_{20}) A_i t}\right\}\hat S^{-1}_c(t | A_i,Z_i)\left\{A_i -\hat \pi( Z_i)\right\}\\&&\times\left\{dN_{ji}(t)-Y_i(t)\beta_jA_idt-Y_i(t)d\hat \Lambda_{j}(t, Z_i ; \beta)\right\}\nonumber\\
&&+\frac{1}{n}\sum_{i=1}^n\int_0^\tau e^{ (\beta_{10}  + \beta_{20}) A_i t}\hat S^{-1}_c(t | A_i,Z_i)\left\{A_i -\hat \pi( Z_i)\right\}\nonumber\\&&\times\left\{dN_{ji}(t)-Y_i(t)\beta_jA_idt-Y_i(t)d\hat \Lambda_{j}(t, Z_i ; \beta)\right\}\nonumber\\
&&-\frac{1}{n}\sum_{i=1}^n\int_0^\tau e^{ (\beta_{10}  + \beta_{20}) A_i t}\hat S^{-1}_c(t | A_i,Z_i)\left\{A_i -\hat \pi( Z_i)\right\}\nonumber\\&&\times\left\{dN_{ji}(t)-Y_i(t)\beta_{j0}A_idt-Y_i(t)d\hat \Lambda_{j}(t, Z_i ; \beta_0)\right\}.\nonumber\een

Therefore:
\eqnn
Q^{(1)}_j&=&(\beta_1+\beta_2-\beta_{10}-\beta_{20})\frac{1}{n}\sum_{i=1}^n\int_0^\tau e^{ (\beta^{*}_1  + \beta^{*}_2)A_i t}A_i t K^{(3)}_{ji}(t, \beta, \hat S_c \hat \pi)\\
&&-(\beta_j-\beta_{j0})K^{(1)}(\beta_0,\hat S_c,\hat \pi)-K^{(2)}_j(\beta,\hat S_c,\hat \pi),\een
for some $\beta^{*}$ between $\beta_0$ and $\beta$.

Moreover:
\be\label{3}
Q^{(1)}_j
&=&(\beta_1+\beta_2-\beta_{10}-\beta_{20})\frac{1}{n}\sum_{i=1}^n\int_0^\tau e^{ (\beta^{*}_1  + \beta^{*}_2)A_i t}A_i t K^{(3)}_{ji}(t, \beta,S^{*}_c,\pi^{*})\\
&&+(\beta_1+\beta_2-\beta_{10}-\beta_{20})^2\frac{1}{n}\sum_{i=1}^n\int_0^\tau e^{ (\beta^{**}_1  + \beta^{**}_2)A_i t}A_i t^2 \left\{K^{(3)}_{ji}(t, \beta,\hat S_c, \hat \pi)-K^{(3)}_{ji}(t, \beta,S^{*}_c, \pi^{*})\right\}\nonumber\\
&&+(\beta_1+\beta_2-\beta_{10}-\beta_{20})\frac{1}{n}\sum_{i=1}^n\int_0^\tau e^{ (\beta_{10}  + \beta_{20})A_i t}A_i t \left\{K^{(3)}_{ji}(t, \beta,\hat S_c, \hat \pi)-K^{(3)}_{ji}(t, \beta, S^{*}_c,\pi^{*})\right\}\nonumber\\
&&-(\beta_j-\beta_{j0})K^{(1)}(\beta_0,S^{*}_c,\pi^{*})\nonumber\\
&&-(\beta_j-\beta_{j0})\left\{K^{(1)}(\beta,\hat S_c, \hat\pi)-K^{(1)}(\beta_0,S^{*}_c,\pi^{*})\right\}\nonumber\\
&&-K^{(2)}_j(\beta,  S^{*}_c,\pi^{*})\nonumber\\
&&+K^{(2)}_j(\beta, \hat S_c,\hat \pi)-K^{(2)}_j(\beta, S^{*}_c,\pi^{*}),\nonumber
\ee
where $\beta^{**}_j$ is a point between $\beta^{*}$ and $\beta_0$.
We remind the reader that the quantities \\$K^{(1)}, K^{(2)}_{j}, K^{(3)}_j,K^{(4)}_{j}$ are defined in equations \eqref{K1}-\eqref{K4}.

We work now on term $\frac{1}{n}\sum_{i=1}^n\int_0^\tau e^{ (\beta^{*}_1  + \beta^{*}_2)A_i t}A_i t K^{(3)}_{ji}(t, \beta,S^{*}_c,\pi^{*})$. By algebra and by Lemma \ref{mvt}:
\eqnn
&&\frac{1}{n}\sum_{i=1}^n\int_0^\tau e^{ (\beta^{*}_1  + \beta^{*}_2)A_i t}A_i t K^{(3)}_{ji}(t, \beta,S^{*}_c,\pi^{*})\\&=&\frac{1}{n}\sum_{i=1}^n\int_0^\tau e^{ (\beta_{10}  + \beta_{20})A_i t}A_i t \frac{A_i -\pi^{*}( Z_i)}{S^{*}_c(t | A_i,Z_i)}dM_{ji}(t ; \beta, \hat \Lambda)\nonumber\\
&&+(\beta_1+\beta_2-\beta_{10}-\beta_{20})\frac{1}{n}\sum_{i=1}^n\int_0^\tau e^{ (\beta^{**}_1  + \beta^{**}_2)A_i t}A_i t^2 \frac{A_i -\pi^{*}( Z_i)}{S^{*}_c(t | A_i,Z_i)}dM_{ji}(t ; \beta, \hat \Lambda).\nonumber\een
Therefore:
\be\label{31}
&=&\frac{1}{n}\sum_{i=1}^n\int_0^\tau e^{ (\beta_{10}  + \beta_{20})A_i t}A_i t \frac{A_i -\pi^{*}( Z_i)}{S^{*}_c(t | A_i,Z_i)}dM_{ji}(t)\nonumber\\
&&-K^{(4)}_j(\beta_0, \pi^{*}, S^{*}_c, \Lambda^{*})-(\beta_j-\beta_{j0})K^{(1)}(\beta_0,S^{*}_c,\pi^{*})\nonumber\\&&-\frac{1}{n}\sum_{i=1}^n\int_0^\tau e^{ (\beta_{10}  + \beta_{20})A_i t}A_i t \frac{A_i -\pi^{*}( Z_i)}{S^{*}_c(t | A_i,Z_i)}d\left\{\hat \Lambda_j(t,Z_i ; \beta)- \hat \Lambda_j(t,Z_i ; \beta_0)\right\}\nonumber\\
&&-\frac{1}{n}\sum_{i=1}^n\int_0^\tau e^{ (\beta_{10}  + \beta_{20})A_i t}A_i t \frac{A_i -\pi^{*}( Z_i)}{S^{*}_c(t | A_i,Z_i)}d\left\{\hat \Lambda_j(t,Z_i ; \beta_0)- \Lambda^{*}_j(t,Z_i)\right\}\nonumber\\
&&+(\beta_1+\beta_2-\beta_{10}-\beta_{20})\frac{1}{n}\sum_{i=1}^n\int_0^\tau e^{ (\beta^{**}_1  + \beta^{**}_2)A_i t}A_i t^2 \frac{A_i -\pi^{*}( Z_i)}{S^{*}_c(t | A_i,Z_i)}dM_{ji}(t ; \beta, \hat \Lambda).\nonumber
\ee
We work now on term $\frac{1}{n}\sum_{i=1}^n\int_0^\tau e^{ (\beta_{10}  + \beta_{20})A_i t}A_i t \left\{K^{(3)}_{ji}(t, \beta, \hat S_c,\hat \pi)-K^{(3)}_{ji}(t, \beta,S^{*}_c,\pi^{*})\right\}$. By algebra and by Lemma \ref{mvt}:
\eqnn
&&\frac{1}{n}\sum_{i=1}^n\int_0^\tau e^{ (\beta_{10}  + \beta_{20})A_i t}A_i t \left\{K^{(3)}_{ji}(t, \beta,\hat S_c, \hat \pi)-K^{(3)}_{ji}(t, \beta, S^{*}_c, \pi^{*})\right\}
\\&=&\frac{1}{n}\sum_{i=1}^n\int_0^\tau e^{ (\beta_{10}  + \beta_{20})A_i t}A_i t \left[\frac{A_i -\hat \pi( Z_i)}{\hat S_c(t | A_i,Z_i)}-\frac{A_i -\pi^{*}( Z_i)}{S^{*}_c(t | A_i,Z_i)}\right]dM_{ji}(t ; \beta, \hat \Lambda)\nonumber\\
\een
Therefore
\be\label{32}
&=&\frac{1}{n}\sum_{i=1}^n\int_0^\tau e^{ (\beta_{10}  + \beta_{20})A_i t}A_i t \left[\frac{A_i -\hat \pi( Z_i)}{\hat S_c(t | A_i,Z_i)}-\frac{A_i -\pi^{*}( Z_i)}{S^{*}_c(t | A_i,Z_i)}\right]dM_{ji}(t ; \beta_{0}, \Lambda^{*})\nonumber\\
&&-(\beta_j-\beta_{j0})\left\{K^{(1)}(\beta_0,\hat S_c, \hat \pi)-K^{(1)}(\beta_0,S^{*}_c, \pi^{*})\right\}\nonumber\\
&&-\frac{1}{n}\sum_{i=1}^n\int_0^\tau e^{ (\beta_{10}  + \beta_{20})A_i t}A_i t \left\{\frac{A_i -\hat \pi( Z_i)}{\hat S_c(t | A_i,Z_i)}-\frac{A_i -\pi^{*}( Z_i)}{S^{*}_c(t | A_i,Z_i)}\right\}Y_i(t)
\nonumber\\
&&\times d\left\{\hat \Lambda_j(t,Z_i ; \beta)-\Lambda^{*}_j(t,Z_i; \beta_0)\right\}\nonumber\\
&&-\frac{1}{n}\sum_{i=1}^n\int_0^\tau e^{ (\beta_{10}  + \beta_{20})A_i t}A_i t \left[\frac{A_i -\hat \pi( Z_i)}{\hat S_c(t | A_i,Z_i)}-\frac{A_i -\pi^{*}( Z_i)}{S^{*}_c(t | A_i,Z_i)}\right]Y_i(t)
\nonumber\\&&\times d\left\{\hat \Lambda_j(t,Z_i ; \beta_0)-\Lambda_j^{*}(t,Z_i)\right\}.\nonumber\\
\ee

Therefore putting together \eqref{3}, \eqref{31} and \eqref{32}, we get:
\begin{align*}
&Q^{(1)}_j\\&=Q^{(11)}_j+Q^{(12)}_j+Q^{(13)}_j+Q^{(14)}_j+Q^{(15)}_j+
Q^{(16)}_j+
Q^{(17)}_j+Q^{(18)}_j+Q^{(19)}_j+Q^{(110)}_j+
Q^{(111)}_j+Q^{(112)}_j,
\end{align*}
where 
\begin{align*}
Q^{(11)}_j&=(\beta_1+\beta_2-\beta_{10}-\beta_{20})\frac{1}{n}\sum_{i=1}^n\int_0^\tau e^{ (\beta_{10}  + \beta_{20})A_i t}A_i t \left\{\frac{A_i -\hat \pi( Z_i)}{\hat S_c(t | A_i,Z_i)}-\frac{A_i -\pi^{*}( Z_i)}{S^{*}_c(t | A_i,Z_i)}\right\}dM_{ji}(t),\nonumber\\
Q^{(12)}_j&=-(\beta_1+\beta_2-\beta_{10}-\beta_{20})K^{(4)}_j(\beta_0, S^{*}_c,\pi^{*}, \Lambda^{*})
\\&\quad-(\beta_1+\beta_2-\beta_{10}-\beta_{20})(\beta_j-\beta_{j0})K^{(1)}(\beta_0,S^{*}_c, \pi^{*}),\\
Q^{(13)}_j&=-(\beta_1+\beta_2-\beta_{10}-\beta_{20})\frac{1}{n}\sum_{i=1}^n\int_0^\tau e^{ (\beta_{10}  + \beta_{20})A_i t}A_i t \frac{A_i -\pi^{*}( Z_i)}{S^{*}_c(t | A_i,Z_i)}
\\&\quad \times d\left\{\hat \Lambda_j(t,Z_i ; \beta_j)- \Lambda^{*}_j(t,Z_i; \beta_{j0})\right\},\nonumber\\
Q^{(14)}_j&=-(\beta_1+\beta_2-\beta_{10}-\beta_{20})\frac{1}{n}\sum_{i=1}^n\int_0^\tau e^{ (\beta_{10}  + \beta_{20})A_i t}A_i t \frac{A_i -\pi^{*}( Z_i)}{S^{*}_c(t | A_i,Z_i)}
\\&\quad \times d\left\{\hat \Lambda_j(t,Z_i ; \beta_{j0})- \Lambda_j^{*}(t,Z_i)\right\},\\
Q^{(15)}_j&=(\beta_1+\beta_2-\beta_{10}-\beta_{20})^2\frac{1}{n}\sum_{i=1}^n\int_0^\tau e^{ (\beta^{**}_1  + \beta^{**}_2)A_i t}A_i t^2 \frac{A_i -\pi^{*}( Z_i)}{S^{*}_c(t | A_i,Z_i)}dM_{ji}(t ; \beta, \hat \Lambda),\\
Q^{(16)}_j&=+\frac{1}{n}(\beta_1+\beta_2-\beta_{10}-\beta_{20})^2\sum_{i=1}^n\int_0^\tau e^{ (\beta^{**}_1  + \beta^{**}_2)A_i t}A_i t^2 \\ & \quad\times\left\{K^{(3)}_{ji}(t, \beta,\hat S_c. \hat \pi)-K^{(3)}_{ji}(t, \beta,S^{*}_c, \pi^{*})\right\},\\
Q^{(17)}_j&=+(\beta_1+\beta_2-\beta_{10}-\beta_{20})\frac{1}{n}\sum_{i=1}^n\int_0^\tau e^{ (\beta_{10}  + \beta_{20})A_i t}A_i t  \left\{\frac{A_i -\hat \pi( Z_i)}{\hat S_c(t | A_i,Z_i)}-\frac{A_i -\pi^{*}( Z_i)}{S^{*}_c(t | A_i,Z_i)}\right\}
\\ & \quad \times
dM_{ji}(t ; \beta_{0}, \Lambda^{*}),\nonumber\\
Q^{(18)}_j=&-(\beta_j-\beta_{j0})(\beta_1+\beta_2-\beta_{10}-\beta_{20})\left\{K^{(1)}(\beta_0,\hat S_c, \hat \pi)-K^{(1)}(\beta_0,S^{*}_c, \pi^{*})\right\},\\
Q^{(19)}_j=&-(\beta_1+\beta_2-\beta_{10}-\beta_{20})\frac{1}{n}\sum_{i=1}^n\int_0^\tau e^{ (\beta_{10}  + \beta_{20})A_i t}A_i t \left\{\frac{A_i -\hat \pi( Z_i)}{\hat S_c(t | A_i,Z_i)}-\frac{A_i -\pi^{*}( Z_i)}{S^{*}_c(t | A_i,Z_i)}\right\}\\&\;\;\;\;\times Y_i(t)d\left\{\hat \Lambda_j(t,Z_i ; \beta)-\Lambda_j^{*}(t,Z_i; \beta_0)\right\},\nonumber
\end{align*}
\begin{align*}
Q^{(110)}_j=&-(\beta_1+\beta_2-\beta_{10}-\beta_{20})\frac{1}{n}\sum_{i=1}^n\int_0^\tau e^{ (\beta_{10}  + \beta_{20})A_i t}A_i t \left\{\frac{A_i -\hat \pi( Z_i)}{\hat S_c(t | A_i,Z_i)}-\frac{A_i -\pi^{*}( Z_i)}{S^{*}_c(t | A_i,Z_i)}\right\}\\&\;\;\;\;\times Y_i(t)d\left\{\hat \Lambda_j(t,Z_i ; \beta_0)-\Lambda_j^{*}(t,Z_i)\right\},\nonumber\\
Q^{(111)}_j=&-(\beta_j-\beta_{j0})K^{(1)}(\beta_0,S^{*}_c, \pi^{*})-(\beta_j-\beta_{j0})\left\{K^{(1)}(\beta,  \hat S_c, \hat \pi)-K^{(1)}(\beta_0,S^{*}_c, \pi^{*})\right\},\\
Q^{(112)}_j=&-K^{(2)}_j(\beta, S^{*}_c, \pi^{*})+K^{(2)}_j(\beta,\hat S_c, \hat \pi)-K^{(2)}_j(\beta, S^{*}_c, \pi^{*}).
\end{align*}

$Q^{(11)}_j$ is a martingale integral, therefore, by Lemma \ref{ci}, we have $Q^{(11)}_j=O_p(n^{-1/2}|\beta_1+\beta_2-\beta_{10}-\beta_{20}|)$.

By Assumptions \ref{ass0n} and \ref{ass4n}, we have $Q^{(12)}_j=O_p\left(\left|\beta_1+\beta_2-\beta_{10}-\beta_{20}\right|\left|\beta_j-\beta_{j0}\right|\right)$ and $Q^{(16)}_j=O_p\left(\left|\beta_1+\beta_2-\beta_{10}-\beta_{20}\right|^2\right)$.

By Assumptions \ref{ass0n}, \ref{ass4n} and \ref{variationn} we have $Q^{(13)}_j=O_p\left(\left|\beta_1+\beta_2-\beta_{10}-\beta_{20}\right|\left|\beta_j-\beta_{j0}\right|\right)$.

By Assumptions \ref{assumption}, \ref{ass0n}, \ref{ass4n}  and  \eqref{nuisnew} we have $Q^{(14)}_j=o_p(\left|\beta_1+\beta_2-\beta_{10}-\beta_{20}\right|)$, $Q^{(17)}_j=o_p\left(\left|\beta_1+\beta_2-\beta_{10}-\beta_{20}\right|^2\right)$ and $Q^{(18)}_j=o_p\left(\left|\beta_1+\beta_2-\beta_{10}-\beta_{20}\right|\left|\beta_j-\beta_{j0}\right|\right)$.

By Assumptions \ref{ass0n} and \eqref{nuisnew} we have $Q^{(17)}_j=o_p\left(\left|\beta_1+\beta_2-\beta_{10}-\beta_{20}\right|\right)$. Moreover, we notice that, if $\Lambda^{*}(t,Z)=\Lambda^{0}(t,Z)$, we would have $Q^{(17)}_j=o_p\left(n^{-1/2}\left|\beta_1+\beta_2-\beta_{10}-\beta_{20}\right|\right)$ since $dM(t ; \beta_0, \Lambda^{*})$ would be a martingale.

By Assumption \ref{assumption} and \ref{variationn}, we have $Q^{(19)}_j=o_p\left(\left|\beta_1+\beta_2-\beta_{10}-\beta_{20}\right|\left|\beta_j-\beta_{j0}\right|\right)$.
Moreover, if $a_nc_n=o_p(n^{-1/2})$ and $b_nc_n=o_p(n^{-1/2})$ (case a), b), c) of the Lemma), by Cauchy Schwartz inequality, together with Assumption \ref{variationn} we get \\$Q^{(110)}_j=o_p\left(n^{-1/2}\left|\beta_1+\beta_2-\beta_{10}-\beta_{20}\right|\left|\beta_j-\beta_{j0}\right|\right)$.

By Assumption \ref{assumption}, we have $Q^{(111)}_j=-(\beta_j-\beta_{j0})K^{(1)}(\beta_0,S^{*}_c, \pi^{*})+o_p\left(\left|\beta_j-\beta_{j0}\right|\right)$.

Moreover, by Assumptions \ref{variationn} and S\eqref{nuisnew}, we have $Q^{(112)}_j=-K^{(2)}_j(\beta, S^{*}_c, \pi^{*})+o_p\left(\left|\beta_j-\beta_{j0}\right|\right)$. We moreover notice that, if $S^{*}_c(\cdot | \cdot,\cdot)=S_{0c}(\cdot | \cdot,\cdot)$ and $\pi^{*}(\cdot)=\pi_0(\cdot)$, by Lemma \ref{marquis}, we have $Q^{(112)}_j=-K^{(2)}_j(\beta, S^{*}_c, \pi^{*})+O_p\left(n^{-1/2}\left|\beta-\beta_{0}\right|\right)$.

Therefore:
\eqnn
Q^{(1)}_j&=&-(\beta_1+\beta_2-\beta_{10}-\beta_{20})K^{(4)}_j(\beta_0, S^{*}_c,\pi^{*}, \Lambda^{*})
\\&&-(\beta_j-\beta_{j0})K^{(1)}_j(\beta_0,S^{*}_c,\pi^{*})-K^{(2)}_j(\beta, S^{*}_c,\pi^{*})\\
&&+O_p\left(n^{-1/2}\left|\beta_1+\beta_2-\beta_{10}-\beta_{20}\right|+\left|\beta_1+\beta_2-\beta_{10}-\beta_{20}\right|^2\right)\\
&&+o_p(\left|\beta_1+\beta_2-\beta_{10}-\beta_{20}\right|+n^{-1/2}\left|\beta_1+\beta_2-\beta_{10}-\beta_{20}\right|\left|\beta_j-\beta_{j0}\right|).
\een

\begin{itemize}
\item{Term $Q^{(2)}$:}
\end{itemize}
Adding and subtracting we have:
\eqnn
Q^{(2)}_j&=&\frac{1}{n}\sum_{i=1}^n\int_0^\tau e^{ (\beta_{10}  + \beta_{20}) A_i t}\left\{S^{*}_c(t | A_i,Z_i)\right\}^{-1}\left\{A_i -\pi^{*}( Z_i)\right\}Y_i(t)d\left\{\hat \Lambda_{j}(t, Z_i ; \beta_0)-\Lambda^{*}_{j}(t, Z_i)\right\}\\
&&+\frac{1}{n}\sum_{i=1}^n\int_0^\tau e^{ (\beta_{10}  + \beta_{20}) A_i t}\left[\hat S^{-1}_c(t | A_i,Z_i)\left\{A_i -\hat \pi( Z_i)\right\}-\left\{S^{*}_c(t | A_i,Z_i)\right\}^{-1}\left\{A_i -\pi^{*}( Z_i)\right\}\right]\nonumber\\&&\times Y_i(t)d\left\{\hat \Lambda_{j}(t, Z_i ; \beta_0)-\Lambda^{*}_{j}(t, Z_i)\right\}\\
&=&Q^{(21)}_j+Q^{(22)}_j.
\een

By Assumption \ref{assumption}, \ref{ass0n}, \ref{ass4n}, we have:
\eqnn
\left|Q^{(21)}_j\right|&\leq&  \left|\int_0^\tau 
\left[ \frac{1}{n} \sum_{i=1}^ne^{(\beta_{10}+\beta_{20})A_it}\left\{S^{*}_c(t | A_i,Z_i)\right\}^{-1}\left\{A_i - \pi^{*}( Z_i)\right\}Y_i(t) \right]
\right.
\\
&&\left.\sup_{Z \in \mathcal{Z}}d\left\{\hat \Lambda_{j}(t,Z ; \beta_0)- \Lambda_j^{*}(t,Z)\right\}\right|
\\&=&o_p(1).\nonumber
\een

We moreover notice that, if $S^{*}_c(\cdot | \cdot, \cdot)=S_{c0}(\cdot | \cdot,\cdot)$ and $\pi^{*}(\cdot)=\pi_0(\cdot)$ (case a) and c)), by Lemma \ref{marquis}, we have $Q^{(21)}_j=o_p(n^{-1/2})$.
Otherwise, under case b), $Q^{(21)}_j=O_p(n^{-1/2})$.

Moreover by Cauchy-Schwartz inequality we have:
\eqnn
&&\left|Q^{(22)}_j\right|
\\&\leq&\frac{1}{n}e^{(\beta_{10}+\beta_{20})\tau}\sqrt{\sum_{i=1}^n\sup_{t \in [0,\tau]}\left[\hat S^{-1}_c(t | A_i,Z_i)\left\{A_i -\hat \pi( Z_i)\right\}- \left\{S^{*}_c(t | A_i,Z_i)\right\}^{-1}\left\{A_i - \pi^{*}( Z_i)\right\}\right]^2} \nonumber\\&&\cdot \sqrt{\sum_{i=1}^n\left[\int_0^\tau d\left\{\hat \Lambda_{j}(t,Z_i; \beta_0)-\Lambda^{*}_{j}(t,Z_i)\right\}\right]^2}.
\een
Therefore:
\eqnn
&&\left|Q^{(22)}_j\right|\\
&\leq&e^{(\beta_{10}+\beta_{20})\tau}\sqrt{\frac{1}{n}\sum_{i=1}^n\sup_{t \in [0,\tau]}\left[\hat S^{-1}_c(t | A_i,Z_i)\left\{A_i -\hat \pi( Z_i)\right\}- \left\{S^{*}_c(t | A_i,Z_i)\right\}^{-1}\left\{A_i - \pi^{*}( Z_i)\right\}\right]^2} \nonumber\\&&\cdot \left\{\sup_{t\in[0,\tau],Z\in\mathcal{Z}}\left|\hat \Lambda_j(\tau,Z; \beta_0)- \Lambda^{*}_j(\tau,Z)\right|\right\}\nonumber.\\
\een
Therefore, by Assumption \ref{assumption} $Q^{(22)}_j=o_p(1)$ and so $Q^{(2)}_j=o_p(1)$.  However since $a_nc_n=o_p(n^{-1/2})$ and $b_nc_n=o_p(n^{-1/2})$ (case a), b), c) of the Lemma), we have
$
Q^{(22)}_j=o_p(n^{-1/2}),
$
and so $Q^{(2)}_j=O_p(n^{-1/2})$.

\begin{itemize}
\item{Term $Q^{(3)}$:}
\end{itemize}
If $\Lambda^{*}(\cdot,\cdot)=\Lambda^{0}(\cdot,\cdot)$ (case b), c)), we have $Q^{(3)}=o_p(n^{-1/2})$ since it would be a martingale integral with integrand converging to zero. Otherwise, under case a), $Q^{(3)}=O_p(n^{-1/2})$.

Putting all of these steps together we have:
\eqnn
{S}_{1,n}(\beta, \hat S_c,  \hat \pi, \hat \Lambda)&=&{S}_{1,n}(\beta_0, S^{*}_c,  \pi^{*}, \Lambda^{*})\\
&&-(\beta_1+\beta_2-\beta_{10}-\beta_{20})K^{(4)}(\beta_0, S^{*}_c, \pi^{*}, \Lambda^{*})\\
&&-(\beta_j-\beta_{j0})K^{(1)}(\beta_0,S^{*}_c,\pi^{*})-K^{(2)}(\beta,  S^{*}_c,\pi^{*})\\
&&+O_p\left(n^{-1/2}\left|\beta_1+\beta_2-\beta_{10}-\beta_{20}\right|+\left|\beta_1+\beta_2-\beta_{10}-\beta_{20}\right|^2\right)+Q^{(21)}+Q^{(3)},
\een
where $Q^{(21)}=o_p(n^{-1/2}),Q^{(3)}=o_p(n^{-1/2})$ under case c) of the Theorem, $Q^{(21)}=o_p(n^{-1/2}),Q^{(3)}=O_p(n^{-1/2})$ under case a) of the Theorem, and under case c), $Q^{(21)}=O_p(n^{-1/2}),Q^{(3)}=o_p(n^{-1/2})$.

If $\hat \Lambda(\cdot,\cdot)$ depends on the unknown $\beta$, by Assumption \ref{variationn}, we have:
\eqnn
K^{(2)}_j(\beta,S^{*}_c, \pi^{*})&=&(\beta_j-\beta_{j0})\frac{1}{n}\sum_{i=1}^n\int_0^\tau e^{ (\beta_{10}  + \beta_{20}) A_i t}\left\{S^{*}_c(t | A_i,Z_i)\right\}^{-1}\left\{A_i -\pi^{*}( Z_i)\right\}Y_i(t)
\\&&\times\frac{1}{n}\sum_{l=1}^nq_{jl}(t)dt,\een
where we call $q_{ji}(t)$ a function, such that:
\be
\hat \Lambda_{j}(t, Z ; \beta)-\hat \Lambda_{j}(t, Z ; \beta_0)&=&(\beta_j-\beta_{j0})*\frac{1}{n}\sum_{i=1}^n\int_0^\tau q_{ji}(t).
\ee

Therefore, we can conclude that:
\eqnn
{S}_{1,n}(\beta, \hat S_c,  \hat \pi, \hat \Lambda)&=&{S}_{1,n}(\beta_0, S^{*}_c,  \pi^{*}, \Lambda^{*})+Q^{(21)}+Q^{(3)}+{K}({\beta}-{\beta}_0)\\
&&+O_p\left(n^{-1/2}\left|\beta_1+\beta_2-\beta_{10}-\beta_{20}\right|+\left|\beta_1+\beta_2-\beta_{10}-\beta_{20}\right|^2\right),\een
where
${K}$ is a $2\times2$ matrix with the following components:
\eqnn
{K}_{jj}&=&-K^{(1)}(\beta_0,S^{*}_c, \pi^{*})-K^{(2)}_j(\beta_0,S^{*}_c,\pi^{*})/(\beta_j-\beta_{j0})-K^{(4)}_j(\beta_0,S^{*}_c,\pi^{*},\Lambda^{*}),
\een
and
\eqnn
{K}_{12}=-K^{(1)}_{4}(\beta_0,S^{*}_c,\pi^{*},\Lambda^{*}),\;\;\;{K}_{21}=-K^{(4)}_j(\beta_0,S^{*}_c,\pi^{*},\Lambda^{*}).
\een
\begin{remark}
If the estimator $\hat \Lambda(\cdot,\cdot)$ does not depend on $\beta$ or it depends on some initial estimator of it the decomposition simplifies.
Specifically, terms  $K^{(2)}_j(\beta, S^{*}_c, \pi^{*}),Q^{(12)},Q^{(18)},Q^{(112)},Q^{(113)}$ cancels.\end{remark}
\end{newproof7}

\newenvironment{newproof8}{\begin{proof}\textsc{\emph{of Lemma \ref{ifQ4}.}}}{\end{proof}}

\begin{newproof8}
We remind the reader that \\$Q^{(3)}_j=\sqrt{n}\left[{S}_j^n(\beta_0, \hat S_c,  \hat \pi,   \Lambda^{*})-\sqrt{n}{S}_j^n(\beta_0, S_{c0}, \pi_0,\Lambda^{*})\right]$.
Using the fact that $\hat\pi(z)=\expit(\hat\alpha^\top z)$, $\hat S_c(t | a,z)=\exp\left(-\hat\Lambda_{c}(t)e^{\hat\eta^\top d}\right)$ we have, by Taylor expansion:
\eqnn
Q^{(3)}_j&=&\sqrt{n}\left[{S}_j^n(\beta_0, \hat\eta, \hat \Lambda_{c} , \hat \alpha,  \Lambda^{*})-\sqrt{n}{S}_j^n(\beta_0, \eta_0, \Lambda_{c0}, \alpha_0, \Lambda^{*})\right]\\
&=&\frac{1}{\sqrt{n}}\sum_{i=1}^n\int_0^\tau e^{(\beta_{10}+\beta_{20})A_it}\left\{{D}_{ij}^n(t,\eta_{0}, \Lambda_{c0}, \alpha_0\right\}^\top{\Delta} dM_{ji}(t; \beta_{j0}, \Lambda^{*}_j)+o_p(1),
\een
where
\eqnn
{D}_{ij}^n(t, \eta,\Lambda_{c},\alpha):=[\partial_{\eta} f (t,A_i,Z_i;\eta,\Lambda_{c},\alpha),
\partial_{\Lambda_{c}} f(t,A_i,Z_i;\eta,\Lambda_{c},\alpha),
\partial_\alpha f(t,A_i,Z_i;\eta,\Lambda_{c},\alpha)
]^\top,
\een
\eqnn
f (t,A_i,Z_i; \eta,\Lambda_{c},\alpha):=\left\{A_i-\expit(\alpha^\top Z)\right\}\exp\left(\Lambda_{c}(t)e^{\eta^\top D}\right),\een
and 
$
{\Delta}:=[\hat\eta-\eta_0,
\hat\Lambda_{c}(t)-\Lambda_{c0}(t),
\hat\alpha-\alpha_0]^\top.
$

Standard algebra gives us:
\eqnn
{D}_{ij}^n(t, \eta,\Lambda_{c},\alpha)&=&[ \left\{A_i-\expit(\alpha^\top Z_i)\right\})\exp\left(\Lambda_{c}(t)e^{\eta^\top D_i}\right)\Lambda_{c}(t)e^{\eta^\top D_i}D_i,\\&&
\left\{A_i-\expit(\alpha^\top Z_i)\right\})\exp\left(\Lambda_{c}(t)e^{\eta^\top D_i}\right)e^{\eta^\top D_i},\\
&&,-\exp\left(\Lambda_{c}(t)e^{\eta^\top D_i}\right)\expit(\alpha^\top Z_i)e^{\alpha^\top Z_i}Z_i
]^\top.
\een

Moreover, we know by traditional theory that
\eqnn
\hat \alpha-\alpha_0=O_p(n^{-1/2}),\;\;\;\;\;\;\hat \eta-\eta_0=O_p(n^{-1/2}),\;\;\;\;\;\;\;\sup_{t \in [0,\tau]} \left\{\hat\Lambda_{c}(t)-\Lambda_{c0}(t)\right\}=O_p(n^{-1/2}).
\een

Therefore, by the above and by Assumption A*\ref{convnn} we have:
\be\label{Q4n}
Q^{(3)}_j&=&\sqrt{n}\int_0^\tau\left[ \left\{P^{(a')}_{1}\right\}^\top(t) (\hat\eta-\eta_0)+P^{(a')}_{2}(t) (\hat\Lambda_{c}(t;  \hat \eta)-\Lambda_{c0}(t))-\left\{P^{(a')}_{3}\right\}^\top(t) (\hat\alpha-\alpha_0)\right]dt\nonumber \\
&=&\sqrt{n}\int_0^\tau \left[ \left\{p^{(a')}_{1}\right\}^\top(t) (\hat\eta-\eta_0)+p^{(a')}_{2}(t) (\hat\Lambda_{c}(t;  \hat \eta)-\Lambda_{c0}(t))-\left\{p^{(a')}_{3}\right\}^\top(t) (\hat\alpha-\alpha_0)\right]dt\nonumber \\&&+o_p(1).
\ee

Lemma \ref{ifQ4a} and \ref{ifQ4b} provide the influence functions of $\hat \alpha, \hat \eta, \hat \Lambda_c$.
Therefore, plugging them in \eqref{Q4n} we can conclude that:

\eqnn
\sqrt{n}Q^{(3)}_j
&=& \left(\int_0^\tau\left[\frac{s_d^{(2)}(t)}{s_d^{(0)}(t)}-\left\{\frac{s_d^{(1)}(t)}{s_d^{(0)}(t)}\right\}^2\right]s_d^{(0)}(t)d\Lambda_{c0}(t)\right)^{-1}\frac{1}{\sqrt{n}}\sum_{i=1}^n\int_0^\tau \left\{D_i-\frac{s_d^{(1)}(t)}{s_d^{(0)}(t)}\right\}dM^c_i(t)
\\
&&\times\left[\int_0^\tau (p^{(a')}_{1})^\top(t)-[\int_0^\tau p^{(a')}_{2}(t)\int_0^t d\Lambda_{c0}(u; \eta_0)\frac{s_d^{(1)}(u)}{s_d^{(0)}(u)}dt\right]
\\
&& +\int_0^\tau p^{(a')}_{2}(t)\frac{1}{\sqrt{n}}\sum_{i=1}^n\int_0^t \left\{s_d^{(0)}(u)\right\}^{-1}dM^c_i(u)
\\
&&- \int_0^\tau(p^{(a')}_{3})^\top(t) \left(E\left[Z^\top Z \pi_0(Z_i)\left\{1-\pi_0(Z_i)\right\}\right]\right)^{-1}\frac{1}{\sqrt{n}}\sum_{i=1}^n Z_i\left\{A_i-\pi_0(Z_i)\right\}dt+o_p(1).\een
\end{newproof8}

\newenvironment{newproof9}{\begin{proof}\textsc{\emph{of Lemma \ref{ifQ21a}.}}}{\end{proof}}

\begin{newproof9}
We have:
\eqnn
\sqrt{n}Q^{(21)}_j&=&-\frac{1}{\sqrt{n}}\sum_{i=1}^n\int_0^\tau e^{ (\beta_{10}  + \beta_{20}) A_i t} \left\{S^{*}_c(t | A_i,Z_i)\right\}^{-1}\left\{A_i -\pi^{*}( Z_i)\right\}Y_i(t)\\&&\times \left[(\hat\gamma_j-\gamma_{j0})^\top Z_idt+d\left\{\hat G_j(t ; \beta_{j0},\hat \gamma_j)-G_{j0}(t)\right\}\right].
\een

We notice that, by \cite{ly}, under regularity Assumptions, we have, for each $t,z$:
\be\label{basn}
\left\{\hat \Lambda_j(t,z; \beta_{j0}, \hat\gamma_j)-\Lambda_{j0}(t,z)\right\}=O_p(n^{-1/2}),
\ee
and
\be\label{gamman}
\left\{\hat \gamma_{j}-\gamma_{j0}\right\}=O_p(n^{-1/2}).
\ee

Therefore, by the above and by Assumption B*\ref{if1nnn}, we have:
\be\label{3ifn}
Q^{(21)}_j=\sqrt{n}\int_0^\tau \left[(\hat\gamma_j-\gamma_{j0})^\top p^{(b')}_{1}(t)dt +p^{(b')}_{0}(t)d\left\{\hat G_j(t ; \beta_{j0},\hat \gamma_j)-G_{j0}(t)\right\}\right]+o_p(1).
\ee

Lemma \ref{ifQ21aab} and \ref{ifQ21ab} provide influence functions for $\hat \gamma_j$ and $\hat G_j(t ; \beta_{j0},\hat \gamma_j)$.

Therefore, plugging them into \eqref{3ifn} we have;
\eqnn\label{ifQn}
Q^{(21)}_j&=&\int_0^\tau \left(\left[\frac{1}{n}\sum_{i=1}^n\int_0^{\tau}\left\{D_i-\frac{s_d^{(1)}(t)}{s_d^{(0)}(t)}\right\}^{\otimes 2}Y_i(t)dt\right]^{-1}\frac{1}{\sqrt{n}}\sum_{i=1}^n
\int_0^\tau \left\{Z_i-\frac{s_z^{(1)}(t)}{s_z^{(0)}(t)}\right\}dM_{ji}(t)\right)^\top 
\\
&&
\cdot
\left\{p^{(b')}_{1}(t)dt -p^{(b')}_{0}(t)\frac{s_z^{(1)}(t)}{s_z^{(0)}(t)}dt\right\}
\\
&&+\int_0^\tau p^{(b')}_{0}(t)\left\{s_z^{(0)}(t)\right\}^{-1}\frac{1}{\sqrt{n}}\sum_{i=1}^ndM_{ji}(t)+o_p(1).
\een
\end{newproof9}

\newenvironment{newproof20}{\begin{proof}\textsc{\emph{of Lemma \ref{ifQ21b}.}}}{\end{proof}}

\begin{newproof20}
The proof is similar to the proof of Lemma \ref{ifQ21a}, using Lemma \ref{ifQ21bb} instead of \ref{ifQ21ab} and we leave it to the reader.
\end{newproof20}


\subsection{Additional Lemmas and Proofs}

\begin{lemmas}\label{mvt}
A simple application of the multidimensional mean value theorem gives us
\eqnn
e^{(\beta_1+\beta_2)t}-e^{(\beta_{10}+\beta_{20})t}=e^{(\beta^*_1+\beta^*_2)t}t\left(\beta_1+\beta_2-\beta_{10}-\beta_{20}\right),
\een
where $\beta^{*}_j$ is a point between $\beta_j$ and $\beta_{j0}$ for $j=1,2$.
\end{lemmas}

\begin{lemmas}\label{ci}
Let $H(t)$ be a stochastic process such that $P\left(\sup_{t \in [0,\tau]}|H(t)|\leq K\right)=1$ for some $K<\infty$. 
We have, for any bounded $\beta_j$:
\be\label{resci}
\frac{1}{n}\sum_{i=1}^n\int_0^\tau H_i(t)dM_{ji}(t ; \beta_{j}, \Lambda_{j})=E\left[\int_0^\tau H(t)dM_{j}(t ; \beta_{j}, \Lambda_{j})\right]+O_p(n^{-1/2})
\ee
\end{lemmas}
\newenvironment{newproof11}{\begin{proof}\textsc{\emph{of Lemma \ref{ci}.}}}{\end{proof}}

\begin{newproof11}
By definition of $dM_{ji}$, we have:
\eqnn
\frac{1}{n}\sum_{i=1}^n\int_0^\tau H_i(t)dM_{ji}(t ; \beta_{j}, \Lambda_{j})&=&\frac{1}{n}\sum_{i=1}^n\int_0^\tau H_i(t)\left[dN_{ji}(t)-Y_i(t)\left\{\beta_{j}A_idt+d\Lambda_{j}(t,Z_i)\right\}\right]\\
&=&\frac{1}{n}\sum_{i=1}^n\delta_i H_i(X_i)-X_i\beta_j A_i -\int_0^{X_i}H_i(t)d\Lambda_{j}(t,Z_i).
\een

We have, by Assumptions \ref{ass0n} and \ref{boundln}
\eqnn
\left|\delta_i H_i(X_i)-X_i\beta_j A_i -\int_0^{X_i}H_i(t)d\Lambda_{j}(t,Z_i)\right|\leq K+\tau |\beta_j|+K|\Lambda_j(\tau,Z_i)|<\infty
\een

Therefore, by Hoeffding's inequality we have \eqref{resci}.
\end{newproof11}

\newenvironment{newproof12}{\begin{proof}\textsc{\emph{of Lemma \ref{marquis}.}}}{\end{proof}}

\begin{lemmas}\label{marquis}
It holds:
\eqnn
\sup_{t \in [0,\tau]} \left|\frac{1}{n}\sum_{i=1}^n \left\{A_i-\pi_0(Z_i)\right\}\left\{S_{c0}(t | A_i,Z_i)\right\}^{-1}Y_i(t)e^{(\beta_{10}+\beta_{20})A_i t}\right|=O_p\left(n^{-1/2}\right).
\een
\end{lemmas}
\begin{newproof12}
This is a slightly modified version of Lemma A13 of \cite{hou2019estimating}, adapted to include the survival of the censoring. We leave the proof to the reader.
\end{newproof12}

\begin{lemmas}\label{ifQ4a}
Let $\pi(Z ; \alpha)=\expit{(\alpha^\top Z)}$ and 
let $\hat \alpha$ be the MLE estimator for $\alpha$.
We have:
\eqnn
\sqrt{n}(\hat \alpha-\alpha_0)=\left(E\left[Z^\top Z \pi_0(Z)\left\{1-\pi_0(Z)\right\}\right]\right)^{-1}\frac{1}{\sqrt{n}}\sum_{i=1}^n Z_i\left\{A_i-\pi_0(Z_i)\right\}+o_p(1).
\een
\end{lemmas}

\newenvironment{newproof13}{\begin{proof}\textsc{\emph{of Lemma \ref{ifQ4a}.}}}{\end{proof}}

\begin{newproof13}
Estimation of parameter $\alpha$ is done through classical MLE method. By classical MLE argument we have (proved in \cite{zeng2010adjustment}):
\eqnn
\sqrt{n}(\hat \alpha-\alpha_0)&=&\left\{-\frac{1}{n}\sum_{i=1}^n\frac{e^{-(\alpha_0)^\top Z_i}}{\left\{1+e^{-(\alpha_0)^\top Z_i}\right\}^2}Z_i^\top Z_i\right\}^{-1}\frac{1}{\sqrt{n}}\sum_{i=1}^n \frac{A_ie^{-(\alpha_0)^\top Z_i}-1+A_i}{1+e^{-(\alpha_0)^\top Z_i}}Z_i\\
&=&\left(E\left[Z^\top Z \pi_0(Z)\left\{1-\pi_0(Z)\right\}\right]\right)^{-1}\frac{1}{\sqrt{n}}\sum_{i=1}^n Z_i\left\{A_i-\pi_0(Z_i)\right\}+o_p(1).
\een
\end{newproof13}

\begin{lemmas}\label{ifQ4b}
Let $S_c(t | A,Z)=g(t | A,Z ; \eta, \Lambda_c)=\exp\left(-\Lambda_{c}e^{\eta^\top D}\right)$ and 
let $\hat \eta$ and $\hat \Lambda_{c}(t)$ be the Cox estimators.
Under Assumptions \ref{boundln}, \ref{boundZn} and  A*\ref{ifn} we have:
\eqnn
&&\sqrt{n}\left\{\hat\eta-\eta_0\right\}\\&=&\left(\int_0^\tau\left[\frac{s_d^{(2)}(t)}{s_d^{(0)}(t)}-\left\{\frac{s_d^{(1)}(t)}{s_d^{(0)}(t)}\right\}^2\right]s_d^{(0)}(t)d\Lambda_{c0}(t)\right)^{-1}\frac{1}{\sqrt{n}}\sum_{i=1}^n\int_0^\tau \left\{D_i-\frac{s_d^{(1)}(t)}{s_d^{(0)}(t)}\right\}dM^c_i(t)+o_p(1),\een
and
\be\label{3if}
&&
\sqrt{n}\left\{\hat\Lambda_{c}(t ; \hat \eta)-\Lambda_{c0}(t)\right\}\\
&=&\left\{\left(\int_0^\tau\left[\frac{s_d^{(2)}(t)}{s_d^{(0)}(t)}-\left\{\frac{s_d^{(1)}(t)}{s_d^{(0)}(t)}\right\}^2\right]s_d^{(0)}(t)d\Lambda_{c0}(t)\right)^{-1}\frac{1}{\sqrt{n}}\sum_{i=1}^n\int_0^\tau \left\{D_i-\frac{s_d^{(1)}(t)}{s_d^{(0)}(t)}\right\}dM^c_i(t)\right\}^\top\nonumber\\
&&\cdot\int_0^t  -d\Lambda_{c0}(u; \eta_0)\frac{s_d^{(1)}(u)}{s_d^{(0)}(u)}du +\frac{1}{\sqrt{n}}\sum_{i=1}^n\int_0^t \left\{s_d^{(0)}(u)\right\}^{-1}dM^c_i(u)+o_p(1).\nonumber
\ee\end{lemmas}

\newenvironment{newproof14}{\begin{proof}\textsc{\emph{of Lemma \ref{ifQ4b}.}}}{\end{proof}}

\begin{newproof14}
Estimation of parameter $\eta$ uses the following score:
\eqnn
U_1(\eta)&=&\frac{1}{n}\sum_{i=1}^n\int_0^\tau \left\{D_i-\frac{\sum_{j=1}^nY_j(t)D_je^{\eta^\top D_j}}{\sum_{j=1}^nY_j(t)e^{\eta^\top D_j}}\right\}dN^c_i(t)\\
&=&\frac{1}{n}\sum_{i=1}^n\int_0^\tau \left\{D_i-\frac{\sum_{j=1}^nY_j(t)D_je^{\eta^\top D_j}}{\sum_{j=1}^nY_j(t)e^{\eta^\top D_j}}\right\}\left[dM^c_i(t)+Y_i(t)d\Lambda_{c0}(t)e^{(\eta_0)^\top D_i}dt\right]\\
&=&\frac{1}{n}\sum_{i=1}^n\int_0^\tau \left\{D_i-\frac{\sum_{j=1}^nY_j(t)D_je^{\eta^\top D_j}}{\sum_{j=1}^nY_j(t)e^{\eta^\top D_j}}\right\}dM^c_i(t),
\een
where 
$M^c_i(t)=N_i^c(t)-Y_i(t)\Lambda_{c0}(t)e^{(\eta_0)^\top D_i}$,
and
$N^c(t):=\mathbf{1}\left\{X\leq t, \delta=0\right\}$.
By Taylor expansion we have:
\eqnn
U_1(\eta_0)&=&U_1(\eta_0)-U_1(\hat\eta)\\
&=&\frac{1}{n}\sum_{i=1}^n\int_0^\tau \left\{\hat\eta-\eta_0\right\}^\top\left[\frac{S_d^{(2)}(t)}{S_d^{(0)}(t)}-\left\{\frac{S_d^{(1)}(t)}{S_d^{(0)}(t)}\right\}^2\right]dM^c_i(t)\\
&=&\frac{1}{n}\sum_{i=1}^n\int_0^\tau \left\{\hat\eta-\eta_0\right\}^\top\left[\frac{S_d^{(2)}(t)}{S_d^{(0)}(t)}-\left\{\frac{S_d^{(1)}(t)}{S_d^{(0)}(t)}\right\}^2\right]dN^c_i(t).
\een
Therefore, by Assumption A*\ref{ifc}
\be\label{2if}
&&\sqrt{n}\left\{\hat\eta-\eta_0\right\}\\&=&\left(\frac{1}{n}\sum_{i=1}^n\int_0^\tau\left[\frac{S_d^{(2)}(t)}{S_d^{(0)}(t)}-\left\{\frac{S_d^{(1)}(t)}{S_d^{(0)}(t)}\right\}^2\right]dN^c_i(t)\right)^{-1}\nonumber\\
&&\times\frac{1}{\sqrt{n}}\sum_{i=1}^n\int_0^\tau \left\{D_i-\frac{\sum_{j=1}^nY_j(t)D_je^{\eta^\top D_j}}{\sum_{j=1}^nY_j(t)e^{\eta^\top D_j}}\right\}dM^c_i(t)\nonumber\\
&=&\left(\frac{1}{n}\sum_{i=1}^n\int_0^t \left[\frac{s_d^{(2)}(t)}{s_d^{(0)}(t)}-\left\{\frac{s_d^{(1)}(t)}{s_d^{(0)}(t)}\right\}^2\right]dN^c_i(t)\right)^{-1}\frac{1}{\sqrt{n}}\sum_{i=1}^n\int_0^\tau \left\{D_i-\frac{s_d^{(1)}(t)}{s_d^{(0)}(t)}\right\}dM^c_i(t)+o_p(1)\nonumber\\
&=&\left(\int_0^\tau\left[\frac{s_d^{(2)}(t)}{s_d^{(0)}(t)}-\left\{\frac{s_d^{(1)}(t)}{s_d^{(0)}(t)}\right\}^2\right]s_d^{(0)}(t)d\Lambda_{c0}(t)\right)^{-1}\frac{1}{\sqrt{n}}\sum_{i=1}^n\int_0^\tau \left\{D_i-\frac{s_d^{(1)}(t)}{s_d^{(0)}(t)}\right\}dM^c_i(t)\nonumber\\&&+o_p(1).\nonumber.
\ee

We now need to find the influence function of 
$\hat \Lambda_{c}(t ; \hat \eta)-\Lambda_{c0}(t)=\hat \Lambda_{c}(t ; \hat \eta)-\hat \Lambda_{c}(t ;  \eta_0)+\hat \Lambda_{c}(t ;  \eta_0)-\Lambda_{c0}(t).$
Since
$
\hat \Lambda_{c}(t ; \eta)=\int_0^t {\sum_{i=1}^n dN^c_i(u)} / {\sum_{i=1}^n Y_i(u)e^{\eta^\top D_i}},
$
by Taylor expansion and by \eqref{2if} and Assumption A* \ref{ifn} we have:
\begin{align}\label{3ifa}
&\sqrt{n}\left\{\hat\Lambda_{c}(t ; \hat\eta)-\hat\Lambda_{c}(t ; \eta_0)\right\}\\&=-(\hat \eta-\eta_0)^\top\int_0^td \hat \Lambda_{c}(u; \eta_0)\frac{S_d^{(1)}(u)}{S_d^{(0)}(u)}du \nonumber\\
&=\left\{\left(\int_0^\tau\left[\frac{s_d^{(2)}(t)}{s_d^{(0)}(t)}-\left\{\frac{s_d^{(1)}(t)}{s_d^{(0)}(t)}\right\}^2\right]s_d^{(0)}(t)d\Lambda_{c0}(t)\right)^{-1}\frac{1}{\sqrt{n}}\sum_{i=1}^n\int_0^\tau \left\{D_i-\frac{s_d^{(1)}(t)}{s_d^{(0)}(t)}\right\}dM^c_i(t)\right\}^\top\nonumber\\
&\quad\times\int_0^t  -d \hat \Lambda_{c}(u; \eta_0)\frac{s_d^{(1)}(u)}{s_d^{(0)}(u)}du +o_p(1).\nonumber
\end{align}

Estimation of parameter $\Lambda_{c}(t)$ uses the following score:
\eqnn
U_2(\Lambda_{c}(t) ; \eta)&=&\frac{1}{n}\sum_{i=1}^n\int_0^t \left\{dN^c_i(u)-Y_i(t)d\Lambda_{c}(u)e^{\eta^\top D_i}\right\}\\
&=&\frac{1}{n}\sum_{i=1}^n\int_0^t \left[dM^c_i(u)-Y_i(u)d\left\{\Lambda_{c}(u)e^{\eta^\top D_i}-\Lambda_{c0}(u)e^{(\eta_0)^\top D_i}\right\}\right].
\een
Therefore, by construction of $\hat \Lambda_{c}(t ; \eta_0)$ we have
\eqnn
U_2(\Lambda_{c0}(t) ; \eta_0)=U_2(\Lambda_{c0}(t) ; \eta_0)-U_2(\hat \Lambda_{c0}(t ; \eta_0) ; \eta_0)=\int_0^tS_d^{(0)}(t)\left\{d\hat\Lambda_{c0}(t; \eta_0)-d\Lambda_{c0}(t)\right\},\een
and so we have:
\eqnn\label{3ifb}
\hat \Lambda_{c}(t ; \eta_0)-\Lambda_{c0}(t)=\frac{1}{n}\sum_{i=1}^n\int_0^t \left\{s_d^{(0)}(u)\right\}^{-1}dM^c_i(u)+o_p(1),
\een

Therefore, by putting together \eqref{3ifa} and \eqref{3ifb} we get:
\be
&&
\sqrt{n}\left\{\hat\Lambda_{c}(t ; \hat \eta)-\Lambda_{c0}(t)\right\}\\
&=&\left\{\left(\int_0^\tau\left[\frac{s_d^{(2)}(t)}{s_d^{(0)}(t)}-\left\{\frac{s_d^{(1)}(t)}{s_d^{(0)}(t)}\right\}^2\right]s_d^{(0)}(t)d\Lambda_{c0}(t)\right)^{-1}\frac{1}{\sqrt{n}}\sum_{i=1}^n\int_0^\tau \left\{D_i-\frac{s_d^{(1)}(t)}{s_d^{(0)}(t)}\right\}dM^c_i(t)\right\}^\top\nonumber\\
&&\times \int_0^t  -d\Lambda_{c0}(u; \eta_0)\frac{s_d^{(1)}(u)}{s_d^{(0)}(u)}du +\frac{1}{n}\sum_{i=1}^n\int_0^t \left\{s_d^{(0)}(u)\right\}^{-1}dM^c_i(u)+o_p(1).\nonumber
\ee
\end{newproof14}

\begin{lemmas}\label{ifQ21aab}
Let $\Lambda_j(t,Z)=G_j(t) + \gamma_j^{\top} Z t$ and let  $ \gamma_j$ be estimated using \eqref{tradgamma} of the paper.
Under Assumption B*\ref{if2n} it holds:
\eqnn
&&\sqrt{n}(\hat\gamma_j-\gamma_{j0})^\top\\
&=&\left[\frac{1}{n}\sum_{i=1}^n\int_0^{\tau}\left\{D_i-\frac{s_d^{(1)}(t)}{s_d^{(0)}(t)}\right\}^{\otimes 2}Y_i(t)dt\right]^{-1}\frac{1}{\sqrt{n}}\sum_{i=1}^n
\int_0^\tau \left\{Z_i-\frac{s_z^{(1)}(t)}{s_z^{(0)}(t)}\right\}dM_{ji}(t)+o_p(1).\nonumber\\
\een
\end{lemmas}

\newenvironment{newproof15}{\begin{proof}\textsc{\emph{of Lemma \ref{ifQ21aab}.}}}{\end{proof}}

\begin{newproof15}
The parameter $\gamma_j$ is estimated through the following score:
\eqnn
&&U_1\left([ \beta^{in}_{j} ,
           \gamma_j]^\top\right)\\&=&\frac{1}{n}\sum_{i=1}^n\int_0^{\tau}\left\{D_i-\frac{S_d^{(1)}(t)}{S_d^{(0)}(t)}\right\}\left\{dN_{ji}(t)-Y_i(t)\beta^{in}_jA_idt-Y_i(t)\gamma^\top_jZ_idt\right\}\\
           &=&\frac{1}{n}\sum_{i=1}^n\int_0^{\tau}\left\{D_i-\frac{S_d^{(1)}(t)}{S_d^{(0)}(t)}\right\}
           \\
           &&\times\left\{dM_{ji}(t)+Y_i(t)dG_{j0}(t)-Y_i(t)(\beta^{in}_j-\beta_{j0})A_idt-Y_i(t)(\gamma_j-\gamma_{j0})^\top Z_idt\right\}\nonumber\\
           &=&\frac{1}{n}\sum_{i=1}^n\int_0^{\tau}\left\{D_i-\frac{S_d^{(1)}(t)}{S_d^{(0)}(t)}\right\}\left\{dM_{ji}(t)-Y_i(t)(\beta^{in}_j-\beta_{j0})A_idt-Y_i(t)(\gamma_j-\gamma_{j0})^\top Z_idt\right\}. \nonumber
\een
Here $\beta^{in}_j$ is just some initial $\beta_j$ that we need for technical reason.

Therefore, by construction, we have:
\be
U_1\left([ \beta_{j0} ,
           \gamma_{j0}]^\top\right)&=&U_1\left([ \beta^{in}_{j} ,
           \hat \gamma_j]^\top\right)-U_1\left([ \beta_{j0} ,
           \gamma_{j0}]^\top\right)\\
           &=&\frac{1}{n}\sum_{i=1}^n\int_0^{\tau}\left\{D_i-\frac{S_d^{(1)}(t)}{S_d^{(0)}(t)}\right\}\left\{Y_i(t)[
           \beta^{in}_j-\beta_{j0},
           \hat\gamma_j-\gamma_{j0}]^\top D_idt\right\}.
\ee

Therefore, by Assumption B*\ref{if2n}, we have:
\be\label{2ifn}
&&\sqrt{n}(\hat\gamma_j-\gamma_{j0})^\top\\&=&\left[\frac{1}{n}\sum_{i=1}^n\int_0^{\tau}\left\{D_i-\frac{S_d^{(1)}(t)}{S_d^{(0)}(t)}\right\}^{\otimes 2}Y_i(t)dt\right]^{-1}\frac{1}{\sqrt{n}}\sum_{i=1}^n
\int_0^\tau \left\{Z_i-\frac{S_z^{(1)}(t)}{S_z^{(0)}(t)}\right\}dM_{ji}(t)\nonumber\\
&=&\left[\frac{1}{n}\sum_{i=1}^n\int_0^{\tau}\left\{D_i-\frac{s_d^{(1)}(t)}{s_d^{(0)}(t)}\right\}^{\otimes 2}Y_i(t)dt\right]^{-1}\frac{1}{\sqrt{n}}\sum_{i=1}^n
\int_0^\tau \left\{Z_i-\frac{s_z^{(1)}(t)}{s_z^{(0)}(t)}\right\}dM_{ji}(t)+o_p(1).\nonumber\\
\ee
\end{newproof15}

\begin{lemmas}\label{ifQ21ab}
Let $\Lambda_j(t,Z)=G_j(t) + \gamma_j^{\top} Z t$ and let  $ \gamma_j$ be estimated using \eqref{tradgamma} of the paper and $G_j(t)$ be estimated using \eqref{estbas1}.
Under Assumptions B*\ref{if2n} it holds:
\eqnn
&&\sqrt{n}\left\{\hat G_j(t ; \beta_{j0}, \hat \gamma_j)-G_{j0}(t)\right\}
\\&=&\int_0^t\left\{s_z^{(0)}(u)\right\}^{-1}\left[\frac{1}{\sqrt{n}}\sum_{i=1}^ndM_{ji}(u)-\left(\left[\frac{1}{n}\sum_{i=1}^n\int_0^{\tau}\left\{D_i-\frac{s_d^{(1)}(t)}{s_d^{(0)}(t)}\right\}^{\otimes 2}Y_i(t)dt\right]^{-1}
\right.\right.\nonumber\\
&&
\left.\left.\times \frac{1}{\sqrt{n}}\sum_{i=1}^n
\int_0^\tau \left\{Z_i-\frac{s_z^{(1)}(t)}{s_z^{(0)}(t)}\right\}dM_{ji}(t)\right)^\top s_z^{(1)}(u)du\right].\nonumber
\een
\end{lemmas}
\newenvironment{newproof16}{\begin{proof}\textsc{\emph{of Lemma \ref{ifQ21ab}.}}}{\end{proof}}
\begin{newproof16}
The nuisance parameter $G_{j}(t)$ is estimated through the following score:
\eqnn
U_2(G_j(t) ; \beta_{j0}, \hat \gamma_j)&=&\frac{1}{n}\sum_{i=1}^n\int_0^{t}\left\{dN_i(t)-Y_i(t)dG_j(t)-Y_i(t)\beta_{j0}A_idt-Y_i(t)\hat \gamma_jZ_idt\right\}\\
&=&\frac{1}{n}\sum_{i=1}^n\int_0^{t}\left[dM_{ji}(t)-Y_i(t)d\left\{G_j(t)-G_{j0}(t)\right\}-Y_i(t)(\hat \gamma_j-\gamma_{j0})^\top Z_idt\right].
\een
Therefore by construction we have:
\eqnn
U_2(G_{j0}(t) ; \beta_{j0}, \hat \gamma_j)&=&U_2(G_{j0}(t) ; \beta_{j0}, \gamma_{j0})-U_2(\hat G_j(t ; \beta_{j0}, \hat \gamma_j); \beta_{j0}, \hat \gamma_j)\\
&=&\frac{1}{n}\sum_{i=1}^n\int_0^tY_i(t)d\left\{\hat G_j(t ; \beta_{j0}, \hat \gamma_j)-G_{j0}(t)\right\},
\een
and therefore, by Assumption B*\ref{if2n} and Lemma \ref{ifQ21aab}:
\be
\sqrt{n}\left\{\hat G_j(t ; \beta_{j0}, \hat \gamma_j)-G_{j0}(t)\right\}&=&\sqrt{n}\int_0^{t}\left\{S_z^{(0)}\right\}^{-1}\frac{1}{n}\sum_{i=1}^n\left[dM_{ji}(t)-Y_i(t)(\hat \gamma_j-\gamma_{j0})^\top Z_idt\right]\\
&=&\int_0^t\left\{s_z^{(0)}(t)\right\}^{-1}\left[\frac{1}{\sqrt{n}}\sum_{i=1}^ndM_{ji}(t)-\sqrt{n}(\hat \gamma_j-\gamma_{j0})^\top s_z^{(1)}(t)dt\right].\nonumber
\ee
Hence, by \eqref{2ifn}, we have:
\be\label{1ifn}
&&\sqrt{n}\left\{\hat G_j(t ; \beta_{j0}, \hat \gamma_j)-G_{j0}(t)\right\}\\&=&\sqrt{n}\int_0^{t}\left\{S_z^{(0)}(u)\right\}^{-1}\frac{1}{n}\sum_{i=1}^n\left[dM_{ji}(u)-Y_i(u)(\hat \gamma_j-\gamma_{j0})^\top Z_idu\right]\nonumber\\
&=&\int_0^t\left\{s_z^{(0)}(u)\right\}^{-1}\left[\frac{1}{\sqrt{n}}\sum_{i=1}^ndM_{ji}(u)\right.\nonumber\\
&&
\left.-\left(\left[\frac{1}{n}\sum_{i=1}^n\int_0^{\tau}\left\{D_i-\frac{s_d^{(1)}(t)}{s_d^{(0)}(t)}\right\}^{\otimes 2}Y_i(t)dt\right]^{-1}\right.\right. \\ &&\left.\left.\times\frac{1}{\sqrt{n}}\sum_{i=1}^n
\int_0^\tau \left\{Z_i-\frac{s_z^{(1)}(t)}{s_z^{(0)}(t)}\right\}dM_{ji}(t)\right)^\top s_z^{(1)}(u)du\right].\nonumber
\ee
\end{newproof16}

\begin{lemmas}\label{ifQ21bb}
Let $\Lambda_j(t,Z)=G_j(t) + \gamma_j^{\top} Z t$ and let  $ \gamma_j$ be estimated using \eqref{tradgamma} of the paper and $G_j(t)$ be estimated using \eqref{estbas2}.
Under Assumption B*\ref{if3n} it holds:
\eqnn
&&\sqrt{n}\left\{\tilde G_j(t ; \beta_{j0}, \hat \gamma_j)-G_{j0}(t)\right\}\\&=&\int_0^t\left\{s_{wz}^{(0)}(u; S^{*}_c,\pi^{*})\right\}^{-1}\left[\frac{1}{\sqrt{n}}\sum_{i=1}^ndM_{ji}(u)-\left(\left[\frac{1}{n}\sum_{i=1}^n\int_0^{\tau}\left\{D_i-\frac{s_d^{(1)}(t)}{s_d^{(0)}(t)}\right\}^{\otimes 2}Y_i(t)dt\right]^{-1}\right.\right.\nonumber
\\&&\left.\left.\frac{1}{\sqrt{n}}\sum_{i=1}^n \int_0^\tau \left\{Z_i-\frac{s_z^{(1)}(t)}{s_z^{(0)}(t)}\right\}dM_{ji}(t)\right)^\top s_{wz}^{(1)}(u; S^{*}_c,\pi^{*})du\right].\nonumber
\een
\end{lemmas}
\newenvironment{newproof17}{\begin{proof}\textsc{\emph{of Lemma \ref{ifQ21bb}.}}}{\end{proof}}

\begin{newproof17}
The proof is similar to the proof of Lemma \ref{ifQ21ab} and we leave it to the reader.
\end{newproof17}

\newpage
\bibliographystyle{biom}
\bibliography{crdr}


\end{document}